\documentclass[10pt,aps,prb,superscriptaddress,reprint,showpacs,amssymb,amsmath,floatfix,citeautoscript,longbibliography]{revtex4-2}

\usepackage{enumerate}
\usepackage{dcolumn}
 \usepackage{rotating}

\usepackage{graphicx} 
\usepackage{bm} 
\usepackage{amsmath,amssymb,amsthm,mathrsfs,amsfonts,dsfont}
\usepackage{braket}
\usepackage[plainpages=false,colorlinks=true,linkcolor=blue,citecolor=blue,anchorcolor=blue,urlcolor=blue ]{hyperref}

\begin{document}

\title{Investigation of metamagnetism and crystal-field splitting in pseudo-hexagonal CeRh$_3$Si$_2$}

\author{Andrea~Amorese}
\email{Andrea.Amorese@cpfs.mpg.de}
    \affiliation{Max Planck Institute for Chemical Physics of Solids, N{\"o}thnitzer Str. 40, 01187 Dresden, Germany}
\author{Dmitry~Khalyavin}  
     \affiliation{ISIS Facility, Rutherford Appleton Laboratory, Chilton, Didcot Oxon OX11 0QX, United Kingdom}
\author{Kurt~Kummer}
  \affiliation{European Synchrotron Radiation Facility, 71 Avenue des Martyrs, CS40220, F-38043 Grenoble Cedex 9, France}
\author{Nicholas~B.~Brookes}
  \affiliation{European Synchrotron Radiation Facility, 71 Avenue des Martyrs, CS40220, F-38043 Grenoble Cedex 9, France}

  \author{Clemens~Ritter}
  \affiliation{Institut Laue-Langevin, 71 avenue des Martyrs, F-38042 Grenoble Cedex 9, France}
  \author{Oksana~Zaharko}
  \affiliation{Laboratory for Neutron Scattering and Imaging, Paul Scherrer Institut, CH-5232 Villigen PSI, Switzerland}
  \author{Camilla~Buhl~Larsen}
  \affiliation{Laboratory for Neutron Scattering and Imaging, Paul Scherrer Institut, CH-5232 Villigen PSI, Switzerland}
  \author{Orest Pavlosiuk} 
  \affiliation{Institute of Low Temperature and Structure Research, Polish Academy of Sciences, ul. Ok{\'o}lna 2, 50-422 Wroc{\l}aw, Poland}
  \author{Adam~P.~Pikul}
  \affiliation{Institute of Low Temperature and Structure Research, Polish Academy of Sciences, ul. Ok{\'o}lna 2, 50-422 Wroc{\l}aw, Poland}
  \author{Dariusz~Kaczorowski}
   \affiliation{Institute of Low Temperature and Structure Research, Polish Academy of Sciences, ul. Ok{\'o}lna 2, 50-422 Wroc{\l}aw, Poland}
   \author{Matthias Gutmann}  
     \affiliation{ISIS Facility, Rutherford Appleton Laboratory, Chilton, Didcot Oxon OX11 0QX, United Kingdom}
     \author{Andrew T. Boothroyd}
\affiliation{Department of Physics, Clarendon Laboratory, Oxford University, Oxford OX1 3PU, United Kingdom}
\author{Andrea~Severing}
    \affiliation{Institute of Physics II, University of Cologne, Z\"{u}lpicher Str. 77, D-50937 Cologne, Germany}
\author{Devashibhai~T.~Adroja}
\email{devashibhai.adroja@stfc.ac.uk}
  \affiliation{ISIS Facility, Rutherford Appleton Laboratory, Chilton, Didcot Oxon OX11 0QX, United Kingdom}
  \address{Highly Correlated Matter Research Group, Physics Department, University of Johannesburg, PO Box 524, Auckland Park 2006, South Africa}
\date{\today}

\begin{abstract}
CeRh$_3$Si$_2$ has been reported to exhibit metamagnetic transitions below 5~K, a giant crystal field splitting, and anisotropic magnetic properties from  single crystal magnetization and heat capacity measurements. Here we report results of neutron and x-ray scattering studies of the magnetic structure and crystal-field excitations to further understand the magnetism of this compound. Inelastic neutron scattering (INS) and resonant inelastic x-ray scattering (RIXS) reveal a $J_z$\,=\,1/2 groundstate for Ce when considering the crystallographic $a$ direction as quantization axis, thus explaining the anisotropy of the static susceptibility. Furthermore, we find a total splitting of 78\,meV for the $J$\,=\,5/2 multiplet. The neutron diffraction study in zero field reveals that on cooling from the paramagnetic state, the system first orders at $T_{\text{N}_1}=4.7$\,K in a longitudinal spin density wave with ordered Ce moments along the $b$-axis (i.e. the [0 1 0] crystal direction) and an incommensurate propagation vector $\textbf{k}=(0,0.43,0$). Below the lower-temperature transition $T_{\text{N}_2}=4.48$\,K, the propagation vector locks to the commensurate value $\textbf{k}=(0,0.5,0)$, with a so-called lock-in transition. Our neutron diffraction study in applied magnetic field $H\parallel b$-axis shows a change in the commensurate propagation vector and development of a ferromagnetic component at $H=3$\,kOe, followed by a series of transitions before the fully field-induced ferromagnetic phase  is reached at $H = 7$\,kOe. This explains the nature of the steps previously reported in field-dependent magnetization measurements. A very similar behaviour is also observed for the $H\parallel$ [0 1 1] crystal direction.
\end{abstract}


\maketitle

\section{Introduction}

In cerium-based intermetallic compounds, the interplay between crystal-field effects, hybridization, magnetic ordering and Kondo screening is responsible for many electronic and magnetic properties that continute to challenge our understanding. In particular, compounds of the CeT$_3$X$_2$ family (T = Rh, Ir; X = B, Si) have short Ce-Ce distances and extremely large crystal-field splittings (which gained them the name \textit{giant crystal-field} compounds), which are accompanied by some very unusual magnetic behaviors. 
The best known compound in that series is CeRh$_3$B$_2$, which crystallizes in an hexagonal ($P6/mmm$) structure  characterized by chains of Ce ions along the hexagonal $c$-axis, with extremely short intra-chain distances of 3.09\,\AA~\cite{Lawson}. It orders ferromagnetically (with the cerium moments lying in the hexagonal plane) below $T_\text{C}=115$\,K, a record-high Curie temperature among cerium intermetallics, but the saturation magnetic moment of $0.4\mu_{\rm B}$/formula unit is strongly reduced compared to what would be expected for a Ce$^{3+}$ ion in a hexagonal crystal field~\cite{Dhar, Allen1990, Vija1985}. Tentative explanations of such properties have focused on unusually strong hybridization of Ce 4$f$ electrons with the neighboring Rh $4d$ or Ce $5d$ states \cite{Allen1990}, and/or unusually large crystal-field effect that mixes the $J=\frac{5}{2}$ and $J=\frac{7}{2}$ multiplets~\cite{Galatanu}.

\begin{figure}
    \centering
    \includegraphics[width=0.99\columnwidth]{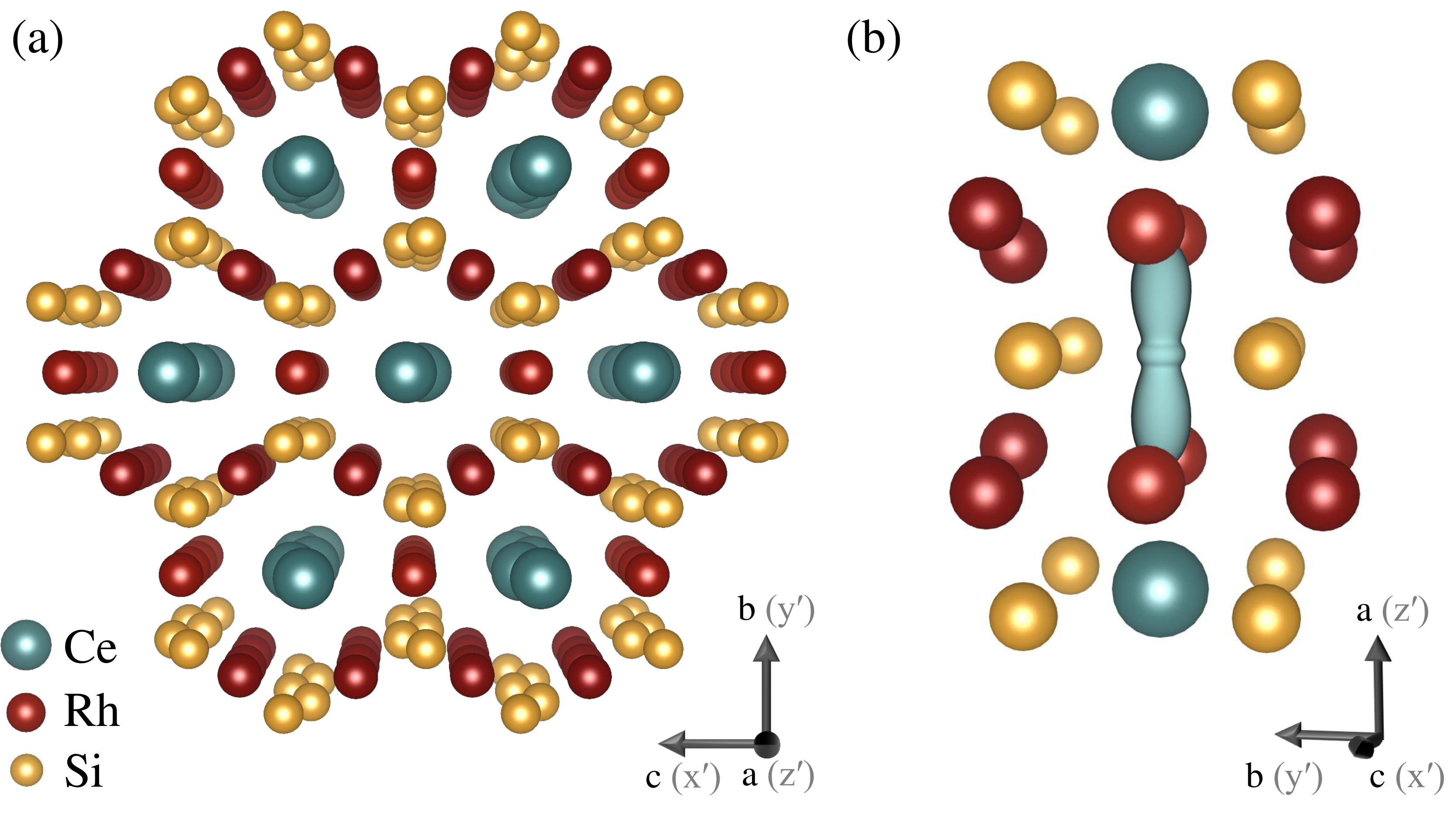}
    \caption{(a) CeRh$_3$Si$_2$ crystal structure, viewed along the orthorhombic $a$ axis, which corresponds to the pseudo-hexagonal $c_h$ axis and is used as the quantization axis $z'$ in the crystal-field analysis. (b) The angular part of the $\ket{\frac{5}{2},\pm\frac{1}{2}}$ groundstate orbital, shown on the Ce site.}
\label{FigStructure}
\end{figure}

The isoelectronic compound CeIr$_3$B$_2$, hexagonal at 473\,K, adopts a slightly distorted monoclinic $C2/m$ structure at room temperature~\cite{Kubota2013}, and is in the valence fluctuating regime, where strong hybridization causes loss of the localized moments~\cite{Yang1984}.

CeRh$_3$Si$_2$ and CeIr$_3$Si$_2$, on the other hand, crystallize in an orthorhombic ($Imma$) structure, which is also a slightly distorted version of that of the hexagonal CeRh$_3$B$_2$, if their crystallographic direction $a$ is taken as the (pseudo-) hexagonal quantization axis (see Fig.~\ref{FigStructure}). Both compounds order antiferromagnetically at low $T<5$\,K, and both show complex behavior that indicates a rearrangement of the moments below $T_\text{N}$ at zero applied magnetic field and a cascade of metamagnetic transitions in applied field.

Given the variety of properties arising from similar crystal structures, the CeT$_3$X$_2$ family is a valuable testing ground to study the interplay between unusually large crystal-field effects, hybridization, and magnetic interactions. In particular, it is important to understand (i) the origin of the \textit{giant} crystal field and its effects on the resulting local Ce states, (ii) the relevant states and mechanisms responsible for the hybridization, and (iii) ultimately the interactions determining the different magnetic structures.

The first step in that direction is the characterization of the electronic states and magnetically ordered phases of Ce. Motivated by this goal, we present microscopic measurements of the $4f$ crystal-field transitions and magnetic structures in CeRh$_3$Si$_2$. Its crystal structure and macroscopic magnetic properties were determined by Pikul \textit{et al.}~\cite{Pikul2010}. It crystallizes in the orthorhombic ErRh$_3$Si$_2$-type structure with lattice constants $a = 7.133$\,\AA, $b = 9.734$\,\AA, and $c = 5.606$\,\AA\ at 293\,K. According to Cenzual \textit{et al.}~\cite{Cenzual1988}, the structure can be treated as a deformed hexagonal structure with lattice constants $a_h$ and $c_h$ defined by the relations: $a = c_h$, $b = \sqrt{3}a_h$, and $c = a_h$. Figure~\ref{FigStructure} shows that the Ce ions in CeRh$_3$Si$_2$ are surrounded by a distorted hexagonal environment with the hexagonal axis perpendicular to the $bc$ plane, and with zigzag chains of Ce atoms along the orthorhombic $a$ axis. The interchain Ce--Ce distances (3.58\,\AA) are fairly short among the intermetallic Ce compounds~\cite{Miyahara2018}. The magnetic properties of the system are strongly anisotropic, with moments confined in the orthorhombic $bc$ plane with a small in-plane anisotropy that favors $b$ as the easy axis~\cite{Pikul2010}. The $4f$ electrons are localized, and the effective magnetic moment at high $T$ is close to that calculated for a free Ce$^{3+}$ ion. Analysis of the experimental magnetic susceptibility curves in terms of a crystal-field model suggests splittings as large as 60\,meV for the groundstate $J=\frac{5}{2}$ multiplet, and 240\,meV for the higher $J=\frac{7}{2}$ multiplet.

In zero field, CeRh$_3$Si$_2$ orders antiferromagnetically at $T_{\text{N}_1}=4.7$\,K with a subsequent first-order phase transition at $T_{\text{N}_2}=4.48$\,K, which was ascribed to a rearrangement of the antiferromagnetic structure~\cite{Pikul2010}. In the presence of applied magnetic field, these transitions split and evolve independently with temperature~\cite{Kaczorowski2008}, and in addition a broad ferromagnetic-like hump appears in the susceptibility plots below $T_{\text{N}_2}=4.48$\,K. The complex magnetic behavior manifests itself in the isothermal magnetization curves at $T<T_{\text{N}_2}$ with field in the $bc$ plane as a sequence of distinct steps, which suggests the presence of a cascade of metamagnetic transitions, with a saturation moment of 1.16\,$\mu_{\rm B}$ per Ce ion~\cite{Pikul2010}.   

To fully understand the nature of the magnetic ground state and obtain direct information on the magnetic structure and crystal-field energy levels in  CeRh$_3$Si$_2$ we performed neutron and x-ray scattering experiments.  Neutron diffraction gives insight into the size and direction of the ordered magnetic moment as a function of temperature and magnetic field, and the combination of inelastic neutron scattering (INS) and high-resolution resonant inelastic X-ray scattering (RIXS) provides a complete picture of the crystal-field split multiplet scheme of the Ce$^{3+}$ configuration~\cite{BURLET199410,PhysRevB.97.184422, PhysRevB.88.104421, AmoreseCeRIXS, AmoreseCeRh2Si2,AmoreseCeRu4Sn6, AmoreseSmB6}. The results obtained from the present study will allow direct comparison with the existing data on CeRh$_3$B$_2$~(Ref.~\onlinecite{Givord_2007}) and CeIr$_3$Si$_2$ (Refs.~\onlinecite{Shigetoh,Muro}).

\begin{figure*}
    \centering
    \includegraphics[width=1.9\columnwidth]{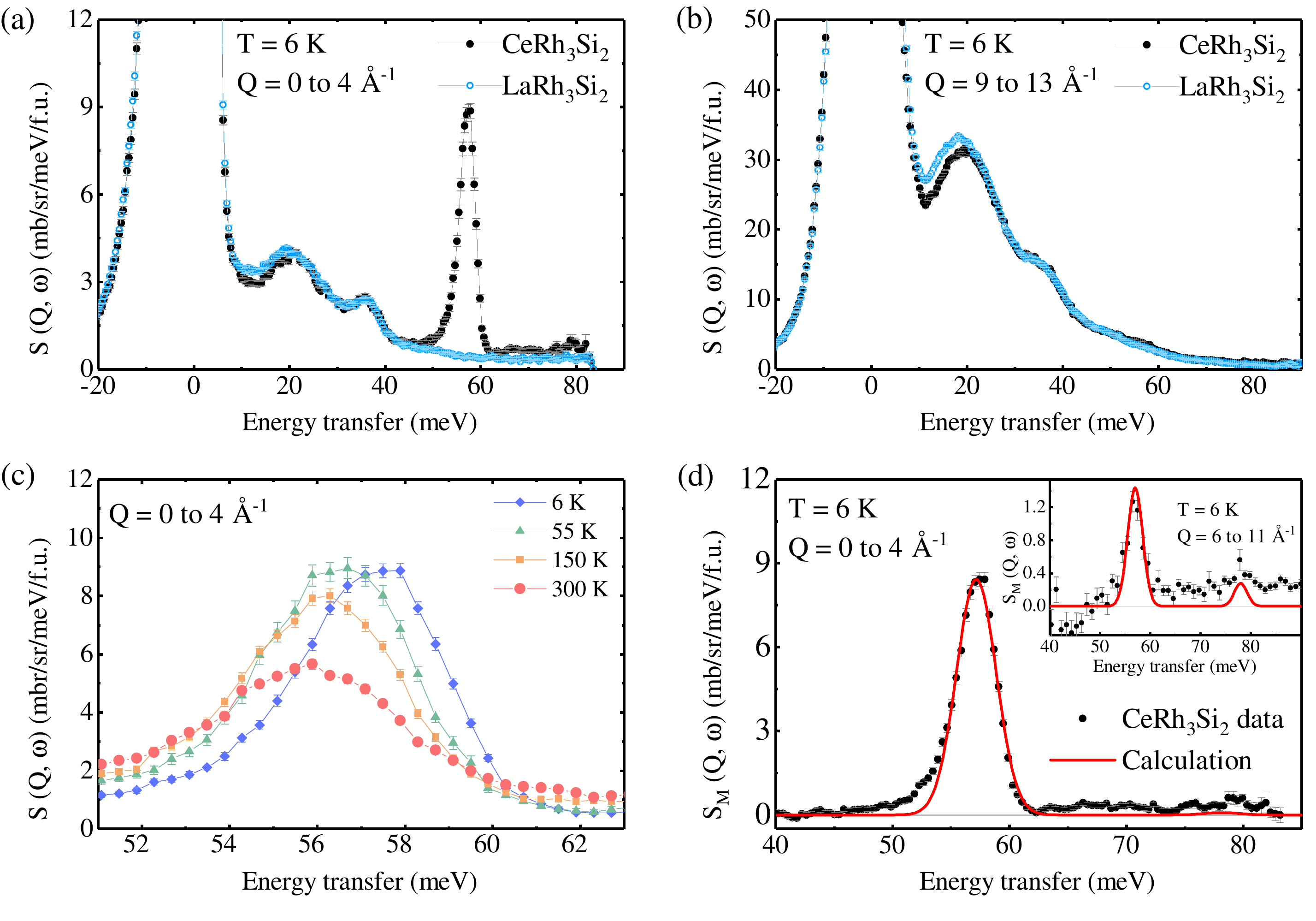}
    \caption{Inelastic neutron scattering data of CeRh$_3$Si$_2$ (black symbols) and LaRh$_3$Si$_2$ (light blue) measured with 100\,meV incident energy at 6\,K with (a) small and (b) large momentum transfers $Q$. (c) Temperature dependence of the crystal-field excitation at 57.5\,meV. (d) Magnetic scattering from CeRh$_3$Si$_2$ at low-$Q$ (panel) and high-$Q$ (inset) obtained after subtraction of the phonon contribution estimated from the spectrum of LaRh$_3$Si$_2$. The red lines show the crystal-field calculation in the pseudo-hexagonal approximation (see below) taking into account the dipole and quadrupole cross-sections.}
\label{INS}
\end{figure*}

\section{Experiment}
\subsection{Sample Preparation}
The polycrystalline samples of CeRh$_3$Si$_2$ and LaRh$_3$Si$_2$ used for inelastic neutron scattering and neutron powder diffraction studies were prepared by conventional arc melting of stoichiometric amounts of the constituents  Ce (3N purity, Ames Laboratory), Rh ingot (3N purity, Chempur), and Si chips (6N purity, Chempur). A single crystal of CeRh$_3$Si$_2$ was grown by the Czochralski pulling method employing a tetra-arc furnace under protective ultrapure argon atmosphere for RIXS and single crystal neutron diffraction. The starting polycrystalline melt was prepared as mentioned above. The pulling rate was 10 mm/h and the copper heart’s rotation speed was 3 rpm. The final ingot of CeRh$_3$Si$_2$ was about 4\,mm in diameter and 40\,mm in length. The single crystal as well as the polycrystalline samples were wrapped in Ta foil, sealed in an evacuated silica tube, and annealed at 900$^\circ$C for 2 weeks.  From this single-crystalline rod, a fragment of about 4\,mm length, 3\,mm width and 2\,mm thickness was cut for neutron diffraction experiments. The quality of the product was verified by means of X-ray powder-diffraction measurements and single crystal Laue diffraction measurements. 

\subsection{Inelastic Neutron Scattering (INS)}

Inelastic neutron scattering measurements on 6.5\,g polycrystalline samples of CeRh$_3$Si$_2$ and LaRh$_3$Si$_2$ were performed on the high flux time-of-flight (TOF) spectrometer MERLIN~\cite{BEWLEY20061029} at the ISIS Facility, UK. The samples were placed in an annular geometry thin-wall Al-can with outer diameter 40\,mm and height 40\,mm. A closed-cycle refrigerator (CCR) with helium exchange gas was used to cool the samples down to 5\,K. A sloppy chopper was used to select the incident neutron energies of $E_{\rm i} = 100$\,meV with 350\,Hz frequency, and 600\,meV with 600\,Hz frequency. The elastic resolution (FWHM) for $E_{\rm i} = 100$\,meV is $\Delta \hbar\omega = 6.5$\,meV (and $\Delta \hbar\omega = 3.3$\,meV at $\hbar\omega = 57.0$\,meV energy transfer) and for $E_{\rm i} = 600$\,meV is $\Delta \hbar\omega = 51$\,meV (and $\Delta \hbar\omega =24.0$\,meV at $\hbar\omega = 350$\,meV). We measured a standard vanadium  sample under identical conditions to normalize the samples data in absolute units of  mb/sr/meV/f.u.

\subsection{Resonant Inelastic X-ray Scattering (RIXS)}
The RIXS experiment was performed at the state-of-the-art soft X-ray beamline ID32 at the European Synchrotron Radiation Facility (ESRF) in Grenoble, France~\cite{ID32}. The experiment was performed at the Ce $M_5$ edge detecting the $3d\rightarrow4f\rightarrow 3d$ process, with resonant energy around 882\,eV~\cite{AmoreseCeRIXS}. The combined beamline resolution during the experiment varied between 30\,meV and 37\,meV and was monitored before and after each acquisition by measuring the (non-resonant) elastic signal from conductive carbon adhesive tape placed close to the sample. The carbon tape elastic signal was also used to determine the zero energy loss position for each spectrum, with an accuracy of $\pm 2$\,meV. The spectra were acquired in the single photon centroiding mode and treated with the RIXSToolBox~\cite{RixsToolBox}; the energy step size was set to 6.2 meV, corresponding to splitting each 15\,$\mu$m-pixel of the Andor iKon-L CCD detector in 2.7 points, a value tuned to maximize the performance of the centroiding algorithm~\cite{AmoreseCCD1,AmoreseCCD2}. The scattering geometry, the incident photon polarization, and the incident photon energies were varied for each acquisition to gain more information by exploiting the cross-section dependence of the different RIXS excitations. The incident photon energy was fixed at the central energy of the Ce M$_5$ edge ($E_{\rm c}=882.2$\,eV) for most acquisitions, and some spectra (see Appendix) were measured with 1\,eV higher photon energies, halfway down the high energy tail. A total of 22 RIXS spectra were acquired with various experimental settings on three samples. The samples were post-cleaved in the cleaving chamber ($10^{-8}$\,mbar) shortly before being transferred to the main measuring chamber with a base pressure of $\approx 10^{-9}$\,mbar. 18 spectra were acquired at the sample temperature of 20\,K, and 4 more at 300\,K, to search for temperature dependencies of the features and to search for excitations from thermally populated low-lying excited states. The full set of acquired spectra is shown in the Supplemental Material \cite{SuppMat}. 

\begin{figure*}
    \centering
    \includegraphics[width=0.49\textwidth]{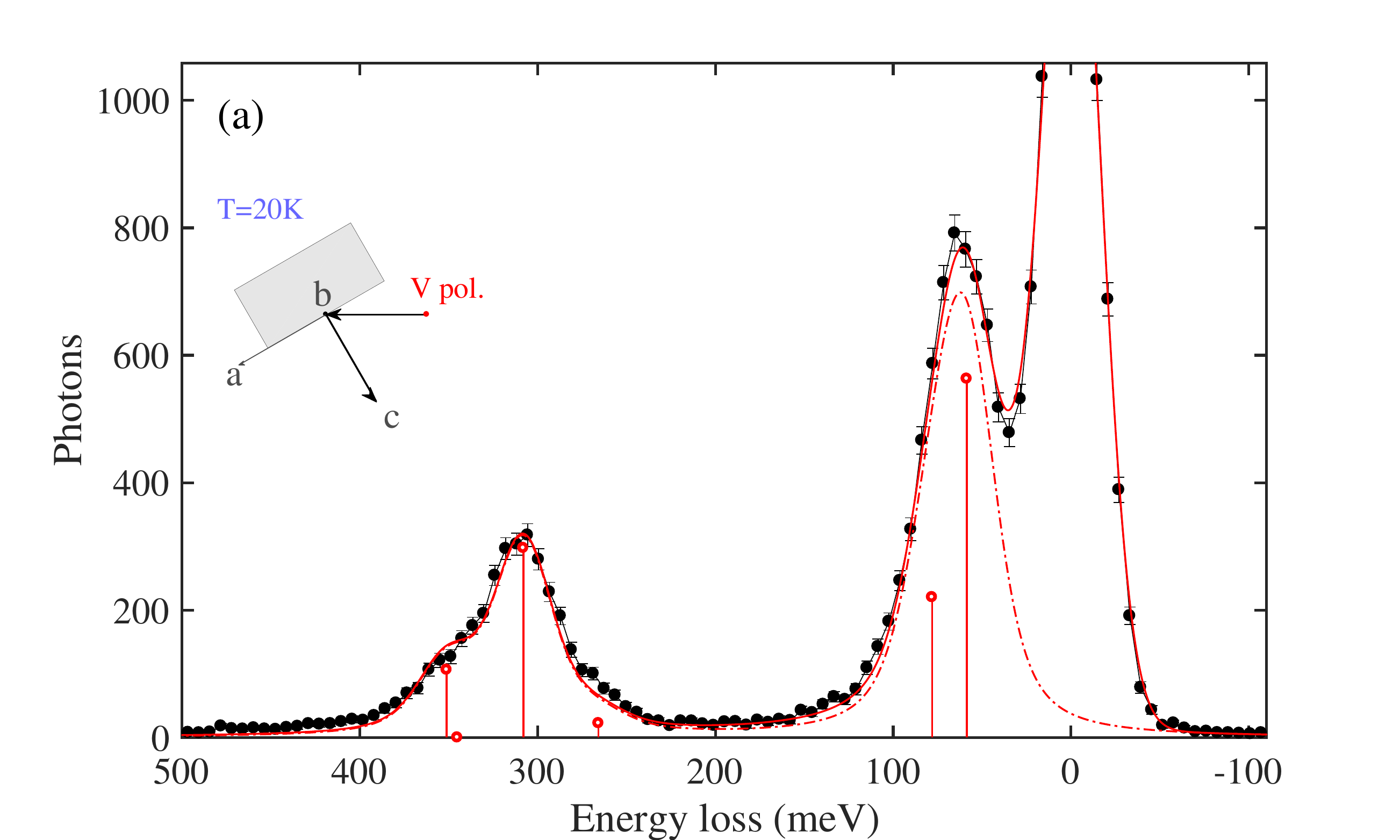} \put(-84.5,77.5){\llap{\includegraphics[width=0.32\columnwidth]{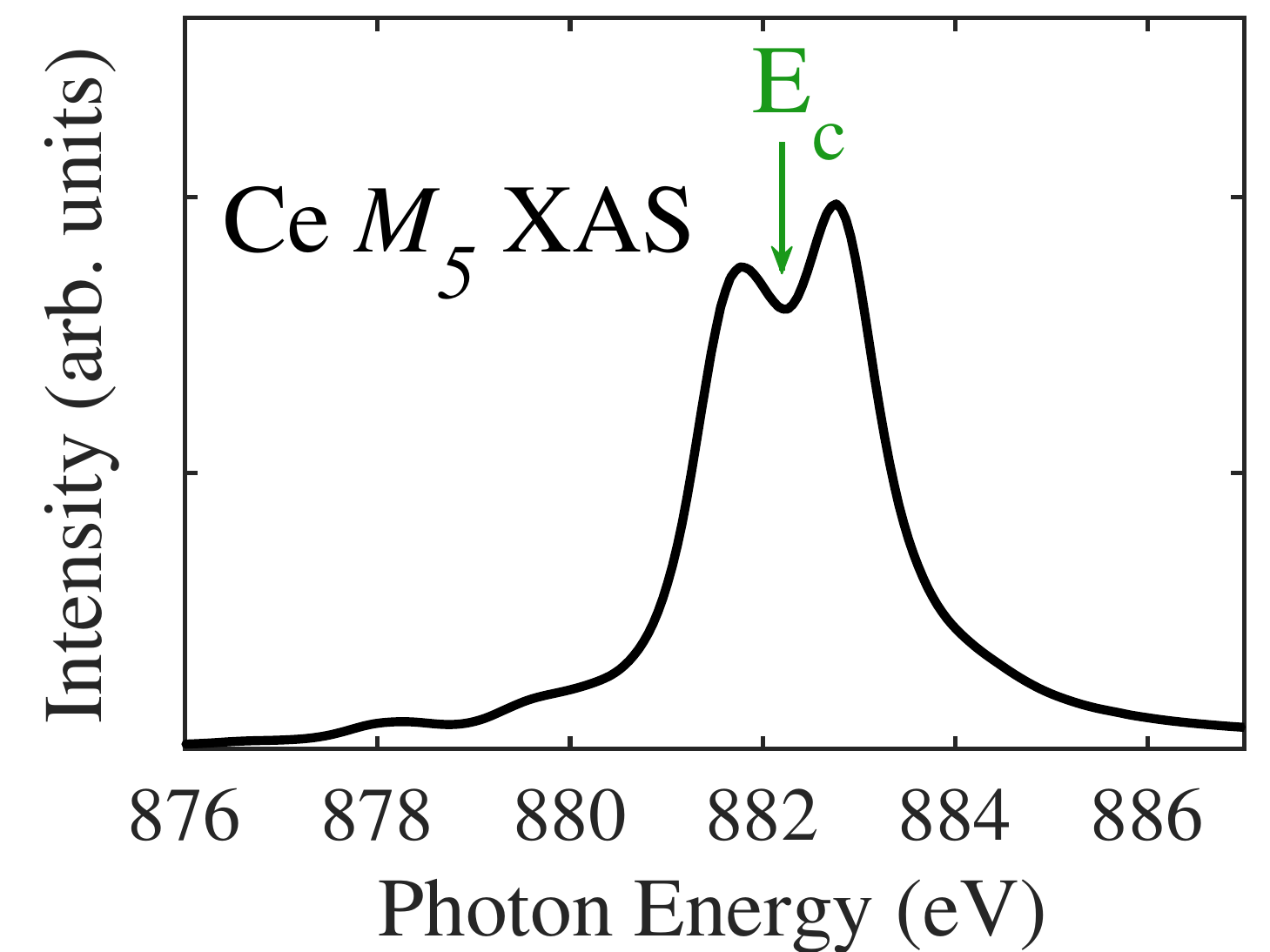}}}
    \includegraphics[width=0.49\textwidth]{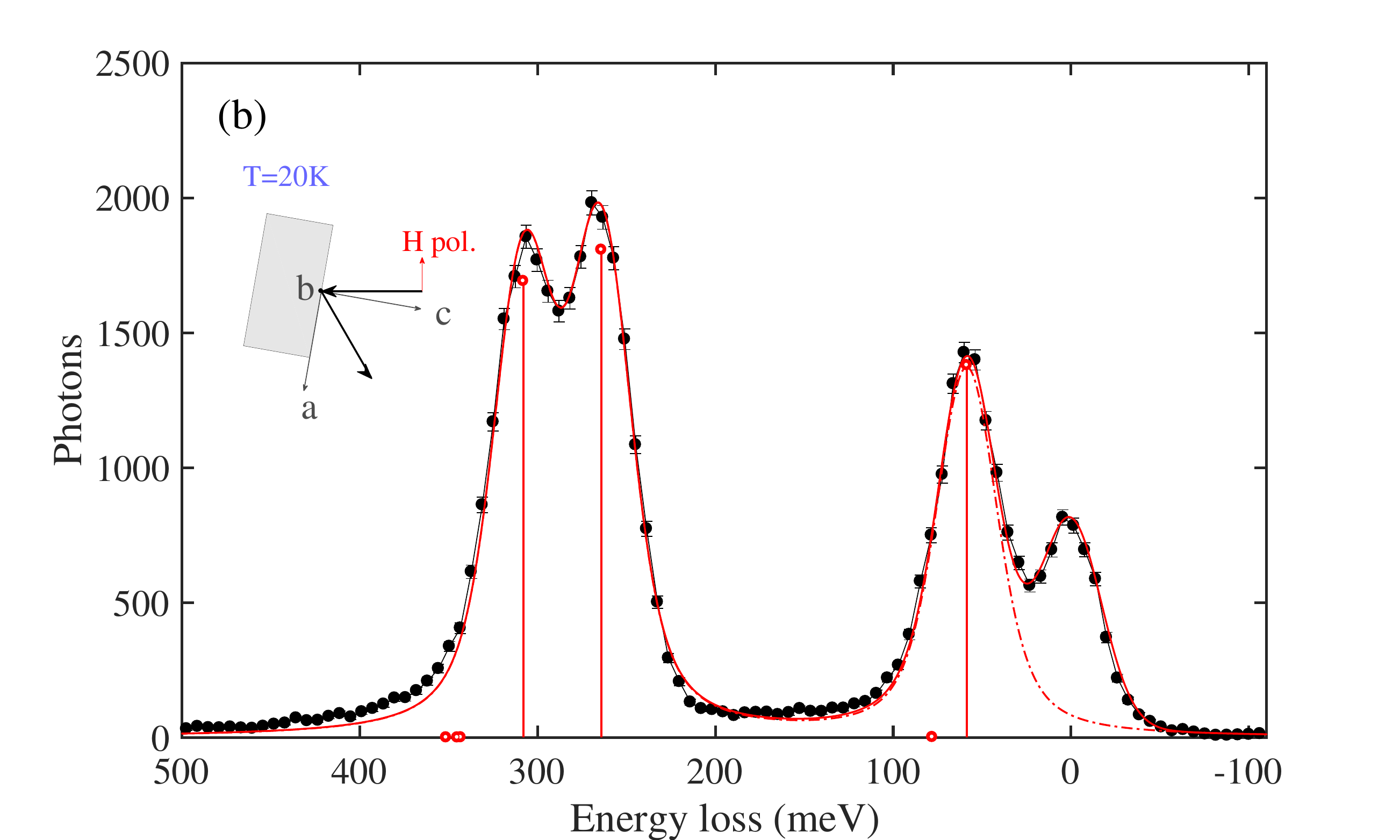}\\
    \includegraphics[width=0.49\textwidth]{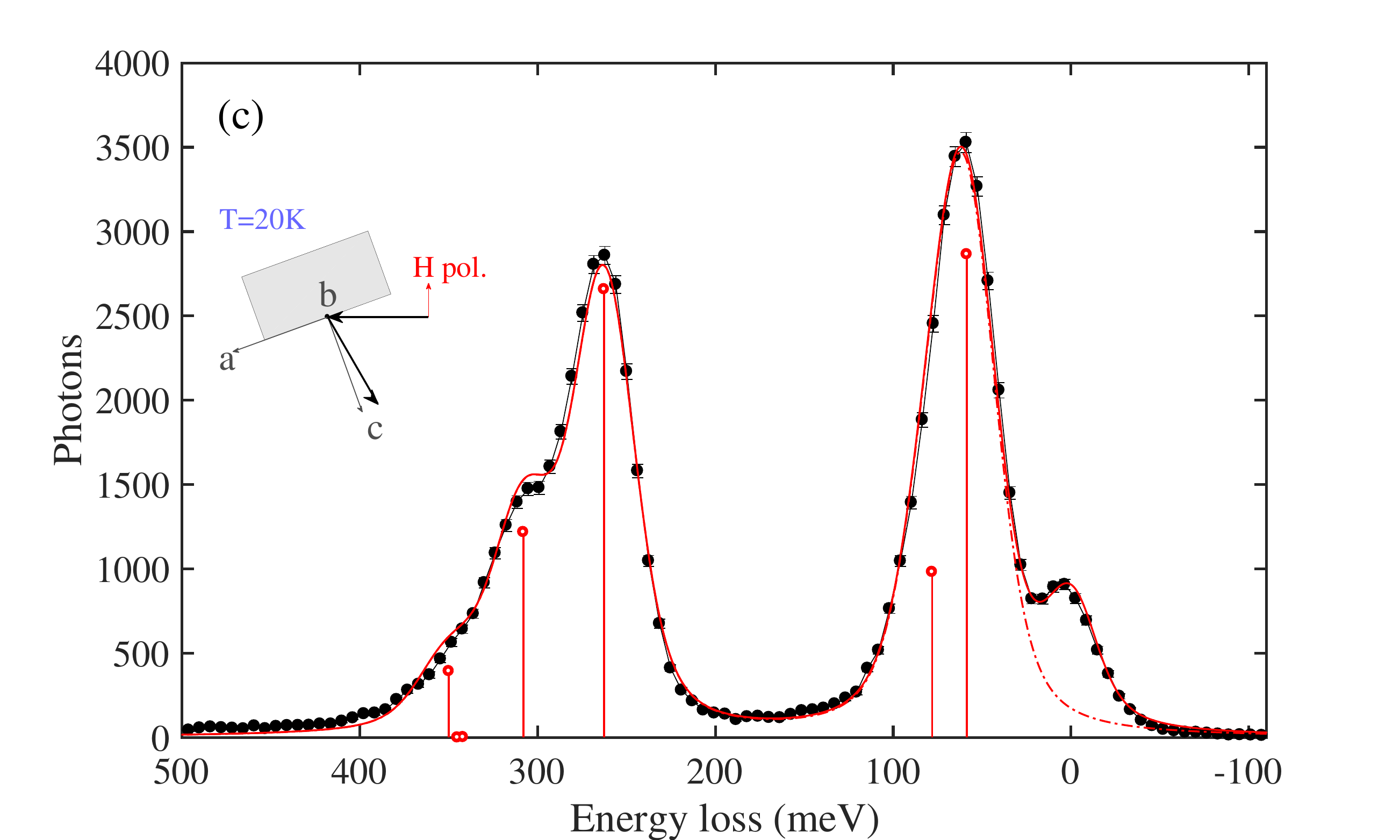}
    \includegraphics[width=0.49\textwidth]{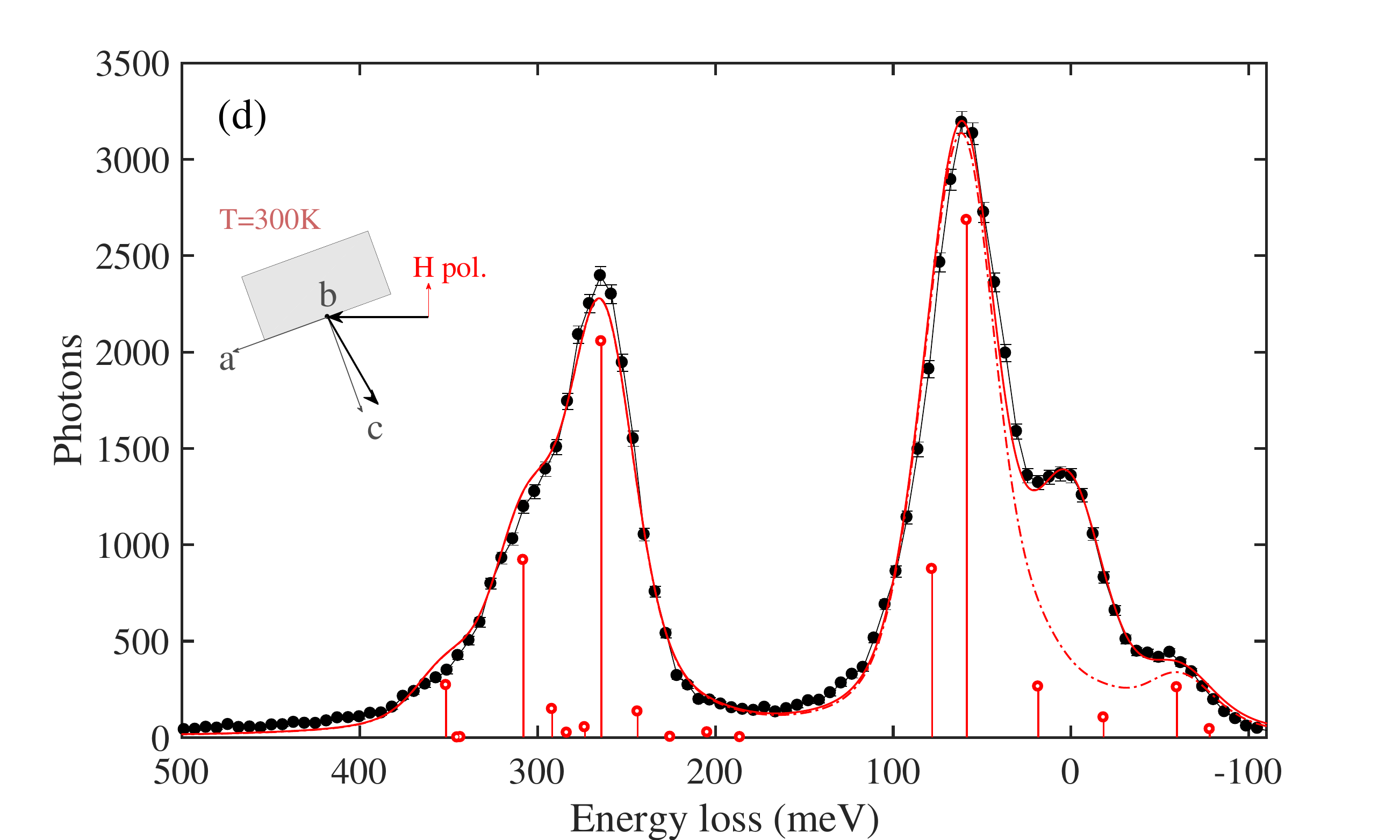}
    \caption{Experimental RIXS spectra (black) acquired with different experimental configurations, (a)--(c) at 20\,K, and (d) at 300\,K. Spectra (c) and (d) correspond to the same geometry. Inset in (a): Isotropic Ce $M_5$ spectrum, showing the incident photon energy used for RIXS. The red dots indicate the transition intensities resulting from the crystal-field analysis, and the full red line shows the calculated spectrum considering the present experimental resolution. The elastic intensity at 0 eV, not obtainable with simulations, was adjusted to fit the spectra.}
\label{FigRIXS}
\end{figure*}

\subsection{Neutron Diffraction (ND)}
Neutron powder diffraction measurements in zero field and neutron single crystal diffraction measurements in applied magnetic field on CeRh$_3$Si$_2$ were carried out using the time-of-flight (TOF) neutron diffractometer WISH~\cite{doi:10.1080/10448632.2011.569650} at the ISIS Facility, UK. In order to reduce the absorption from Rh, an annular vanadium can of 8 mm outer diameter was used. 6.5\,g of powder sample was placed in the annular can with thickness of 1 mm. The sample was cooled to a base temperature of 1.5\,K using a so-called orange cryostat under helium exchange gas to thermalize the sample. Data were collected at several temperatures between 1.5 and 6\,K. Rietveld refinements of the crystal and magnetic structures were performed with the Fullprof program~\cite{FullProf} on the data measured in detector banks at average $2\theta$ values of $58^\circ$, $90^\circ$, $122^\circ$ and $154^\circ$, each covering $32^\circ$ of the scattering plane. 

For single crystal neutron diffraction measurements on the WISH diffractometer in applied magnetic field, the CeRh$_3$Si$_2$ crystal (with dimension of 4\,mm $\times$ 3\,mm $\times$ 2\,mm and a mass of 90\,mg) was mounted on an aluminium holder with the $b$-axis vertical and the field was applied parallel to the $b$-axis. The single crystal data at 1.5\,K  were collected with 5 orientations of the $a$-axis of the crystal with respect to the incident beam and then projected in the reciprocal plane to identify the magnetic Bragg peaks. We collected data with applied magnetic fields of 0, 3, 4, 6 and 7\,kOe. Further, we performed field-dependent neutron diffraction measurements with magnetic fields applied parallel to the [0 1 1] crystal direction (i.e. field in the $bc$-plane) using the thermal neutron single crystal instrument ZEBRA at SINQ, PSI, Switzerland. The experiment was performed with neutron wavelength 2.3\,\AA\ in the normal beam geometry. We also performed neutron powder diffraction measurements between 2 and 300\,K with wavelength 2.41\,\AA\ on the D20 diffractometer at Institut Laue--Langevin, Grenoble, to investigate the temperature dependence of the lattice parameters of CeRh$_3$Si$_2$ (Appendix\,C).

\subsection{Crystal-field model used for the analysis}
The spin--orbit interaction splits the 14-fold degenerate energy levels of Ce$^{3+}$ $f^1$ configuration into the sixfold degenerate Hund's rule ground state $^2F_\frac{5}{2}$ and the eight-fold degenerate $^2F_\frac{7}{2}$ multiplet at about 280\,meV above the ground state. The crystal field lifts the remaining degeneracies further  by splitting the ground state multiplet into three and the excited multiplet into four Kramers doublets. Since the crystal-field potential must reflect the point symmetry of the Ce ion site, it is convenient to express the crystal-field Hamiltonian as a sum of tensor operators $C_n^m$ which transform in the same way as the (renormalized) spherical harmonics, $C_n^m(\theta,\phi)=\sqrt{\frac{4\pi}{2n+1}}Y_n^m(\theta,\phi)$, with parameters $  A_n^m$, that are coefficients of $C_n^{-m}  + (-1)^m C_n^{+m}$ and have the same normalization as that used by  Wybourne in Ref.~\onlinecite{Wybourne1965} (see also Refs.~\onlinecite{McPhase,Boothroyd} for more information about the conversion between different crystal field conventions). For the orthorhombic C$_{2v}$ point symmetry of Ce in CeRh$_3$Si$_2$, nine parameters ($  A_2^0$, $  A_2^2$, $  A_4^0$, $  A_4^2$, $  A_4^4$, $  A_6^0$, $  A_6^2$, $  A_6^4$ and $  A_6^6$) have to be determined experimentally, and the resulting crystal-field wave functions are given by linear combinations of the $\ket{J,J_z}$ components that fulfill $\Delta J_z$\,=\,2 and $\Delta J_z$\,=\,4.

The crystal field analysis can be simplified considerably when treating the structure as pseudo-hexagonal, as described above (see Fig.\,\ref{FigStructure}). Treating the CeRh$_3$Si$_2$ as an hexagonal system implies that the $a$ axis, with its pseudo-sixfold rotational symmetry, is used as the quantization axis. We, therefore, express the new coordinate system as $(x',y',z')=(-z, y, x)$ (where the minus can be neglected thanks to the mirror symmetry about the hexagonal plane) in order to avoid confusion with the usual reference system that uses the $c$ axis as quantization axis (e.g.~Ref.~\onlinecite{Pikul2010}). The hexagonal crystal field only mixes $\ket{J,J_{z'}}$ basis functions that fulfill $\Delta J_{z'} = 6$, so that the eigenfunctions are the pure $J_{z'}$ states $\ket{\frac{5}{2},\pm\frac{1}{2}}$, $\ket{\frac{5}{2},\pm\frac{3}{2}}$, $\ket{\frac{5}{2},\pm\frac{5}{2}}$ for the lowest $^2F_\frac{5}{2}$ multiplet, and  $\ket{\frac{7}{2},\pm\frac{1}{2}}$, $\ket{\frac{7}{2},\pm\frac{3}{2}}$, $\alpha \ket{\frac{7}{2},\pm\frac{5}{2}} + \beta \ket{\frac{7}{2},\mp\frac{7}{2}}$ and $\beta \ket{\frac{7}{2},\pm\frac{5}{2}} - \alpha \ket{\frac{7}{2},\mp\frac{7}{2}}$  for the excited $^2F_\frac{7}{2}$ multiplet. The active crystal-field parameters are reduced to $  A_2^0$, $  A_4^0$, $  A_6^0$, and $  A_6^6$. The $  A_6^6$ parameter determines the $\alpha$ parameter of the mixed $J_{z'}$ states and is known to have a minor effect on the splitting scheme and on the measurable properties so that it can often be neglected. This further reduces the number of crystal-field parameters to 3. We used inelastic neutron scattering (INS) and resonant inelastic x-ray scattering (RIXS) to determine the $  A_n^m$ parameters experimentally. \\

\section{Results}
\subsection{Crystal-Field Scheme}

Figure~\ref{INS} shows the INS data of CeRh$_3$Si$_2$ and of the non-magnetic reference sample LaRh$_3$Si$_2$, taken just above the magnetic ordering transition. The spectrum of CeRh$_3$Si$_3$ at small momentum transfers, Fig.~\ref{INS}(a), exhibits one sharp excitation at 57.5\,meV, which is not visible in the La data nor at large momentum transfer, Fig.~\ref{INS}(b), and must therefore be attributed to a crystal-field excitation. The narrow line width ($\approx3.87$\,meV) is resolution-limited and in agreement with the localized character of the $f$ electrons in CeRh$_3$Si$_2$. Fig.\,\ref{INS}(c) shows the temperature dependence of that crystal-field excitation: its intensity decreases, the excitation broadens, and its energy moves towards 55.5\,meV as the temperature is increased up to 300\,K. These changes are reversed when cooling the sample back down to 6\,K. Figure~\ref{INS}(d) shows the data in Fig.~\ref{INS}(a) after subtracting the phonon and background scattering, both estimated from the non-magnetic reference compound LaRh$_3$Si$_2$. There is tentative evidence for a magnetic peak near 78\,meV (see inset to Fig.~\ref{INS}(d)), and also for a magnetic signal between 250 and 450\,meV where transitions between the $J=5/2$ and $J=7/2$ multiplets are expected (see Appendix\,D). These weak signals in the INS data have insufficient strength to be interpreted conclusively. This is due to limitations of the INS cross-section, as discussed below and in Appendix\,D.

RIXS, which has recently proved its ability to overcome the cross-section limitations of INS~\cite{AmoreseCeRu4Sn6} can measure the  crystal-field splitting scheme of both the $J=5/2$ and the $J=7/2$ multiplets of Ce$^{3+}$ multiplets~\cite{AmoreseCeRh2Si2,AmoreseCeRu4Sn6}, provided the splittings are comparable or larger than the experimental resolution that currently amounts to about 30\,meV at state-of-the-art beamlines. A selection of three RIXS spectra acquired at sample temperature of 20\,K with various scattering geometries and photon polarizations are shown in Fig.\,\ref{FigRIXS}\,(a)--(c) (black dots). Excitations from the ground state at zero energy transfer into both multiplets are clearly visible: at first glance one peak at about 60\,meV, and more than one excitation around 300\,meV. Fig.\,\ref{FigRIXS}\,(d) shows a spectrum acquired in the same geometry as that in (c), but at the higher temperature of 300\,K. 

The hexagonal crystal-field potential splits the $^2F_\frac{5}{2}$ multiplet into three doublets, one of which will be the ground state, and the $^2F_\frac{7}{2}$ multiplet, centered at about 300\,meV, into four doublets, hence we expect to see two excitations from the ground state into the low energy multiplet, and four into the higher multiplet. After fitting Voigt profiles to the 18 RIXS spectra measured at 20\,K, we found that the transition energies cluster at 59\,meV, 78\,meV, 265\,meV, 308\,meV and 353\,meV (see Appendix\,A and Fig.~\ref{FigSFits} for details). We could not find the sixth transition, which must either be buried under another close-lying excitation or it must have a next-to-zero cross-section. Comparing the spectrum at 300\,K with the corresponding one at 20\,K [Fig.~\ref{FigRIXS}\,(c) and\,(d)] only minor differences are visible in the positive energy loss side, while an anti-Stokes peak appears at about 59\,meV negative energy loss. This confirms that the state at 59\,meV is the first excited crystal-field state because, if present, the lower lying states would have been significantly populated at 300\,K, thus, leading to additional excitations from the excited states. In RIXS we cannot study the temperature dependence of the excitation energies observed in our INS data due to the experimental accuracy of about $\pm2$\,meV on the determination of the peak positions.

Having determined the crystal-field transition energies, we now describe the RIXS spectra with a full multiplet RIXS calculation using the Quanty code~\cite{Haverkort2012}, including the hexagonal crystal-field approximation. Details about the calculations (reduction factors, optimization of the crystal-field parameters, etc.) can be found in Appendix\,B. The following crystal-field parameters best describe the majority of the RIXS spectra:  $  A_2^0=-135$\,meV, $  A_4^0=-172$\,meV, $  A_6^0=30$\,meV and $  A_6^6=0$\,meV, with a spin-obit constant $\zeta_{SO}= 76.56$\,meV, obtained by reducing the Hartree-Fock atomic parameter to 88\% of its value. The resulting crystal-field energy levels and wave functions are listed in Table~\ref{tab:Table1}. The ground state is a virtually pure $\ket {J_{z'}=\frac{1}{2}}$, followed by the almost pure states $\ket{J_{z'}=\frac{3}{2}}$ and $\ket{J_{z'}=\frac{5}{2}}$. The mixing between the two $J=\frac{5}{2}$ and $J=\frac{7}{2}$ multiplets was found to be very small. For example, the squared amplitudes of the $\ket{\frac{7}{2},J_{z'}}$ components in the first three crystal-field wave functions amount to 0.0018 for the ground state, 0.02 for the first excited state and 0.00015 for the second excited state, respectively (see Table~IV in Appendix~B). We recall that the notation $J_{z'}$  indicates that the pseudo-hexagonal axis $z'\parallel a$ is the quantization axis used to express the crystal field, while previous literature as Ref. \cite{Pikul2010} expressed the crystal field scheme along $z\parallel c$ (see Fig.~\ref{FigStructure} and Supplemental Material \cite{SuppMat} for further details). The vertical lines in Figs.~\ref{FigRIXS}\,(a)--(c) represent the peak position and intensities obtained from the crystal-field fits, and the red lines show the calculated crystal-field excitation spectra convoluted with the Voigt lineshape. The expected excitation at 345\,meV corresponds to the transition  $\ket{\frac{5}{2},\pm\frac{1}{2}}$\,$\rightarrow$\,$\ket{\frac{7}{2},\pm\frac{7}{2}}$ i.e. it is forbidden by the RIXS selection rules $\Delta J_{z'}=0,\pm1,\pm2$ and therefore not observable. The elastic intensity (i.e. the difference between the full red line and the dashed red line), being affected by other parameters not included in the calculation (such as the surface roughness), was tuned to best fit the data. The effect of the surface roughness can be clearly seen as an increase in the elastic intensity in the spectrum in Fig.~\ref{FigRIXS}\,(a), which was acquired on a cleaved surface different from that of the spectra in the other panels. The same crystal-field model used to fit the spectra at 20\,K also describes very well the data at 300\,K, including the anti-Stokes intensity, further confirming the validity of our analysis, see Fig.~\ref{FigRIXS}\,(d). 

The proposed crystal-field scheme (Table~\ref{tab:Table1}) also agrees well with the INS data. The red line in Fig.~\ref{INS}(d) shows the calculated INS cross-section including both dipole and quadrupole contributions \cite{ROTTER2004E481,SPECTRE}, using the same crystal-field scheme obtained by the RIXS analysis.
The calculation reproduces the shape of the spectrum, with only one clearly visible peak at 58\,meV, but the calculated intensity had to be scaled down by a factor of 2.5 to be comparable with the measured spectrum. This discrepancy in the intensity is due to sample attenuation effects, caused mainly by the relatively large neutron absorption by Rh. 
 
The state at 78\,meV has a negligible neutron cross-section at low-$Q$ because it corresponds to a $\ket{J=\frac{5}{2},J_{z'\parallel a}=\pm\frac{1}{2}}\rightarrow\ket{J=\frac{5}{2},J_{z'\parallel a}=\pm\frac{5}{2}}$ transition for the assumed hexagonal point symmetry, and such a transition is dipole-forbidden~\cite{Zaliznyak}. Therefore, the small but non-zero signal at energies around 78\,meV in the low-$Q$ INS spectrum [see Fig.~\ref{INS}(d)] must be either due to the imperfect subtraction of the non-magnetic (phonon) background, or because the true point-symmetry is orthorhombic, and therefore some $J_{z'}$ mixing of the states could connect this excitation to the ground state via the dipole interaction. At the same time, the low intensity at 78\,meV confirms the validity of the hexagonal approximation for the crystal-field analysis, meaning that the orthorhombic $J_{z'}$ mixing, if present, is small. To further compare with the RIXS crystal-field scheme, we show INS data at larger momentum transfer [see the inset to Fig.~\ref{INS}(d)], at which quadrupole transitions are expected to gain intensity in INS. The 78\,meV transition appears in the beyond-dipole INS simulations (red line in the inset), while it is just barely detectable in the experimental data due to the relatively large noise and background (see Appendix~D for more details).

\begin{table}[]
\def\arraystretch{1.8}
\caption{Calculated crystal-field eigenenergies and eigenfunctions in the hexagonal crystal-field approximation with quantization axis  $z'$\,$\parallel$\,$c_h$\,$\parallel$\,$a$.  The state at 345\,meV is in parenthesis because it has zero or negligible intensity in the RIXS calculations and is not visible in the experimental spectra.}
\begin{tabular}{c|c}
\hline 
\rule[-1ex]{0pt}{2.5ex} $\ket{J,J_{z'}}$                   							& Energy (meV)    \\ 
\hline\hline	 
\rule[-1ex]{0pt}{2.5ex} $\ket{\frac{5}{2},\pm\frac{1}{2}}$   						& 0 			\\
\hline                                                                                     
\rule[-1ex]{0pt}{2.5ex} $\ket{\frac{5}{2},\pm\frac{3}{2}}$ 							& 59			\\ 
\hline                                                                                                                    
\rule[-1ex]{0pt}{2.5ex} $\ket{\frac{5}{2},\pm\frac{5}{2}}$   						& 78			\\                                        
\hline                                                                               
\rule[-1ex]{0pt}{2.5ex} $\ket{\frac{7}{2},\pm\frac{1}{2}}$   						& 265 		\\
\hline                                                                             
\rule[-1ex]{0pt}{2.5ex} $\ket{\frac{7}{2},\pm \frac{3}{2}}$							& 308     \\ 
\hline                                                                                                                    
\rule[-1ex]{0pt}{2.5ex} $\ket{\frac{7}{2},\pm \frac{7}{2}}$							& (345)  \\                                        
\hline                                                                             
\rule[-1ex]{0pt}{2.5ex} $\ket{\frac{7}{2},\pm \frac{5}{2}}$							& 353      \\ 
\hline                                                
\end{tabular}                                         
\label{tab:Table1}
\end{table}

\subsection{Magnetic Structure}
\begin{figure}
    \centering
    \includegraphics[width=1.0\columnwidth, trim=80 200 45 40, clip]{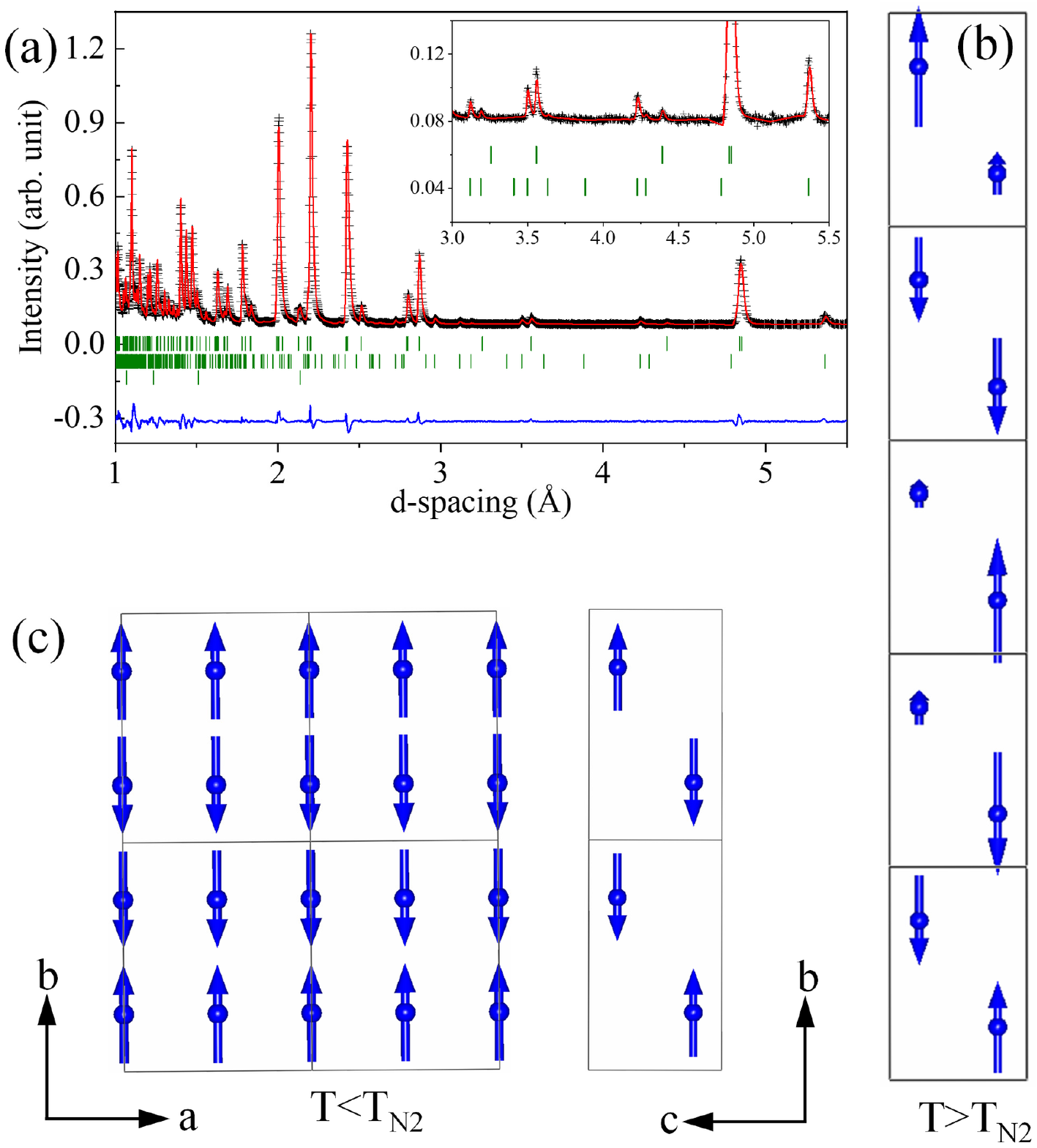}        
    \caption{(a) Rietveld refinement of the neutron diffraction intensity collected on WISH at 1.5\,K from the detector bank with average $2\theta=58.33^{\circ}$. The cross symbols and solid line (red) represent the experimental and calculated intensities, respectively, and the line below (blue) is the difference between them. Tick marks (green) indicate the positions of Bragg peaks: nuclear (top) and magnetic (bottom). The inset shows the specific $d$-spacing range where strong magnetic Bragg peaks are observed. (b) Incommensurate magnetic structure above the lock-in transition. (c) Commenusrate magnetic structure below the lock-in transition. The arrows denote the ordered Ce$^{3+}$ magnetic moment directions.}
\label{FigND_1}
\end{figure}

The powder neutron diffraction patterns collected above $T_{\text{N}_1}=4.7$\,K can be satisfactorily fitted in the orthorhombic structural model proposed by Cenzual, Chabot and Parthe for ErRh$_{3}$Si$_{2}$ \cite{Cenzual1988}. Below $T_{\text{N}_1}$, a set of additional diffraction peaks appears, indicating the onset of a long-range magnetic ordering in agreement with the magnetization and specific heat data \cite{Pikul2010}. The magnetic peaks can be indexed using an incommensurate propagation vector $\mathbf{k}=(0,k_{y},0)$ with $k_{y} \sim 0.439(3)$ at $T=4.5$\,K. A quantitative magnetic structure refinement revealed that the Ce moments are aligned along the $b$-axis forming amplitude modulated longitudinal spin density waves with ferromagnetic coupling along the $a$- and $c$-directions (Fig.\,\ref{FigND_1}(b)). The structure is associated with the two-dimensional $m\Delta_2(\eta,\eta^*)$ irreducible representation of the paramagnetic $Imma$ space group, where $\eta$ and $\eta*$ are components of the order parameter in the representation space. In the present case, the order parameter is the long-range ordering of magnetic moments localized on the Ce sites and described by a Fourier series as specified below. The symmetry of the magnetic structure is described by the $Icmm1'(0,0,\gamma)ss0s$ superspace group with the basis vectors and origin related to the paramagnetic structure as $(0,0,-1,0)$, $(1,0,0,0)$, $(0,-1,0,0)$, $(0,0,0,1)$ and $(-1/4, 1/4, 1/4, 1/4)$, respectively.
Below the second magnetic transition at $T_{\text{N}_2} = 4.48$\,K, the propagation vector locks to the commensurate value $\mathbf{k}=(0,1/2,0)$ (Fig.\,\ref{FigND_1}(a)) giving rise to an equal-moment magnetic structure with up-up-down-down (UUDD) stacking of the ferromagnetic $(ac)$ planes as shown in Fig.~\ref{FigND_1}(c). The moment size refined at $T=1.5$\,K is 1.23(3)\,$\mu_{\rm B}$. This value of the moment along the $b\parallel y'$ axis is in good agreement with the value $1.286\,\mu_{\rm B}$ calculated for a pure $\ket{\frac{5}{2},\pm\frac{1}{2}}$ ground state (the expectation value of the magnetic moments along the $b$ axis in presence of a magnetic or exchange field that aligns the moments along $b \parallel y' \perp z'$  is $\frac{3}{2} g_J$ for a  $\ket{J = \frac{5}{2},J_{z'\parallel a} = \pm\frac{1}{2}}$ ground state doublet). This value is further reduced to about $1.26\,\mu_{\rm B}$ if the small intermultiplet mixing is taken into account. The magnetic space group corresponding to the magnetic structure found is 52.314 $P_anna$. This space group appears with the basis vectors and origin related to the paramagnetic $Imma$ structure as $(0,2,0)$, $(0,0,1)$, $(1,0,0)$ and $(0,1/2,0)$, respectively, in the list of $k=(0,1/2,0)$, generated by the ISODISTORT \cite{ISODISPLACE} and MAXSYM \cite{MAXSYM} software. The low temperature transition from the incommensurate to commensurate magnetic order is a so-called lock-in transition associated with free-energy terms activated at specific values of propagation vectors. In particular, in the present case with the propagation vector along the $\Delta$-line of symmetry of the $Imma$ space group, the free-energy term, $\eta^4 + \eta^{*4}$ is allowed when the $k_{y}$ component takes the commensurate value 1/2. The invariance of this term can be verified using the matrix operators for the generating symmetry elements of the paramagnetic $Imma$ space group, listed in Table~\ref{tab:Table2}. This type of magnetic phase transitions has also been observed in some other Ce-based intermetallic compounds such as CeIrGe$_3$ \cite{Anand2018} and CeRhGe$_3$ \cite{Hillier2012}.

\begin{table*}[]
\def\arraystretch{1.8}
\caption{Matrices of the $m\Delta_2(\eta,\eta^*)$ irreducible representation of the $Imma$ space group, associated with the $\mathbf{k}=(0,1/2,0)$ propagation vector \cite{Aroyo2006}. T is the time-reversal operator. }
\begin{tabular}{c|c|c|c|c|c}
\hline 
\rule[-1ex]{0pt}{2.5ex} $\left \{ 2_z|0,1/2,0\right \}$ & $\left \{ 2_y|0,1/2,0\right \}$ & $\left \{ -1|0,0,0\right \}$ & $\left \{ 1|1/2,1/2,1/2\right \}$ & $\left \{ 1|0,1,0\right \}$ & T    \\ 
\hline\hline	 
\rule[-1ex]{0pt}{2.5ex} $\begin{pmatrix}0 & e^{\frac{3}{2}\pi i} \\ e^{\frac{1}{2}\pi i} & 0 \end{pmatrix}$ & $\begin{pmatrix} e^{\frac{1}{2}\pi i} & 0 \\ 0 & e^{-\frac{1}{2}\pi i} \end{pmatrix}$ & $\begin{pmatrix}0 & 1 \\ 1 & 0 \end{pmatrix}$ & $\begin{pmatrix} e^{\frac{1}{2}\pi i} & 0 \\ 0 & e^{-\frac{1}{2}\pi i} \end{pmatrix}$ & $\begin{pmatrix}-1 & 0 \\ 0 & -1 \end{pmatrix}$ & $\begin{pmatrix}-1 & 0 \\ 0 & -1 \end{pmatrix}$    \\
\hline                                                
\end{tabular}                                         
\label{tab:Table2}
\end{table*}

\begin{figure*}
    \centering
    \includegraphics[width=2\columnwidth, trim=2.9cm 1cm 2.9cm 1cm, clip]{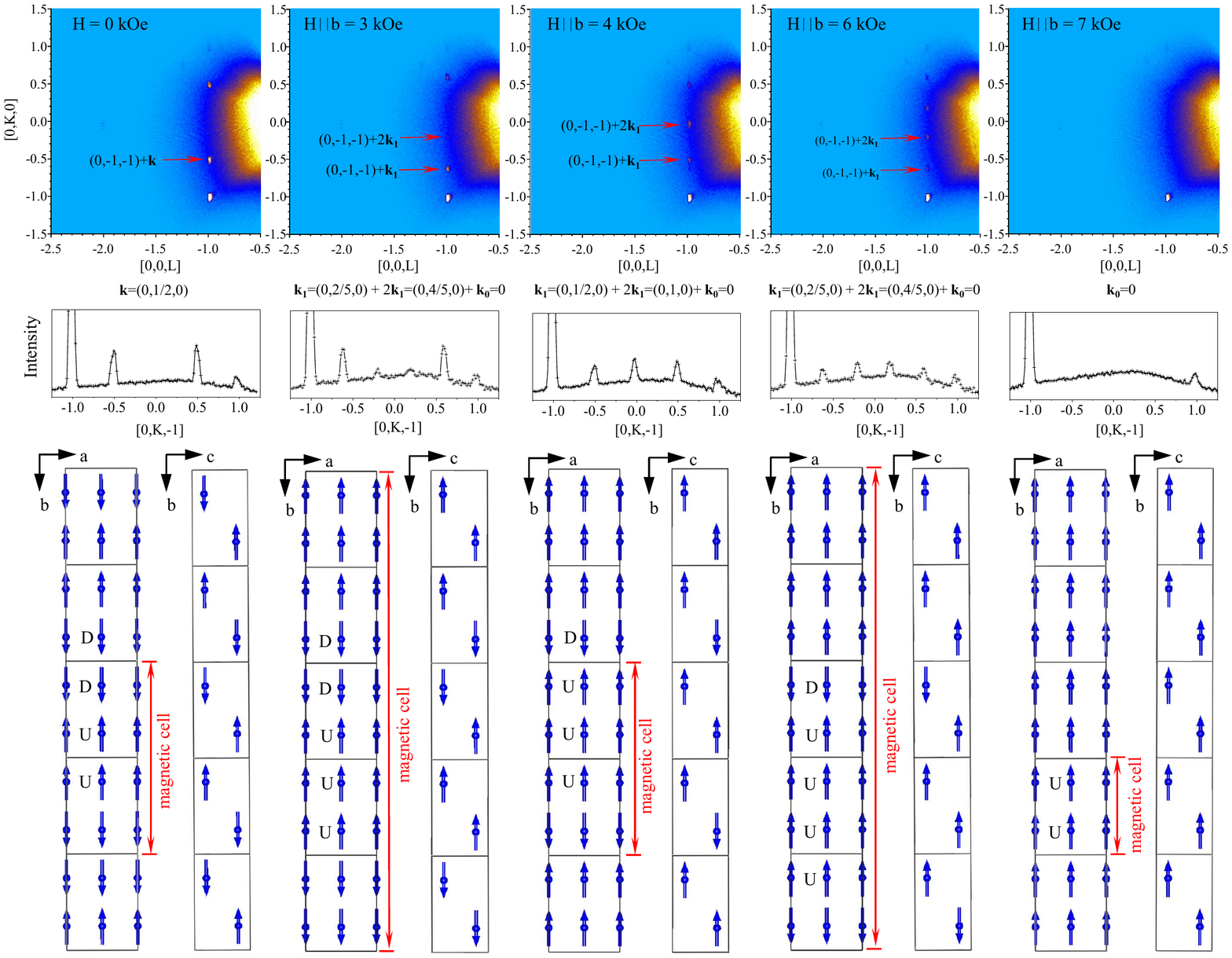}        
    \caption{(top) A portion of the $(0, K, L)$ reciprocal plane measured at 1.5\,K in various magnetic fields applied parallel to the $b$-axis. (middle) 1D cut of the reciprocal plane along the $[0,K,-1]$ direction. (bottom) The corresponding magnetic structures.}
\label{FigNDfield_2}
\end{figure*}

To explore the origin of the magnetization steps reported by Pikul et al.~\cite{Pikul2010}, we performed single-crystal neutron diffraction measurements in magnetic field applied along the $b$-axis. The $(0,k,l)$ reciprocal plane demonstrating the field evolution of magnetic satellites is shown in Fig.~\ref{FigNDfield_2} (top and middle). In agreement with the powder diffraction data, the zero-field magnetic structure is characterized by the $\mathbf{k}=(0,1/2,0)$ propagation vector. The corresponding UUDD magnetic structure with the unit cell twice as big as the nuclear one (see the bottom of Fig.~\ref{FigNDfield_2}) can be presented as a Fourier series $\mathbf{M}_{j\mathbf{t}}=\sum_{\mathbf{k}}\mathbf{S}_{j\mathbf{k}} e^{-2\pi i(\mathbf{k}\cdot\mathbf{t}+\phi_{j\mathbf{k}})}$  with special choice of the magnetic phases $\phi_{j\mathbf{k}}$  as shown in Table\,\ref{tab:Table3}.  The series specifies a magnetic moment, $\mathbf{M}_{j\mathbf{t}}$, localized on $j$-th Ce ion in a unit cell with the lattice translation $\mathbf{t}$. This is the standard way to describe magnetic ordering in crystalline solids \cite{IzumovBook} and it is a convenient way to discuss the field-induced magnetic structures. Application of the magnetic field $H$\,=\,3\,kOe results in a change of the magnetic satellites (Fig.~\ref{FigNDfield_2}). The new satellites reveal the presence of two commensurate propagation vectors $\mathbf{k}_1=(0,2/5,0)$ and $\mathbf{k}_2=2\mathbf{k}_1=(0,4/5,0)$. The corresponding magnetic ordering can be interpreted as a UUUDD structure whose Fourier decomposition involves these two propagation vectors along with the $\mathbf{k}=0$ ferromagnetic component (see Table~\ref{tab:Table3}). Increasing the field up to $H$\,=\,4\,kOe re-establishes the $\mathbf{k}=(0,1/2,0)$ propagation vector but in addition the second harmonic $\mathbf{k}_2=2\mathbf{k}_1=(0,1,0)$, which violates the $I$-centring lattice condition, is also observed. This experimental result reveals another unusual magnetic structure with UUUD stacking of the ferromagnetic layers (Fig.~\ref{FigNDfield_2}). Further increasing the field results in a reappearance of the magnetic satellites with the $\mathbf{k}_1=(0,2/5,0)$ and $\mathbf{k}_2=2\mathbf{k}_1=(0,4/5,0)$ propagation vectors and roughly the same intensity at $H = 6$\,kOe. The Fourier analysis indicates that these two components, combined with equal amplitudes and an additional $\mathbf{k}=0$ ferromagnetic moment, constitute a UUUUD magnetic structure (Table~\ref{tab:Table3}) with the magnetic unit cell five times bigger than the nuclear one (Fig.~\ref{FigNDfield_2}). Finally, increasing the field up to $H = 7$\,kOe results in the disappearance of the magnetic satellites, and only a ferromagnetic contribution to the fundamental nuclear reflections satisfying the $I$-centring condition is observed.

\begin{table*}[]
\def\arraystretch{1.8}
\caption{Fourier decomposition $\mathbf{M}_{j\mathbf{t}}=\sum_{\mathbf{k}}\mathbf{S}_{j\mathbf{k}} e^{-2\pi i(\mathbf{k}\cdot\mathbf{t}+\phi_{j\mathbf{k}})}$  of the magnetic structures in zero and applied magnetic fields, where $\mathbf{S}_{j\mathbf{k}}$ is the amplitude of the Fourier component with propagation vector $\mathbf{k}$, localized on the $j$-th Ce atom $(j=1,2)$,  $\phi_{j\mathbf{k}}$ is the magnetic phase and $\mathbf{t}$ is a lattice translation. }
\begin{tabular}{c|c|c|c}
\hline 
\rule[-1ex]{0pt}{2.5ex} Field (kOe) & Propagation vector & Ce1$(0,1/4,0.78644)$ & Ce2$(0,3/4,0.21356)$  \\ 
\hline\hline	 
\rule[-1ex]{0pt}{1.0ex} 0 & $\mathbf{k}_1=(0,1/2,0)$ & $\mathbf{S}_{1\mathbf{k}_1}=1.75\, \mu_{\rm B}$ & $\mathbf{S}_{2\mathbf{k}_1}=1.75\, \mu_{\rm B}$  \\
\rule[-1ex]{0pt}{1.0ex}  & & $\phi_{1\mathbf{k}_1}=1/4$ & $\phi_{2\mathbf{k}_1}=-1/4$  \\
\hline  
\rule[-1ex]{0pt}{1.0ex} 3 & $\mathbf{k}_1=(0,2/5,0)$ & $\mathbf{S}_{1\mathbf{k}_1}=1.59\, \mu_{\rm B}$ & $\mathbf{S}_{2\mathbf{k}_1}=1.59\, \mu_{\rm B}$  \\
\rule[-1ex]{0pt}{1.0ex}  & & $\phi_{1\mathbf{k}_1}=0$ & $\phi_{2\mathbf{k}_1}=1/5$  \\
\rule[-1ex]{0pt}{1.0ex}  & $\mathbf{k}_2=2\mathbf{k}_1=(0,4/5,0)$ & $\mathbf{S}_{1\mathbf{k}_2}=-0.61\, \mu_{\rm B}$ & $\mathbf{S}_{2\mathbf{k}_2}=-0.61\, \mu_{\rm B}$  \\
\rule[-1ex]{0pt}{1.0ex}  & & $\phi_{1\mathbf{k}_2}=0$ & $\phi_{2\mathbf{k}_2}=2/5$  \\
\rule[-1ex]{0pt}{1.0ex}  & $\mathbf{k}_3=0$ & $\mathbf{S}_{1\mathbf{k}_3}=0.25\, \mu_{\rm B}$ & $\mathbf{S}_{2\mathbf{k}_3}=0.25\, \mu_{\rm B}$  \\
\rule[-1ex]{0pt}{1.0ex}  & & $\phi_{1\mathbf{k}_3}=0$ & $\phi_{2\mathbf{k}_3}=0$  \\
\hline
\rule[-1ex]{0pt}{1.0ex} 4 & $\mathbf{k}_1=(0,1/2,0)$ & $\mathbf{S}_{1\mathbf{k}_1}=1.23\, \mu_{\rm B}$ & $\mathbf{S}_{2\mathbf{k}_1}=1.23\, \mu_{\rm B}$  \\
\rule[-1ex]{0pt}{1.0ex}  & & $\phi_{1\mathbf{k}_1}=0$ & $\phi_{2\mathbf{k}_1}=1/4$  \\
\rule[-1ex]{0pt}{1.0ex}  & $\mathbf{k}_2=2\mathbf{k}_1=(0,1,0)$ & $\mathbf{S}_{1\mathbf{k}_2}=-0.62\, \mu_{\rm B}$ & $\mathbf{S}_{2\mathbf{k}_2}=-0.62\, \mu_{\rm B}$  \\
\rule[-1ex]{0pt}{1.0ex}  & & $\phi_{1\mathbf{k}_2}=0$ & $\phi_{2\mathbf{k}_2}=1/4$  \\
\rule[-1ex]{0pt}{1.0ex}  & $\mathbf{k}_3=0$ & $\mathbf{S}_{1\mathbf{k}_3}=0.62\, \mu_{\rm B}$ & $\mathbf{S}_{2\mathbf{k}_3}=0.62\, \mu_{\rm B}$  \\
\rule[-1ex]{0pt}{1.0ex}  & & $\phi_{1\mathbf{k}_3}=0$ & $\phi_{2\mathbf{k}_3}=0$  \\
\hline
\rule[-1ex]{0pt}{1.0ex} 6 & $\mathbf{k}_1=(0,2/5,0)$ & $\mathbf{S}_{1\mathbf{k}_1}=-0.98\, \mu_{\rm B}$ & $\mathbf{S}_{2\mathbf{k}_1}=-0.98\, \mu_{\rm B}$  \\
\rule[-1ex]{0pt}{1.0ex}  & & $\phi_{1\mathbf{k}_1}=0$ & $\phi_{2\mathbf{k}_1}=1/5$  \\
\rule[-1ex]{0pt}{1.0ex}  & $\mathbf{k}_2=2\mathbf{k}_1=(0,4/5,0)$ & $\mathbf{S}_{1\mathbf{k}_2}=-0.98\, \mu_{\rm B}$ & $\mathbf{S}_{2\mathbf{k}_2}=-0.98\, \mu_{\rm B}$  \\
\rule[-1ex]{0pt}{1.0ex}  & & $\phi_{1\mathbf{k}_2}=0$ & $\phi_{2\mathbf{k}_2}=2/5$  \\
\rule[-1ex]{0pt}{1.0ex}  & $\mathbf{k}_3=0$ & $\mathbf{S}_{1\mathbf{k}_3}=0.74\, \mu_{\rm B}$ & $\mathbf{S}_{2\mathbf{k}_3}=0.74\, \mu_{\rm B}$  \\
\rule[-1ex]{0pt}{1.0ex}  & & $\phi_{1\mathbf{k}_3}=0$ & $\phi_{2\mathbf{k}_3}=0$  \\
\hline
\rule[-1ex]{0pt}{1.0ex} 7 & $\mathbf{k}_1=0$ & $\mathbf{S}_{1\mathbf{k}_1}=1.23\, \mu_{\rm B}$ & $\mathbf{S}_{2\mathbf{k}_1}=1.23\, \mu_{\rm B}$  \\
\rule[-1ex]{0pt}{1.0ex}  & & $\phi_{1\mathbf{k}_3}=0$ & $\phi_{2\mathbf{k}_3}=0$  \\
\hline

\end{tabular}                                         
\label{tab:Table3}
\end{table*}
The observed sequence of the field-induced transitions indicates progressive flipping of the ferromagnetic layers initially aligned antiparallel to the field direction. The number of layers with the moments polarized up becomes bigger, increasing the magnetization along the field direction and resulting in the magnetization steps \cite{Pikul2010}. The presence of the four distinct field-induced phases with large magnetic unit cells points to the fact that exchange interactions stabilizing these exotic phases have a long-range nature, far beyond the nearest and next nearest neighbour interactions.

\begin{figure}
    \centering
    \includegraphics[width=1.00\columnwidth, trim=2 0 0 0, clip]{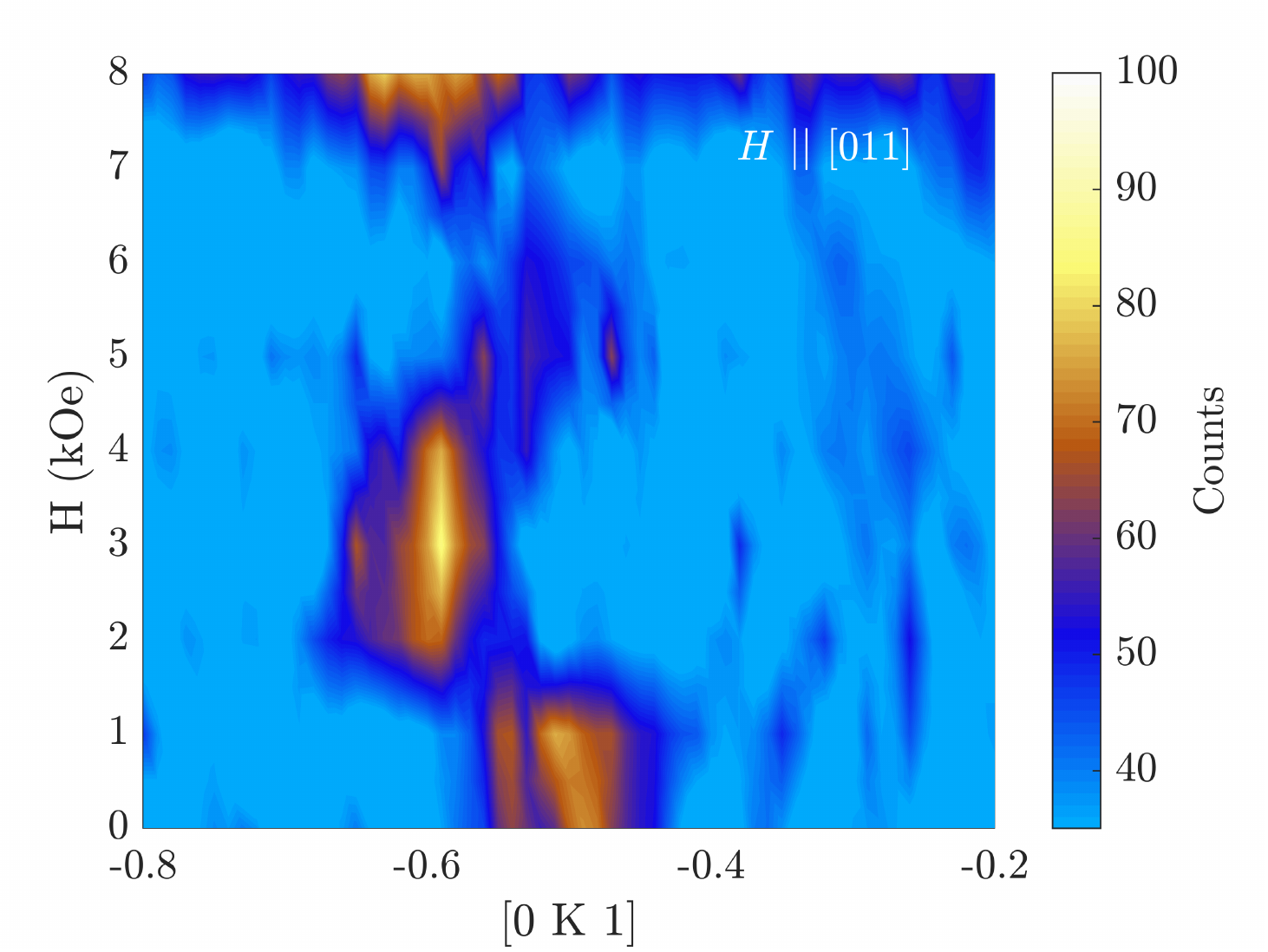}        
    \caption{Neutron diffraction intensity maps in the $[0, K, 1]$ reciprocal direction as a function of magnetic field ($H$) applied parallel to the $[0 1 1]$ direction of CeRh$_3$Si$_2$. The measurements were made on the Zebra diffractometer at $T = 1.5$\,K. It is to be noted that the $[0 1 1]$ direction is at 30$^{\circ}$ to the $b$-axis}
\label{FigNDfield_3}
\end{figure}

The same sequence of magnetic phases were also observed for the field applied along the $[0 1 1]$ direction (see Fig.~\ref{FigNDfield_3}), although the critical fields were slightly higher. The orthorhombic structure is a distorted version of the hexagonal lattice shown in Fig.~\ref{FigStructure}(a) with the $bc$ plane of the orthorhombic cell corresponding to the $ab$ plane of the hexagonal one. The two studied field directions correspond to the $[1 2 0]$ and $[1 0 0]$ directions of the hexagonal lattice and it is not surprising that the response of the crystal is qualitatively similar in both cases. 

\section{Discussion}

Our crystal-field analysis of the RIXS and INS spectra found rather large splitting of the low energy $^2F_{5/2}$ multiplet and a pure $\ket{J=\frac{5}{2},J_{z' \parallel a}\pm\frac{1}{2}}$ ground state with the pseudo-hexagonal $a$ axis as the quantization direction (see Supplemental Material \cite{SuppMat} for the comparison of the two reference systems, $z' \parallel a$ in the present work, and $z \parallel c$ in Ref. \cite{Pikul2010}). Our INS spectra show a resolution-limited crystal-field excitation (Fig.~\ref{INS}) indicating the absence of hybridization of the $4f$ states, in agreement with previous core-level photoemission data \cite{Pikul2010}. The small temperature dependence of the crystal-field excitation at 58\,meV that has been observed with INS is well explained with the thermal expansion of the lattice, as further discussed in Appendix~C.

\begin{figure}
    \centering
    \includegraphics[width=1.00\columnwidth, trim=0.1cm 0 0 0, clip]{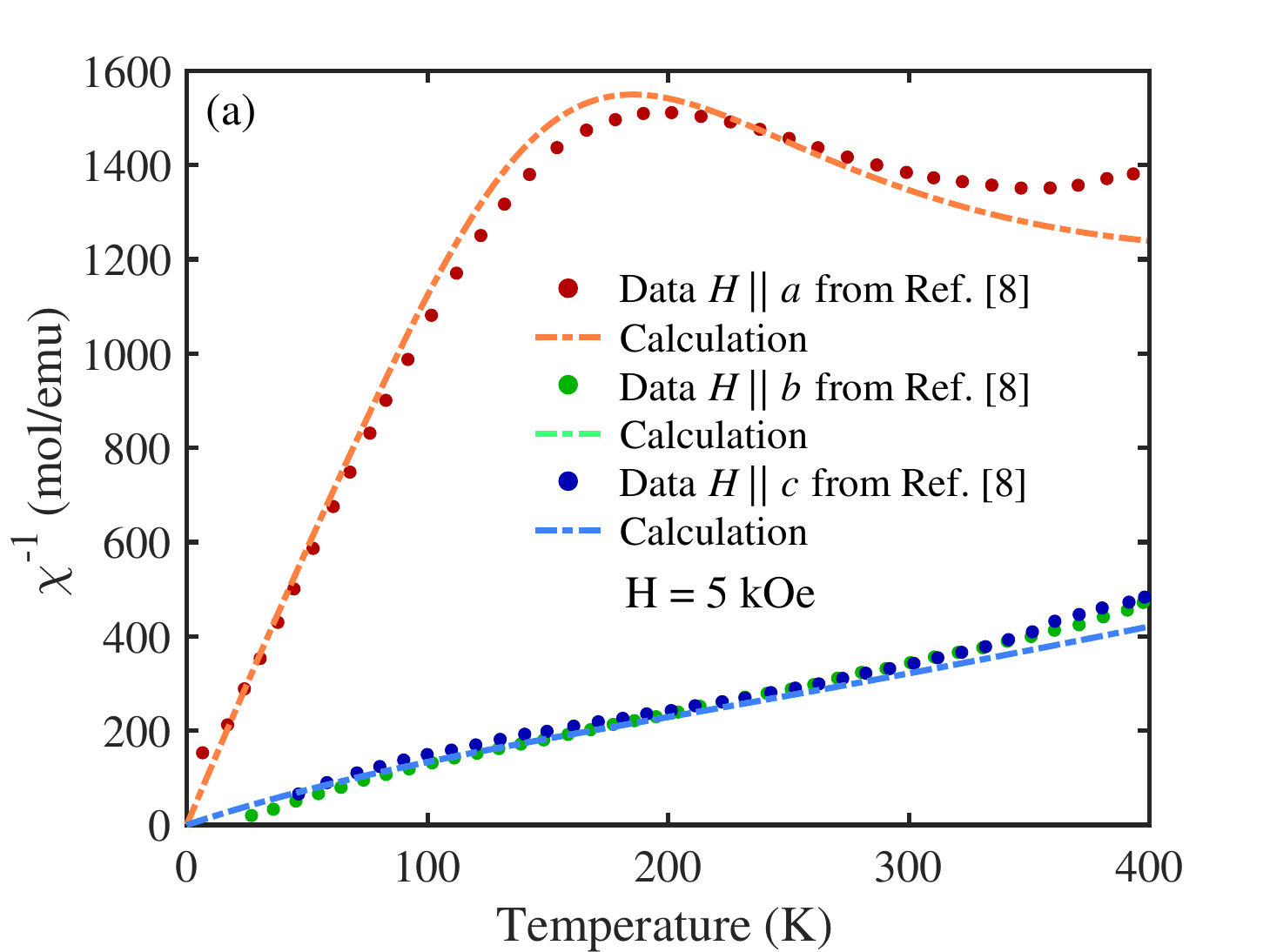}\\
    \includegraphics[width=1.00\columnwidth, trim=-0.2cm 0 0 0, clip]{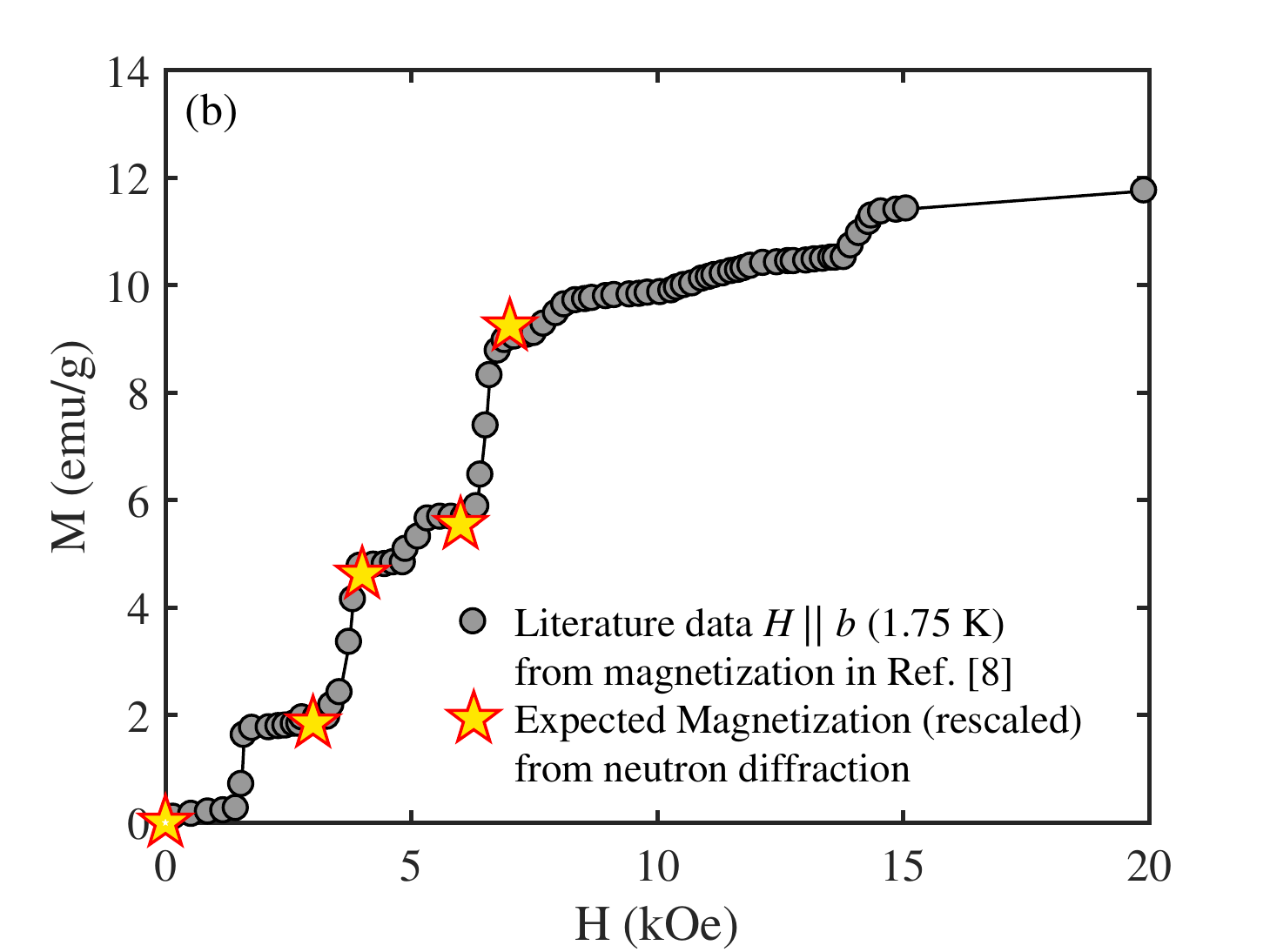}
    \caption{Comparison between the results of this work and the macroscopic measurements from Ref.\,\cite{Pikul2010}. (a) Magnetic susceptibility data and crystal-field calculation based on the results of our analysis. Note: the calculated lines for $H \parallel b$ and $H \parallel c$ are perfectly overlapping in the hexagonal crystal-field approximation. (b) Field-dependent magnetization and the expected steps resulting from the structures shown in Fig.\,\ref{FigNDfield_2}. The absolute value of the expected magnetization is rescaled to compensate for the uncomplete saturation of the moments at 7\,kOe in the macroscopic data.}
\label{FigSusceptibility}
\end{figure}

The deviation from the hexagonal approximation for the crystal field, which would allow some $J_{z'}$ mixing in all crystal field levels, must be small. This can be inferred from (i) the quality of the fits shown in Fig.~\ref{FigRIXS} and in the Supplemental Material \cite{SuppMat}, (ii) the fact that the 78\,meV excitation is barely visible in the INS spectra (Fig.~\ref{INS}), and (iii) the non-observability of the sixth transition in RIXS spectra (see Appendix~A). Furthermore, the proposed crystal-field scheme is in very good agreement with the magnetic susceptibility $\chi(T)$ data from  Pikul \textit{et al.}~\cite{Pikul2010} (Fig.~\ref{FigSusceptibility}\,(a)). It is interesting to note that the orthorhombic crystal-field scheme proposed in that study, once rotated onto the $\{\ket{J,J_{z'}}\}$ basis, corresponds to a 98.2\% pure $\ket{\frac{5}{2}, \pm\frac{1}{2}}$ ground state, meaning that the same almost pure ground state was found independently and without making the hexagonal approximation.

The $\ket{\frac{5}{2}, \pm\frac{1}{2}}$ state is spatially elongated along the quantization axis $z'\parallel a$ as shown in Fig.~\ref{FigStructure}(b) and, in the hexagonal approximation, is symmetric about that axis. Such a ground state explains the strong magnetocrystalline anisotropy that constraints the moments in the $bc$ easy magnetization plane and is in good agreement with the value of the in-plane moment of 1.23\,$\mu_{\rm B}$ measured with neutron diffraction and 1.16\,$\mu_{\rm B}$ estimated from the bulk properties studies~\cite{Pikul2010}. The value calculated based on our crystal-field model is 1.26\,$\mu_{\rm B}$. The good agreement of the measured and crystal-field only moment confirms the absence of Kondo screening and also the absence of frustration that potentially could play a role in a pseudo-hexagonal lattice. 

It is worth noting that the same almost pure $\ket{\frac{5}{2}, \pm\frac{1}{2}}$ ground state is found along the pseudo-hexagonal quantization axis in CeIr$_3$Si$_2$, which has a similar structure and similar magnetic behavior \cite{Shigetoh,Muro} and in the strongly hybridized hexagonal ferromagnet CeRh$_3$B$_2$ \cite{Givord2007a, Givord2007}. The crystal-field scheme resulting from our spectroscopic analysis does not support strong effects due to multiplet intermixing in CeRh$_3$Si$_2$, contrary to the expectation from the crystal-field analysis of the static susceptibility \cite{Pikul2010}. 

Our neutron diffraction data revealed the appearance of an incommensurate (amplitude modulated) antiferromagnetic structure below $T_{\text{N}_1}$, with moments aligned along the orthorhombic $b$ axis, followed by a so-called lock-in of the propagation vector to $\mathbf{k}=(0,1/2,0)$ below $T_{\text{N}_2}$, with a UUDD stacking of the ferromagnetic $ac$ layers. Upon increasing the magnetic field at $T=1.5$\,K, we observe a series of different magnetic structures, alternating between $\mathbf{k}=(0,1/2,0)$ and $\mathbf{k}_1=(0,2/5,0)+\mathbf{k}_2$ with $\mathbf{k}_2=2\mathbf{k}_1=(0,4/5,0)$, with, for each structure, a progressive flip of the layers aligned antiparallel to the field direction. These different structures correspond each to a different step seen in the magnetization measurement, explaining the origin of the stair-like shape up to the highest field measured in our experiment. The ratio between the values of the magnetization calculated on the basis of the disproportion between moments pointing ``up" and ``down" for the different structures at each field value (0 at 0\,kOe, 1/5 at 3\,kOe, 2/4 at 4\,kOe, 3/5 at 6\,kOe and 1 at 7\,kOe), perfectly agrees with the ratio between the steps in the magnetization reported in literature\cite{Pikul2010} (see Fig.\ref{FigSusceptibility}\,(b)). The crystal structure was found to be rigid across all the transitions.

Various rare-earth based systems have been reported to show similar magnetic behavior, with incommensurate antiferromagnetism in the proximity of the N\'{e}el temperature, and/or metamagnetic transitions upon changing field strength \cite{PrGa2, NdGa2, TbNi2Ge2, TbNi2Si2, TmAgGe, TmAgGeFrustration, HoNi2B2C, TmB4, CeSb,ReviewGignoux, ReviewDate}. These phenomena can originate from: strong magnetocrystalline anisotropy, competition between long-range ferromagnetic and antiferromagnetic interactions, energy crossing of crystal-field levels upon increasing fields, locking of the moments in intermediate easy directions, or frustration. Recently, the metamagnetic transitions in the famous metamagnetic compound CeSb have been linked to a reorganization of the electronic structure \cite{CeSb_Electronic}.

Based on our data, we can make some considerations on the mechanisms ruling the magnetism in CeRh$_3$Si$_2$. The crystal-field ground state explains the strong magnetic anisotropy. The large crystal-field splitting, on the other hand, excludes crossing or depopulation of crystal-field states upon cooling or applying a magnetic field to be responsible for the metamagnetic transitions. The collinearity of the magnetism at all temperatures and fields further excludes the locking of the moments in several intermediate easy directions while rotating towards the field.

Having stacks of collinear ferromagnetic planes, and no in-plane propagation vector, one can reduce the problem to a one dimensional Ising chain (along the normal to the planes) with oscillating exchange interactions between nearest and next-nearest neighbor planes, and possibly extending even farther. If limited to the next-nearest neighbors, the model is reduced to the so-called axial next nearest-neighbor Ising model (ANNNI), which was shown able to generate, depending on the parameters, a series of commensurate and incommensurate phases, resulting in the so-called \textit{devil's staircase} shape of the magnetisation and in the incommensurate propagation vector near the paramagnetic transition \cite{Bak, BakIsing, BakIncommensurate}. 

Interestingly, the ANNNI model yields, for a wide range of its parameters, an up-up-down-down ordered ground state at low temperature \cite{Bak, BakIncommensurate, Kaplan}, i.e.~the same magnetic structure found here for CeRh$_3$Si$_2$. Therefore, the metamagnetism is most likely related to oscillating long range exchange interactions, which in rare-earth based compounds can naturally be explained with the Ruderman--Kittel--Kasuya--Yosida interaction.

\section{Conclusion}

We have investigated the electronic and magnetic structure of CeRh$_3$Si$_2$. 
Inelastic neutron scattering and resonant inelastic X-ray scattering allowed us to fully characterize the crystal-field scheme of the Ce$^{3+}$ ions, and measured rather large crystal-field splittings of about 58\,meV and 78\,meV for the Hund's rule ground state multiplet. The INS and RIXS spectra can be well reproduced by simulations using an hexagonal approximation for crystal-field potential (with quantization axis $z'\parallel a$). The details of the spectra have been explained on the basis of the cross-section and its selection rules. The resulting crystal-field scheme with a $J_{z'}=\pm\frac{1}{2}$ ground state explains the strong planar anisotropy previously reported in susceptibility measurement, and provides a good agreement with those data and with the measured value of the saturated moment of about $1.23\,\mu_{\rm B}$ along $b$. The residual in-plane anisotropy must be due to the additional orthorhombicity of the system, not considered in our crystal-field analysis.

Neutron diffraction in zero field revealed an incommensurate antiferromagnetic structure below $T_{\text{N}_1} = 4.7$\,K, followed by a locking of the propagation vector below $T_{\text{N}_{2}} = 4.48$\,K, resulting in an up-up-down-down alternation of the moments, aligned along the $b$ axis. In applied field, we identified four different magnetic structures, each corresponding to a step in the field-dependent magnetization. The size of the moment measured is in agreement with the single-ion crystal-field ground state, further confirming the absence of hybridization and/or frustration.

\section{Appendix} 

\subsection{Finding transition energies in RIXS spectra}

\begin{figure}
    \centering
    \includegraphics[width=1.00\columnwidth]{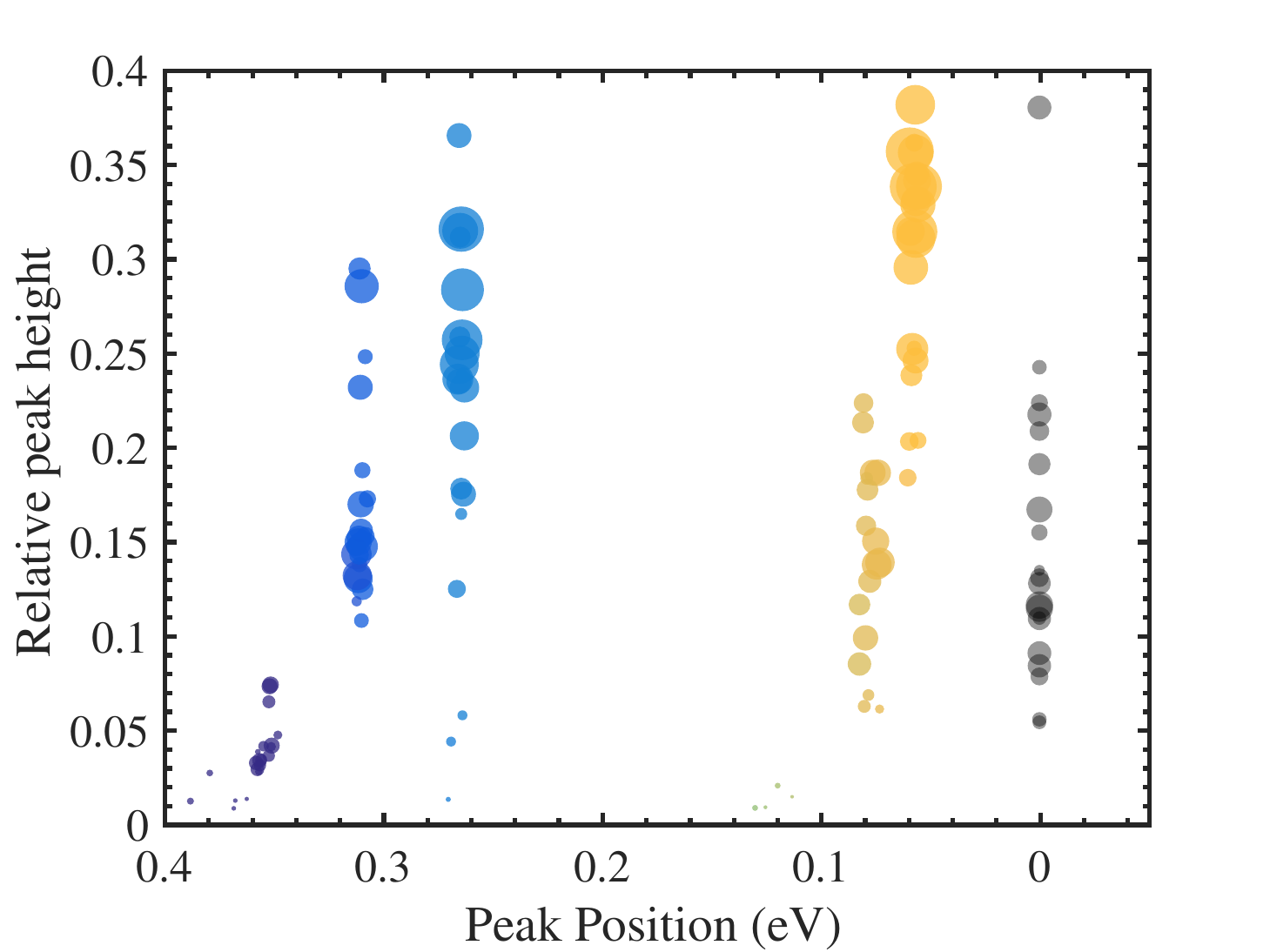}        
    \caption{Peak positions obtained from Voigt fits to the 18 RIXS spectra at 20\,K; yellow symbolizing the transitions into the ground state multiplet and blue the excited multiplet (see Appendix\,A).  The size of the dots corresponds to the relative peak intensity. }
\label{FigSFits}
\end{figure}

In order to determine the crystal-field splittings from the RIXS data, the position of the peaks were determined from a Voigt fit using the \textit{fityk} program for data fitting~\cite{fityk}. Each of the 18 spectra acquired at 20\,K (the full set of spectra is shown in the Supplemental Material \cite{SuppMat}) was fitted independently using 5 Voigt peaks for the crystal-field excitation, plus one Gaussian function located at 0\,meV to account for the the elastic signal. The fit was constrained by assuming the same linewidth for all five peaks, with a Gaussian contribution given by the width of the elastic peak. Two peaks were restricted to energies lower than $150$\,meV, representing the two excited $^2F_{\frac{5}{2}}$ levels, and the other 3 peaks were assigned in the 150--450\,meV region, representing the 3 visible excitations in the $^2F_{\frac{7}{2}}$ multiplet. 

Figure~\ref{FigSFits} shows the peak positions obtained from all these fits, and their relative heights normalized to the sum of the peak heights in each spectrum. The size of the circles is proportional to the absolute intensity of the peaks, to give more importance to the fits of spectra with better statistics. The five inelastic peaks are at about the same positions for all the fitted spectra, their intensity-weighted average positions being: 0\,meV, 59\,meV, 78\,meV, 265\,meV, 308\,meV and 353\,meV. The yellow dots in Fig.\,\ref{FigSFits} refer to transitions into the $^2F_\frac{5}{2}$ multiplet and the blue ones to transitions into $^2F_\frac{7}{2}$.
We can observe that independently on the experimental configuration (geometry, polarization, incident photon energy) the first inelastic peak at 59\,meV is stronger than the peak at 78\,meV, while the last peak at 353\,meV has always very low intensity. As discussed in Appendix\,B, the comparison between these observations and the peak ratios in the calculated RIXS spectra can ease the identification of the crystal field excitations associated to those peaks.

\subsection{Full multiplet calculation of RIXS process including crystal-field model}
We performed full-multiplet calculations using the \textit{Quanty} code by Maurits Haverkort~\cite{Haverkort2012}.
In the calculations, the Slater integrals for the electron--electron interactions in the intermediate $3d^94f^2$ RIXS state, as well as the $3d$ and $4f$ spin-orbit parameters, were obtained from the Robert D. Cowan's Atomic Structure Code\,\cite{CowanBook}. To best reproduce the peak positions, the $4f$ spin-orbit parameter was reduced to 88\% of the value calculated for an isolated ion ($\zeta_{SO}=0.88 \times 0.087$\,eV$= 0.07656$\,eV). The $F_{ff}$ integrals for the intermediate RIXS state were scaled to 55\% and the $F_{df}$ and $G_{df}$ integrals to 75\% of their calculated value, having been tuned by fitting the isotropic XAS spectrum.

\begin{figure}
    \centering
    \includegraphics[width=1.00\columnwidth]{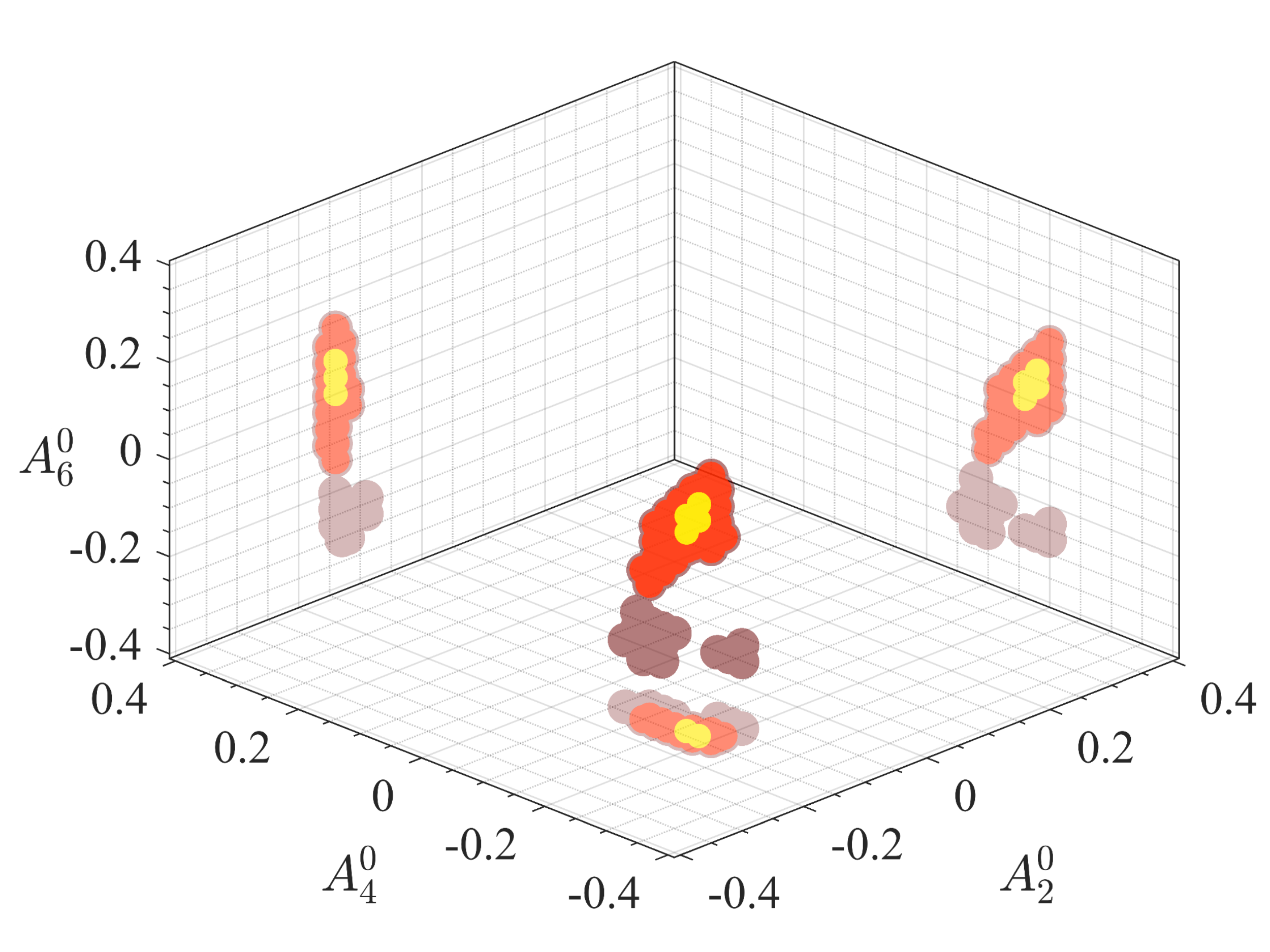}        
    \caption{Regions of the parameter space in agreement with the crystal field splitting measured by RIXS. The brown spheres indicate the set of crystal field parameters that give the measured splittings, with an error of $\pm 7$\,meV, the red spheres are the sets of parameters that respect the further conditions on the relative peak intensities and the yellow spheres respect the stricter condition of having a maximum error of $\pm 2$\,meV. The $  A_6^6 $ parameter (not shown) has a minor effect. To ease the reading of the values, the projection of the spheres are also shown on the planes in lighter colors. The axis unit is eV.}
\label{FigSParameters}
\end{figure}

\begin{table}[]
\def\arraystretch{1.8}
\caption{Full crystal-field wave functions and energy levels calculated with the crystal-field parameters given in the text, considering the pseudo-hexagonal quantization axis $z'$\,$\parallel$\,$c_h$\,$\parallel$\,$a$. Note that the $  A^6_6$ parameter was not determined and kept zero. In case of finite $  A^6_6$, the $J_{z'}=\pm \frac{5}{2}$ and $J_{z'}=\pm \frac{7}{2}$ components are allowed to mix.}
\begin{tabular}{c|c}
\hline 
\rule[-1ex]{0pt}{2.5ex} $\ket{J,J_{z'}}$                   							& Energy (meV)    \\ 
\hline\hline	 
\rule[-1ex]{0pt}{2.5ex} 0.9991$\ket{\frac{5}{2},\pm\frac{1}{2}} \pm  0.0426\ket{\frac{7}{2},\pm\frac{1}{2}}$ 							& 0 			\\
\hline                                                                                     
\rule[-1ex]{0pt}{2.5ex} 0.9899$\ket{\frac{5}{2},\pm\frac{3}{2}} \pm  0.1415 \ket{\frac{7}{2},\pm\frac{3}{2}}$						& 59			\\ 
\hline                                                                                                                    
\rule[-1ex]{0pt}{2.5ex} 0.9999$\ket{\frac{5}{2},\pm\frac{5}{2}} \mp  0.0122 \ket{\frac{7}{2},\pm\frac{5}{2}}$ 							& 78			\\                                        
\hline                                                                               
\rule[-1ex]{0pt}{2.5ex} 
0.9991$\ket{\frac{7}{2},\pm\frac{1}{2}} \mp 0.0426\ket{\frac{5}{2},\pm\frac{1}{2}}$   							& 265 		\\
\hline                                                                             
\rule[-1ex]{0pt}{2.5ex} 
0.9899$\ket{\frac{7}{2},\pm\frac{3}{2}} \mp 0.1415\ket{\frac{5}{2},\pm\frac{3}{2}}$						& 308     \\ 
\hline                                                                                                                    
\rule[-1ex]{0pt}{2.5ex} $\ket{\frac{7}{2},\pm \frac{7}{2}}$							& 345  \\                                        
\hline                                                                             
\rule[-1ex]{0pt}{2.5ex} 
0.9999$\ket{\frac{7}{2},\pm\frac{5}{2}}\pm 0.0122\ket{\frac{5}{2},\pm\frac{5}{2}}$							& 353      \\ 
\hline                                                
\end{tabular}                                         
\label{tab:Table4}
\end{table}

We searched in the phase space ($  A_2^0 $, $  A_4^0 $, $  A_6^0 $, $  A_6^6 $) for sets of crystal-field parameters which correspond to the observed splittings. The parameters were varied in a fine grid of steps and the regions of the phase space compatible with the crystal-field splittings observed in our data are highlighted in Fig.~\ref{FigSParameters}.
In particular, the $  A_2^0 $, $  A_4^0 $, $  A_6^0 $ and $  A_6^6 $ parameters were each varied in the region $[-0.8,+0.8]$\,eV in steps of 0.01, 0.02, 0.03 and 0.05\,eV, respectively.
The brown spheres in Fig.~\ref{FigSParameters} represent the solutions compatible with a $\ket{J=\frac{5}{2},J_{z'\parallel a}=\pm\frac{1}{2}}$ ground state as determined by magnetic susceptibility studies, ~\cite{Pikul2010} (once rotated in the $x',y',z'$ reference system used in this work, see Supplemental Material \cite{SuppMat} for more details), and the measured splittings, allowing for a $\pm 7$\,meV error on the position of the crystal-field levels, in order to be sure to take into account all the sets of parameters that yield similar splittings.  This region can be further reduced by noticing that in the RIXS calculations the $\ket{\frac{5}{2},\pm\frac{1}{2}} \rightarrow \ket{\frac{5}{2},\pm\frac{3}{2}}$ excitation always has higher intensity than the $\ket{\frac{5}{2},\pm\frac{1}{2}} \rightarrow \ket{\frac{5}{2},\pm\frac{5}{2}}$ one, and that the excitation in the $\alpha \ket{\frac{7}{2},\pm\frac{5}{2}} + \beta \ket{\frac{7}{2},\mp\frac{7}{2}}$ and $\beta \ket{\frac{7}{2},\pm\frac{5}{2}} - \alpha \ket{\frac{7}{2},\mp\frac{7}{2}}$ states has in general a lower intensity  relative to the other excitations in the spectrum. Combining this information with the relative peak intensities of Fig.~\ref{FigSFits}, we can assign the 59\,meV peak to the $\ket{\frac{5}{2},\pm\frac{3}{2}}$ state, the 78\,meV peak to the $\ket{\frac{5}{2},\pm\frac{5}{2}}$ and 353\,meV peak to one of the two mixed $\alpha\,(\beta)\cdot\ket{\frac{7}{2},\pm\frac{5}{2}}\,\,+(-)\,\,\beta\,(\alpha)\cdot\ket{\frac{7}{2},\pm\frac{7}{2}}$. The red spheres in Fig.~\ref{FigSFits} indicate the combination of crystal-field parameters complying with these further restrictions. Finally, the yellow spheres show the sets of crystal-field parameters that fit the spectra, obtained by further restricting the condition on the peak positions to a maximum error of $\pm 2$\,meV. The  $  A_6^6 $ parameter (not shown in Fig.~\ref{FigSParameters}) was found to have a minor effect on the calculated spectra. In particular, the most important effect of the $  A_6^6 $ parameter it to mix the $J_{z'}=\pm \frac{5}{2}$ and $\mp \frac{7}{2}$ components in the two states with highest energy. The measured crystal field splittings agree with any value of $  A_6^6$ in the [-0.05, +0.05]\,eV range,  and from this we can estimate that the maximum allowed $J_{z'}$ mixing is of the order of 25\%.

With the crystal-field scheme found, the mixing between the two multiplets is found to be negligible (see Table \ref{tab:Table4}).

\begin{figure}
    \centering
    \includegraphics[width=0.99\columnwidth]{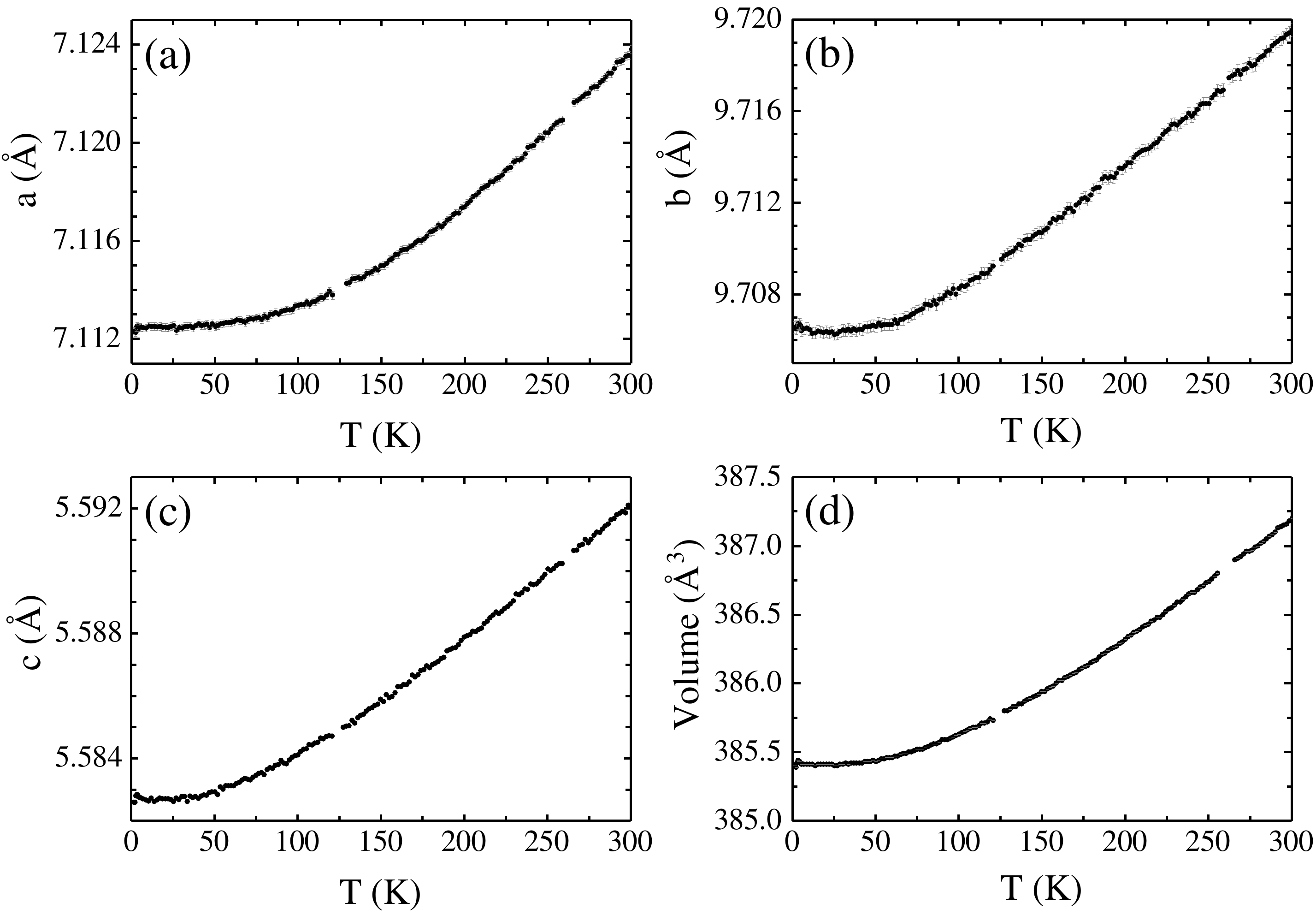}        
    \caption{Lattice parameters and unit cell volume as a function of temperature}
\label{FigD20Lattice}
\end{figure}

\subsection{Temperature dependence of crystal-field excitation}
In order to understand the origin of the temperature dependence of the crystal-field peak positions of CeRh$_3$Si$_2$, we have measured the temperature dependence of lattice parameters with neutron diffraction on the D20 diffractometer at the Institut Laue--Langevin (ILL), Grenoble, France. The results for the temperature dependence of the lattice parameters and unit cell volume are presented in Fig.~\ref{FigD20Lattice}. All lattice parameters decrease with decreasing temperature from 300\,K. Below 20\,K, the $a$ lattice parameter is nearly temperature-independent, while  the $b$ parameter exhibits a small upturn. A smaller upturn was also observed in the $c$ lattice parameter below  10\,K. The unit cell volume decreases with temperature from 300\,K and becomes temperature-independent below 20\,K. Considering there is no obvious anomaly observed in the lattice parameters and unit cell volume with temperature, the temperature dependence of the INS peak position could be associated with gradual changes in the lattice and the volume expansion with increasing temperature. In order to estimate the change in the crystal-field peak position, we have used point change model to calculate the crystal-field energy using the lattice parameters at 7\,K and 300\,K. We used the McPhase program for this calculation. Although the first excited crystal-field level estimated from the point-charge model is very high (412.64\,meV at 7\,K and 410.72\,meV at 300\,K), the difference between 7\,K and 300\,K is 1.92\,meV, which is in good agreement with the experimental results.

\begin{figure}[]
    \centering
    \includegraphics[width=0.99\columnwidth]{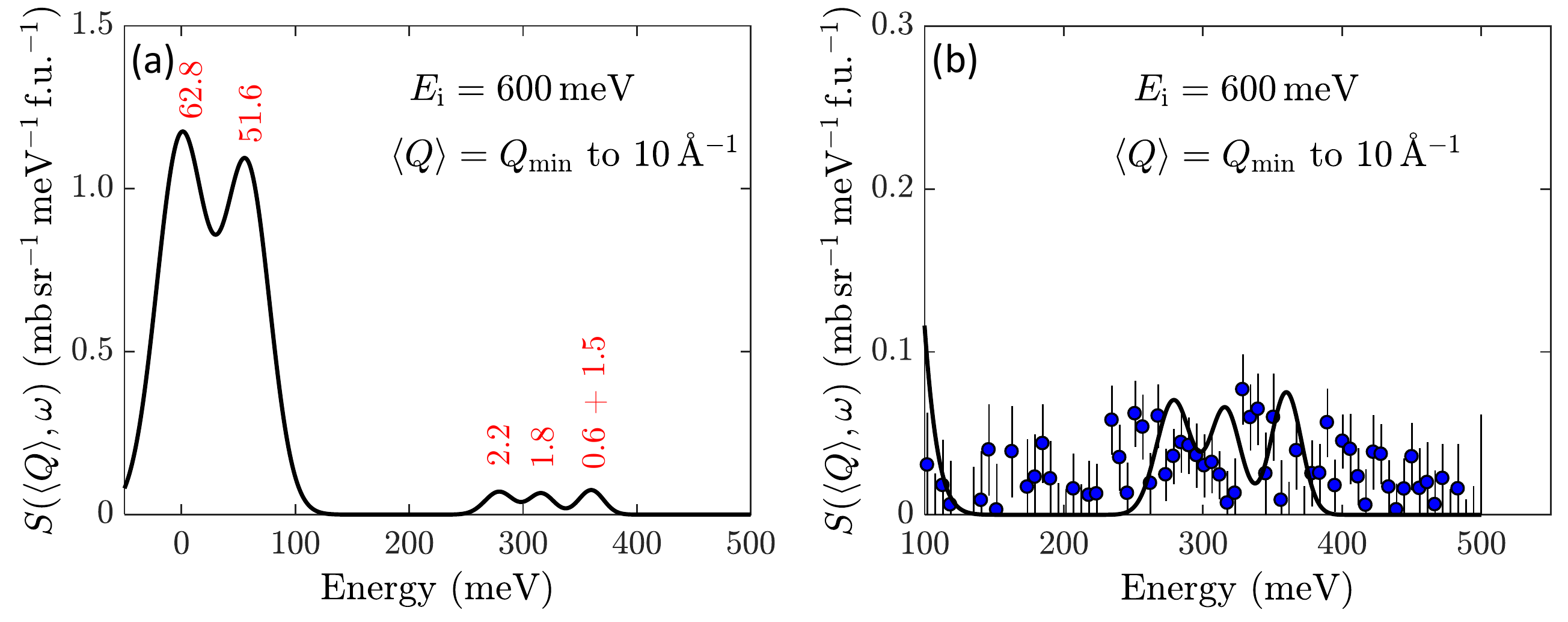}        
    \caption{(a) Simulated crystal-field excitation spectrum including multipole interactions beyond the dipole interaction. The spectrum is averaged over a $Q$ range extending from $Q_{\rm min}$ to 10\,\AA$^{-1}$, where $Q_{\rm min}$ is determined by the neutron kinematics and the lowest scattering angle in the experiment (about $4^{\circ}$). The integrated intensity of each crystal-field transition is written in red above the peak. (b) Comparison between  the experimental data (5\,K, filled symbols) from MERLIN and the simulated crystal-field spectrum (solid line).}
\label{FigSM600meV}
\end{figure}

\subsection{Details of Inelastic Neutron Scattering cross-section and calculations}
In Fig.~\ref{FigSM600meV} we report the INS spectrum acquired with an incident neutron energy of $E_{\rm i} = 600$\,meV, and a calculation of the spectrum from the crystal-field model presented in this work. The calculation includes all relevant multipoles in the neutron--$4f$ scattering potential~\cite{Boothroyd}, and the cross-section is averaged over the same range of $Q$ as the data. The calculation predicts four excitations into the $J = 7/2$ multiplet between 250 and 450\,meV, which are only barely visible in our INS data due to low counting statistics, see Fig.~\ref{FigSM600meV}(b). The detectability of this part of the spectrum is strongly suppressed by the magnetic form factor and by the broadening given by the resolution, which at these energies transfers is about 28\,meV. However, our calculation shows that with larger samples and longer counting time, it is in principle possible to observe these $J=5/2$ to $J=7/2$ excitations by INS.

As discussed in the main text, the excitation near 78\,meV is a quadrupolar transition, and, therefore, it is not expected to be visible in the experimental data at the low Q values ($Q <  4$\,\AA$^{-1}$) of the data in Fig.~\ref{INS}(d), or in the calculation of the dipole cross-section using the hexagonal CEF approximation, also shown in Fig.~\ref{INS}(d). 
For larger momentum transfers, quadrupole excitations gain intensity and the structure factor (or form factor) has a maximum at $Q \simeq 8$\,\AA$^{-1}$ for Ce$^{3+}$. To compare with the RIXS results we have plotted the INS data at higher momentum transfers, averaged from $Q=6$ to $11$\,\AA$^{-1}$, and calculated the neutron scattering cross-section going beyond the dipole approximation (Fig.~\ref{INS}(d), inset). The calculation shows how these excitations with $\Delta J_{z'}$\,=$\pm$2 are accessible at these values of $Q$, and some extra intensity is indeed detected in the experiment around 78\,meV, although it is near the detection threshold due to the noise.

\subsection{Gaussian fits of neutron diffraction data}
\begin{figure}[b]
    \centering
    \includegraphics[width=1.00\columnwidth, trim=0 0 0 0, clip]{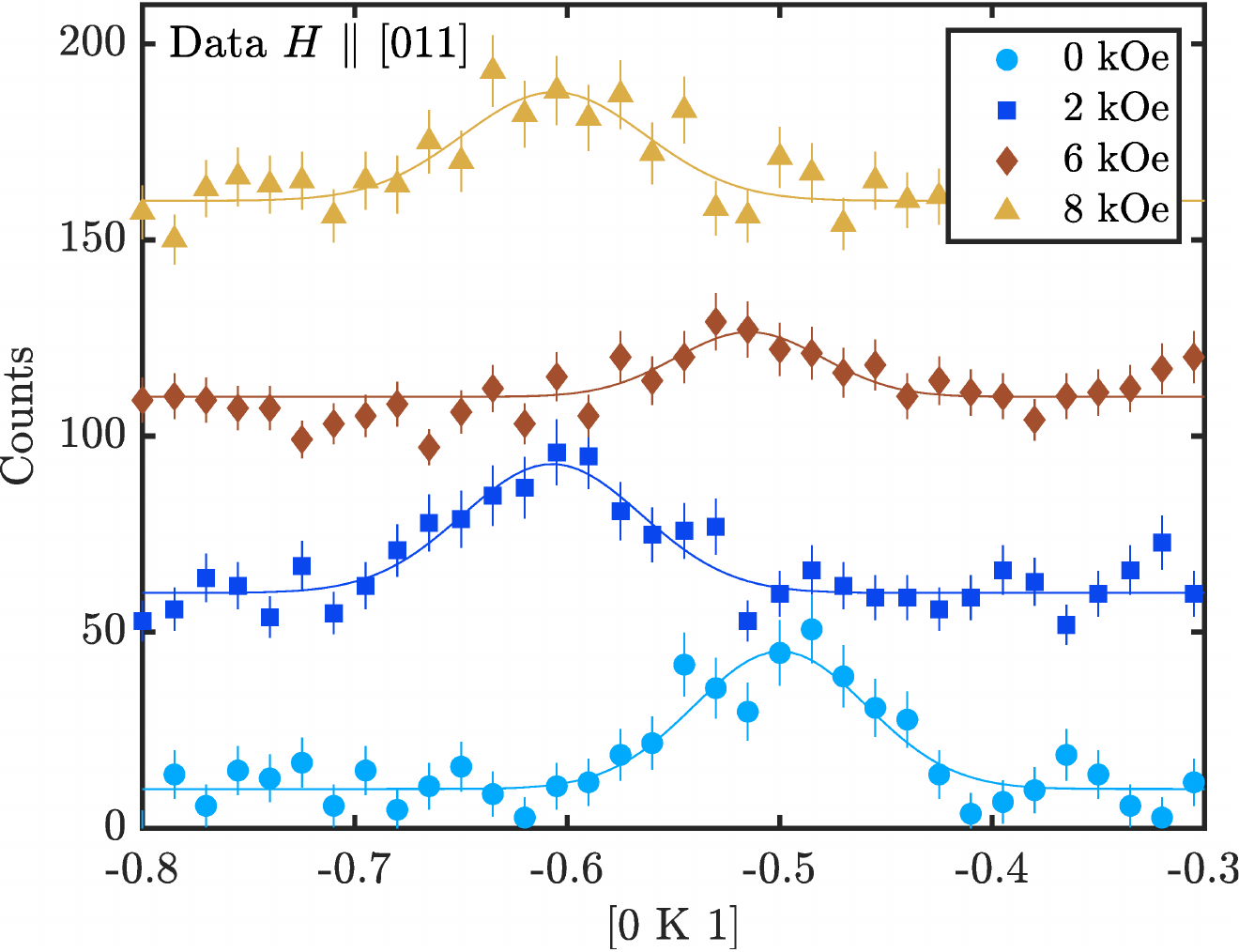}        
    \caption{Neutron diffraction scans in the $[0, K, 1]$ reciprocal direction as a function of magnetic field ($H$) applied parallel to the $[0 1 1]$ direction of CeRh$_3$Si$_2$. The measurements were made on the Zebra diffractometer at $T = 1.5$\,K. It is to be noted that the $[0 1 1]$ direction is at 30$^{\circ}$ to the $b$-axis.}
\label{SFigNDfield_3}
\end{figure}

In Fig.\,\ref{SFigNDfield_3} are shown the 1D cuts of the 2D neutron diffraction intensity map (Fig.\,\ref{FigNDfield_3} the main text) used to determine the magnetic propagation vectors.

\section{Acknowledgements}

We would like to thank Oliver Stockert for many instructive and useful discussions. DTA would like to thank the Royal Society of London for provision of a Newton Advanced Fellowship for exchange between the UK and China, and International exchange funding between the UK and Japan. We  thank the ISIS Facility for beam time on MERLIN, proposal number RB2010784 \cite{Proposal1}, and WISH, proposal number RB2010798 \cite{Proposal2}; SINQ-PSI for beam time on ZEBRA, proposal ID 202100097; and EPSRC UK for the funding grant reference EP/W00562X/1.
AA and AS acknowledge support from the German Research Foundation (DFG) - Project No. 387555779.
O.P. was supported by the Foundation for Polish Science (FNP), program START 66.2020.

%
%
%

\begin{thebibliography}{61}%
\makeatletter
\providecommand \@ifxundefined [1]{%
 \@ifx{#1\undefined}
}%
\providecommand \@ifnum [1]{%
 \ifnum #1\expandafter \@firstoftwo
 \else \expandafter \@secondoftwo
 \fi
}%
\providecommand \@ifx [1]{%
 \ifx #1\expandafter \@firstoftwo
 \else \expandafter \@secondoftwo
 \fi
}%
\providecommand \natexlab [1]{#1}%
\providecommand \enquote  [1]{``#1''}%
\providecommand \bibnamefont  [1]{#1}%
\providecommand \bibfnamefont [1]{#1}%
\providecommand \citenamefont [1]{#1}%
\providecommand \href@noop [0]{\@secondoftwo}%
\providecommand \href [0]{\begingroup \@sanitize@url \@href}%
\providecommand \@href[1]{\@@startlink{#1}\@@href}%
\providecommand \@@href[1]{\endgroup#1\@@endlink}%
\providecommand \@sanitize@url [0]{\catcode `\\12\catcode `\$12\catcode
  `\&12\catcode `\#12\catcode `\^12\catcode `\_12\catcode `\%12\relax}%
\providecommand \@@startlink[1]{}%
\providecommand \@@endlink[0]{}%
\providecommand \url  [0]{\begingroup\@sanitize@url \@url }%
\providecommand \@url [1]{\endgroup\@href {#1}{\urlprefix }}%
\providecommand \urlprefix  [0]{URL }%
\providecommand \Eprint [0]{\href }%
\providecommand \doibase [0]{https://doi.org/}%
\providecommand \selectlanguage [0]{\@gobble}%
\providecommand \bibinfo  [0]{\@secondoftwo}%
\providecommand \bibfield  [0]{\@secondoftwo}%
\providecommand \translation [1]{[#1]}%
\providecommand \BibitemOpen [0]{}%
\providecommand \bibitemStop [0]{}%
\providecommand \bibitemNoStop [0]{.\EOS\space}%
\providecommand \EOS [0]{\spacefactor3000\relax}%
\providecommand \BibitemShut  [1]{\csname bibitem#1\endcsname}%
\let\auto@bib@innerbib\@empty
\bibitem [{\citenamefont {Lawson}\ \emph {et~al.}(1987)\citenamefont {Lawson},
  \citenamefont {Williams},\ and\ \citenamefont {Huber}}]{Lawson}%
  \BibitemOpen
  \bibfield  {author} {\bibinfo {author} {\bibfnamefont {A.~C.}\ \bibnamefont
  {Lawson}}, \bibinfo {author} {\bibfnamefont {A.}~\bibnamefont {Williams}},\
  and\ \bibinfo {author} {\bibfnamefont {J.~G.}\ \bibnamefont {Huber}},\
  }\bibfield  {title} {\bibinfo {title} {Structural properties of
  {CeRh}$_3${B}$_2$},\ }\href
  {https://doi.org/https://doi.org/10.1016/0022-5088(87)90012-9} {\bibfield
  {journal} {\bibinfo  {journal} {Journal of the Less Common Metals}\ }\textbf
  {\bibinfo {volume} {136}},\ \bibinfo {pages} {87 } (\bibinfo {year}
  {1987})}\BibitemShut {NoStop}%
\bibitem [{\citenamefont {Dhar}\ \emph {et~al.}(1981)\citenamefont {Dhar},
  \citenamefont {Malik},\ and\ \citenamefont {Vijayaraghavan}}]{Dhar}%
  \BibitemOpen
  \bibfield  {author} {\bibinfo {author} {\bibfnamefont {S.~K.}\ \bibnamefont
  {Dhar}}, \bibinfo {author} {\bibfnamefont {S.~K.}\ \bibnamefont {Malik}},\
  and\ \bibinfo {author} {\bibfnamefont {R.}~\bibnamefont {Vijayaraghavan}},\
  }\bibfield  {title} {\bibinfo {title} {Strong itinerant magnetism in ternary
  boride {CeRh}$_3${B}$_2$},\ }\href
  {https://doi.org/10.1088/0022-3719/14/11/008} {\bibfield  {journal} {\bibinfo
   {journal} {Journal of Physics C: Solid State Physics}\ }\textbf {\bibinfo
  {volume} {14}},\ \bibinfo {pages} {L321} (\bibinfo {year}
  {1981})}\BibitemShut {NoStop}%
\bibitem [{\citenamefont {Allen}\ \emph {et~al.}(1990)\citenamefont {Allen},
  \citenamefont {Maple}, \citenamefont {Kang}, \citenamefont {Yang},
  \citenamefont {Torikachvili}, \citenamefont {Lassailly}, \citenamefont
  {Ellis}, \citenamefont {Pate},\ and\ \citenamefont {Lindau}}]{Allen1990}%
  \BibitemOpen
  \bibfield  {author} {\bibinfo {author} {\bibfnamefont {J.~W.}\ \bibnamefont
  {Allen}}, \bibinfo {author} {\bibfnamefont {M.~B.}\ \bibnamefont {Maple}},
  \bibinfo {author} {\bibfnamefont {J.-S.}\ \bibnamefont {Kang}}, \bibinfo
  {author} {\bibfnamefont {K.~N.}\ \bibnamefont {Yang}}, \bibinfo {author}
  {\bibfnamefont {M.~S.}\ \bibnamefont {Torikachvili}}, \bibinfo {author}
  {\bibfnamefont {Y.}~\bibnamefont {Lassailly}}, \bibinfo {author}
  {\bibfnamefont {W.~P.}\ \bibnamefont {Ellis}}, \bibinfo {author}
  {\bibfnamefont {B.~B.}\ \bibnamefont {Pate}},\ and\ \bibinfo {author}
  {\bibfnamefont {I.}~\bibnamefont {Lindau}},\ }\bibfield  {title} {\bibinfo
  {title} {{Density-of-states-driven transition from superconductivity to
  ferromagnetism in
  Ce({Ru}$_{1-x}${Rh}$_{x}$)$_{3}${B}$_{2}$:
  Scenario for an exchange-split Kondo resonance}},\ }\href
  {https://doi.org/10.1103/PhysRevB.41.9013} {\bibfield  {journal} {\bibinfo
  {journal} {Phys. Rev. B}\ }\textbf {\bibinfo {volume} {41}},\ \bibinfo
  {pages} {9013} (\bibinfo {year} {1990})}\BibitemShut {NoStop}%
\bibitem [{\citenamefont {Vijayaraghavan}(1985)}]{Vija1985}%
  \BibitemOpen
  \bibfield  {author} {\bibinfo {author} {\bibfnamefont {R.}~\bibnamefont
  {Vijayaraghavan}},\ }\bibfield  {title} {\bibinfo {title} {{NMR} in some
  mixed valent alloys},\ }\href
  {https://doi.org/https://doi.org/10.1016/0304-8853(85)90223-9} {\bibfield
  {journal} {\bibinfo  {journal} {Journal of Magnetism and Magnetic Materials}\
  }\textbf {\bibinfo {volume} {52}},\ \bibinfo {pages} {37 } (\bibinfo {year}
  {1985})}\BibitemShut {NoStop}%
\bibitem [{\citenamefont {Galatanu}\ \emph {et~al.}(2003)\citenamefont
  {Galatanu}, \citenamefont {Yamamoto}, \citenamefont {Okubo}, \citenamefont
  {Yamada}, \citenamefont {Thamizhavel}, \citenamefont {Takeuchi},
  \citenamefont {Sugiyama}, \citenamefont {Inada},\ and\ \citenamefont
  {Onuki}}]{Galatanu}%
  \BibitemOpen
  \bibfield  {author} {\bibinfo {author} {\bibfnamefont {A.}~\bibnamefont
  {Galatanu}}, \bibinfo {author} {\bibfnamefont {E.}~\bibnamefont {Yamamoto}},
  \bibinfo {author} {\bibfnamefont {T.}~\bibnamefont {Okubo}}, \bibinfo
  {author} {\bibfnamefont {M.}~\bibnamefont {Yamada}}, \bibinfo {author}
  {\bibfnamefont {A.}~\bibnamefont {Thamizhavel}}, \bibinfo {author}
  {\bibfnamefont {T.}~\bibnamefont {Takeuchi}}, \bibinfo {author}
  {\bibfnamefont {K.}~\bibnamefont {Sugiyama}}, \bibinfo {author}
  {\bibfnamefont {Y.}~\bibnamefont {Inada}},\ and\ \bibinfo {author}
  {\bibfnamefont {Y.}~\bibnamefont {Onuki}},\ }\bibfield  {title} {\bibinfo
  {title} {On the unusual magnetic behaviour of {CeRh}$_3${B}$_2$},\ }\href
  {https://doi.org/10.1088/0953-8984/15/28/349} {\bibfield  {journal} {\bibinfo
   {journal} {Journal of Physics: Condensed Matter}\ }\textbf {\bibinfo
  {volume} {15}},\ \bibinfo {pages} {S2187} (\bibinfo {year}
  {2003})}\BibitemShut {NoStop}%
\bibitem [{\citenamefont {Kubota}\ \emph {et~al.}(2013)\citenamefont {Kubota},
  \citenamefont {Matsuoka}, \citenamefont {Funasako}, \citenamefont {Mochida},
  \citenamefont {Sakurai}, \citenamefont {Ohta}, \citenamefont {Onimaru},
  \citenamefont {Takabatake},\ and\ \citenamefont {Sugawara}}]{Kubota2013}%
  \BibitemOpen
  \bibfield  {author} {\bibinfo {author} {\bibfnamefont {K.}~\bibnamefont
  {Kubota}}, \bibinfo {author} {\bibfnamefont {E.}~\bibnamefont {Matsuoka}},
  \bibinfo {author} {\bibfnamefont {Y.}~\bibnamefont {Funasako}}, \bibinfo
  {author} {\bibfnamefont {T.}~\bibnamefont {Mochida}}, \bibinfo {author}
  {\bibfnamefont {T.}~\bibnamefont {Sakurai}}, \bibinfo {author} {\bibfnamefont
  {H.}~\bibnamefont {Ohta}}, \bibinfo {author} {\bibfnamefont {T.}~\bibnamefont
  {Onimaru}}, \bibinfo {author} {\bibfnamefont {T.}~\bibnamefont
  {Takabatake}},\ and\ \bibinfo {author} {\bibfnamefont {H.}~\bibnamefont
  {Sugawara}},\ }\bibfield  {title} {\bibinfo {title} {{Weak Ferromagnetism
  below 41 K and Structural Transition at 395 K in {CeIr$_3$B$_2$} Single
  Crystal}},\ }\href {https://doi.org/10.7566/JPSJ.82.104715} {\bibfield
  {journal} {\bibinfo  {journal} {J. Phys. Soc. Jpn.}\ }\textbf {\bibinfo
  {volume} {82}},\ \bibinfo {pages} {104715} (\bibinfo {year}
  {2013})}\BibitemShut {NoStop}%
\bibitem [{\citenamefont {Yang}\ \emph {et~al.}(1984)\citenamefont {Yang},
  \citenamefont {Torikachvili}, \citenamefont {Maple},\ and\ \citenamefont
  {Ku}}]{Yang1984}%
  \BibitemOpen
  \bibfield  {author} {\bibinfo {author} {\bibfnamefont {K.~N.}\ \bibnamefont
  {Yang}}, \bibinfo {author} {\bibfnamefont {M.~S.}\ \bibnamefont
  {Torikachvili}}, \bibinfo {author} {\bibfnamefont {M.~B.}\ \bibnamefont
  {Maple}},\ and\ \bibinfo {author} {\bibfnamefont {H.~C.}\ \bibnamefont
  {Ku}},\ }\bibfield  {title} {\bibinfo {title} {Intermediate valence behavior
  of ternary cerium and uranium transition metal borides},\ }\href
  {https://doi.org/10.1007/bf00681814} {\bibfield  {journal} {\bibinfo
  {journal} {Journal of Low Temperature Physics}\ }\textbf {\bibinfo {volume}
  {56}},\ \bibinfo {pages} {601} (\bibinfo {year} {1984})}\BibitemShut
  {NoStop}%
\bibitem [{\citenamefont {Pikul}\ \emph {et~al.}(2010)\citenamefont {Pikul},
  \citenamefont {Kaczorowski}, \citenamefont {Gajek}, \citenamefont
  {St\c{e}pie\'{n}-Damm}, \citenamefont {\'{S}lebarski}, \citenamefont
  {Werwi\'{n}ski},\ and\ \citenamefont {Szajek}}]{Pikul2010}%
  \BibitemOpen
  \bibfield  {author} {\bibinfo {author} {\bibfnamefont {A.~P.}\ \bibnamefont
  {Pikul}}, \bibinfo {author} {\bibfnamefont {D.}~\bibnamefont {Kaczorowski}},
  \bibinfo {author} {\bibfnamefont {Z.}~\bibnamefont {Gajek}}, \bibinfo
  {author} {\bibfnamefont {J.}~\bibnamefont {St\c{e}pie\'{n}-Damm}}, \bibinfo
  {author} {\bibfnamefont {A.}~\bibnamefont {\'{S}lebarski}}, \bibinfo {author}
  {\bibfnamefont {M.}~\bibnamefont {Werwi\'{n}ski}},\ and\ \bibinfo {author}
  {\bibfnamefont {A.}~\bibnamefont {Szajek}},\ }\bibfield  {title} {\bibinfo
  {title} {Giant crystal-electric-field effect and complex magnetic behavior in
  single-crystalline {CeRh}$_3${Si}$_2$},\ }\href
  {https://doi.org/10.1103/PhysRevB.81.174408} {\bibfield  {journal} {\bibinfo
  {journal} {Phys. Rev. B}\ }\textbf {\bibinfo {volume} {81}},\ \bibinfo
  {pages} {174408} (\bibinfo {year} {2010})}\BibitemShut {NoStop}%
\bibitem [{\citenamefont {Cenzual}\ \emph {et~al.}(1988)\citenamefont
  {Cenzual}, \citenamefont {Chabot},\ and\ \citenamefont
  {Parth{\'{e}}}}]{Cenzual1988}%
  \BibitemOpen
  \bibfield  {author} {\bibinfo {author} {\bibfnamefont {K.}~\bibnamefont
  {Cenzual}}, \bibinfo {author} {\bibfnamefont {B.}~\bibnamefont {Chabot}},\
  and\ \bibinfo {author} {\bibfnamefont {E.}~\bibnamefont {Parth{\'{e}}}},\
  }\bibfield  {title} {\bibinfo {title} {{ErRh}$_3${Si}$_2$ and isotypes with
  an orthorhombic deformation superstructure of the {CeCo}$_3${B}$_2$ type},\
  }\href {https://doi.org/10.1107/S0108270187010357} {\bibfield  {journal}
  {\bibinfo  {journal} {Acta Crystallographica Section C}\ }\textbf {\bibinfo
  {volume} {44}},\ \bibinfo {pages} {221} (\bibinfo {year} {1988})}\BibitemShut
  {NoStop}%
\bibitem [{\citenamefont {Miyahara}\ \emph {et~al.}(2018)\citenamefont
  {Miyahara}, \citenamefont {Shirakawa}, \citenamefont {Setoguchi},
  \citenamefont {Tsubota}, \citenamefont {Kuroiwa},\ and\ \citenamefont
  {Kitagawa}}]{Miyahara2018}%
  \BibitemOpen
  \bibfield  {author} {\bibinfo {author} {\bibfnamefont {J.}~\bibnamefont
  {Miyahara}}, \bibinfo {author} {\bibfnamefont {N.}~\bibnamefont {Shirakawa}},
  \bibinfo {author} {\bibfnamefont {Y.}~\bibnamefont {Setoguchi}}, \bibinfo
  {author} {\bibfnamefont {M.}~\bibnamefont {Tsubota}}, \bibinfo {author}
  {\bibfnamefont {K.}~\bibnamefont {Kuroiwa}},\ and\ \bibinfo {author}
  {\bibfnamefont {J.}~\bibnamefont {Kitagawa}},\ }\bibfield  {title} {\bibinfo
  {title} {Hill plot focusing on {Ce} compounds with high magnetic ordering
  temperatures and consequent study of {Ce$_2$AuP$_3$}},\ }\href
  {https://doi.org/10.1007/s10948-018-4624-9} {\bibfield  {journal} {\bibinfo
  {journal} {J. Superconductivity and Novel Magnetism}\ }\textbf {\bibinfo
  {volume} {31}},\ \bibinfo {pages} {3559} (\bibinfo {year}
  {2018})}\BibitemShut {NoStop}%
\bibitem [{\citenamefont {Kaczorowski}\ and\ \citenamefont
  {Komatsubara}(2008)}]{Kaczorowski2008}%
  \BibitemOpen
  \bibfield  {author} {\bibinfo {author} {\bibfnamefont {D.}~\bibnamefont
  {Kaczorowski}}\ and\ \bibinfo {author} {\bibfnamefont {T.}~\bibnamefont
  {Komatsubara}},\ }\bibfield  {title} {\bibinfo {title} {Complex magnetic
  behavior in single-crystalline {CeRh}$_3${Si}$_2$},\ }\href
  {https://doi.org/10.1016/j.physb.2007.10.150} {\bibfield  {journal} {\bibinfo
   {journal} {Physica B: Condensed Matter}\ }\textbf {\bibinfo {volume}
  {403}},\ \bibinfo {pages} {1362} (\bibinfo {year} {2008})}\BibitemShut
  {NoStop}%
\bibitem [{\citenamefont {Burlet}\ \emph {et~al.}(1994)\citenamefont {Burlet},
  \citenamefont {Regnault}, \citenamefont {Rossat-Mignod}, \citenamefont
  {Vettier},\ and\ \citenamefont {Flouquet}}]{BURLET199410}%
  \BibitemOpen
  \bibfield  {author} {\bibinfo {author} {\bibfnamefont {P.}~\bibnamefont
  {Burlet}}, \bibinfo {author} {\bibfnamefont {L.~P.}\ \bibnamefont
  {Regnault}}, \bibinfo {author} {\bibfnamefont {J.}~\bibnamefont
  {Rossat-Mignod}}, \bibinfo {author} {\bibfnamefont {C.}~\bibnamefont
  {Vettier}},\ and\ \bibinfo {author} {\bibfnamefont {J.}~\bibnamefont
  {Flouquet}},\ }\bibfield  {title} {\bibinfo {title} {Neutron scattering and
  heavy fermion compounds},\ }\href
  {https://doi.org/https://doi.org/10.1016/0304-8853(94)90424-3} {\bibfield
  {journal} {\bibinfo  {journal} {Journal of Magnetism and Magnetic Materials}\
  }\textbf {\bibinfo {volume} {129}},\ \bibinfo {pages} {10} (\bibinfo {year}
  {1994})}\BibitemShut {NoStop}%
\bibitem [{\citenamefont {Anand}\ \emph
  {et~al.}(2018{\natexlab{a}})\citenamefont {Anand}, \citenamefont {Hillier},
  \citenamefont {Adroja}, \citenamefont {Khalyavin}, \citenamefont {Manuel},
  \citenamefont {Andre}, \citenamefont {Rols},\ and\ \citenamefont
  {Koza}}]{PhysRevB.97.184422}%
  \BibitemOpen
  \bibfield  {author} {\bibinfo {author} {\bibfnamefont {V.~K.}\ \bibnamefont
  {Anand}}, \bibinfo {author} {\bibfnamefont {A.~D.}\ \bibnamefont {Hillier}},
  \bibinfo {author} {\bibfnamefont {D.~T.}\ \bibnamefont {Adroja}}, \bibinfo
  {author} {\bibfnamefont {D.~D.}\ \bibnamefont {Khalyavin}}, \bibinfo {author}
  {\bibfnamefont {P.}~\bibnamefont {Manuel}}, \bibinfo {author} {\bibfnamefont
  {G.}~\bibnamefont {Andre}}, \bibinfo {author} {\bibfnamefont
  {S.}~\bibnamefont {Rols}},\ and\ \bibinfo {author} {\bibfnamefont {M.~M.}\
  \bibnamefont {Koza}},\ }\bibfield  {title} {\bibinfo {title} {Understanding
  the magnetism in noncentrosymmetric {CeIrGe}$_{3}$: Muon spin relaxation and
  neutron scattering studies},\ }\href
  {https://doi.org/10.1103/PhysRevB.97.184422} {\bibfield  {journal} {\bibinfo
  {journal} {Phys. Rev. B}\ }\textbf {\bibinfo {volume} {97}},\ \bibinfo
  {pages} {184422} (\bibinfo {year} {2018}{\natexlab{a}})}\BibitemShut
  {NoStop}%
\bibitem [{\citenamefont {Princep}\ \emph {et~al.}(2013)\citenamefont
  {Princep}, \citenamefont {Prabhakaran}, \citenamefont {Boothroyd},\ and\
  \citenamefont {Adroja}}]{PhysRevB.88.104421}%
  \BibitemOpen
  \bibfield  {author} {\bibinfo {author} {\bibfnamefont {A.~J.}\ \bibnamefont
  {Princep}}, \bibinfo {author} {\bibfnamefont {D.}~\bibnamefont
  {Prabhakaran}}, \bibinfo {author} {\bibfnamefont {A.~T.}\ \bibnamefont
  {Boothroyd}},\ and\ \bibinfo {author} {\bibfnamefont {D.~T.}\ \bibnamefont
  {Adroja}},\ }\bibfield  {title} {\bibinfo {title} {Crystal-field states of
  {Pr}$^{3+}$ in the candidate quantum spin ice
  {Pr}$_{2}${Sn}$_{2}${O}$_{7}$},\ }\href
  {https://doi.org/10.1103/PhysRevB.88.104421} {\bibfield  {journal} {\bibinfo
  {journal} {Phys. Rev. B}\ }\textbf {\bibinfo {volume} {88}},\ \bibinfo
  {pages} {104421} (\bibinfo {year} {2013})}\BibitemShut {NoStop}%
\bibitem [{\citenamefont {Amorese}\ \emph {et~al.}(2016)\citenamefont
  {Amorese}, \citenamefont {Dellea}, \citenamefont {Fanciulli}, \citenamefont
  {Seiro}, \citenamefont {Geibel}, \citenamefont {Krellner}, \citenamefont
  {Makarova}, \citenamefont {Braicovich}, \citenamefont {Ghiringhelli},
  \citenamefont {Vyalikh}, \citenamefont {Brookes},\ and\ \citenamefont
  {Kummer}}]{AmoreseCeRIXS}%
  \BibitemOpen
  \bibfield  {author} {\bibinfo {author} {\bibfnamefont {A.}~\bibnamefont
  {Amorese}}, \bibinfo {author} {\bibfnamefont {G.}~\bibnamefont {Dellea}},
  \bibinfo {author} {\bibfnamefont {M.}~\bibnamefont {Fanciulli}}, \bibinfo
  {author} {\bibfnamefont {S.}~\bibnamefont {Seiro}}, \bibinfo {author}
  {\bibfnamefont {C.}~\bibnamefont {Geibel}}, \bibinfo {author} {\bibfnamefont
  {C.}~\bibnamefont {Krellner}}, \bibinfo {author} {\bibfnamefont {I.~P.}\
  \bibnamefont {Makarova}}, \bibinfo {author} {\bibfnamefont {L.}~\bibnamefont
  {Braicovich}}, \bibinfo {author} {\bibfnamefont {G.}~\bibnamefont
  {Ghiringhelli}}, \bibinfo {author} {\bibfnamefont {D.~V.}\ \bibnamefont
  {Vyalikh}}, \bibinfo {author} {\bibfnamefont {N.~B.}\ \bibnamefont
  {Brookes}},\ and\ \bibinfo {author} {\bibfnamefont {K.}~\bibnamefont
  {Kummer}},\ }\bibfield  {title} {\bibinfo {title} {$4f$ excitations in {Ce
  Kondo} lattices studied by resonant inelastic x-ray scattering},\ }\href
  {https://doi.org/10.1103/PhysRevB.93.165134} {\bibfield  {journal} {\bibinfo
  {journal} {Phys. Rev. B}\ }\textbf {\bibinfo {volume} {93}},\ \bibinfo
  {pages} {165134} (\bibinfo {year} {2016})}\BibitemShut {NoStop}%
\bibitem [{\citenamefont {Amorese}\ \emph
  {et~al.}(2018{\natexlab{a}})\citenamefont {Amorese}, \citenamefont
  {Caroca-Canales}, \citenamefont {Seiro}, \citenamefont {Krellner},
  \citenamefont {Ghiringhelli}, \citenamefont {Brookes}, \citenamefont
  {Vyalikh}, \citenamefont {Geibel},\ and\ \citenamefont
  {Kummer}}]{AmoreseCeRh2Si2}%
  \BibitemOpen
  \bibfield  {author} {\bibinfo {author} {\bibfnamefont {A.}~\bibnamefont
  {Amorese}}, \bibinfo {author} {\bibfnamefont {N.}~\bibnamefont
  {Caroca-Canales}}, \bibinfo {author} {\bibfnamefont {S.}~\bibnamefont
  {Seiro}}, \bibinfo {author} {\bibfnamefont {C.}~\bibnamefont {Krellner}},
  \bibinfo {author} {\bibfnamefont {G.}~\bibnamefont {Ghiringhelli}}, \bibinfo
  {author} {\bibfnamefont {N.~B.}\ \bibnamefont {Brookes}}, \bibinfo {author}
  {\bibfnamefont {D.~V.}\ \bibnamefont {Vyalikh}}, \bibinfo {author}
  {\bibfnamefont {C.}~\bibnamefont {Geibel}},\ and\ \bibinfo {author}
  {\bibfnamefont {K.}~\bibnamefont {Kummer}},\ }\bibfield  {title} {\bibinfo
  {title} {{Crystal electric field in CeRh$_{2}$Si$_{2}$ studied with
  high-resolution resonant inelastic soft x-ray scattering}},\ }\href
  {https://doi.org/10.1103/PhysRevB.97.245130} {\bibfield  {journal} {\bibinfo
  {journal} {Phys. Rev. B}\ }\textbf {\bibinfo {volume} {97}},\ \bibinfo
  {pages} {245130} (\bibinfo {year} {2018}{\natexlab{a}})}\BibitemShut
  {NoStop}%
\bibitem [{\citenamefont {Amorese}\ \emph
  {et~al.}(2018{\natexlab{b}})\citenamefont {Amorese}, \citenamefont {Kummer},
  \citenamefont {Brookes}, \citenamefont {Stockert}, \citenamefont {Adroja},
  \citenamefont {Strydom}, \citenamefont {Sidorenko}, \citenamefont {Winkler},
  \citenamefont {Zocco}, \citenamefont {Prokofiev}, \citenamefont {Paschen},
  \citenamefont {Haverkort}, \citenamefont {Tjeng},\ and\ \citenamefont
  {Severing}}]{AmoreseCeRu4Sn6}%
  \BibitemOpen
  \bibfield  {author} {\bibinfo {author} {\bibfnamefont {A.}~\bibnamefont
  {Amorese}}, \bibinfo {author} {\bibfnamefont {K.}~\bibnamefont {Kummer}},
  \bibinfo {author} {\bibfnamefont {N.~B.}\ \bibnamefont {Brookes}}, \bibinfo
  {author} {\bibfnamefont {O.}~\bibnamefont {Stockert}}, \bibinfo {author}
  {\bibfnamefont {D.~T.}\ \bibnamefont {Adroja}}, \bibinfo {author}
  {\bibfnamefont {A.~M.}\ \bibnamefont {Strydom}}, \bibinfo {author}
  {\bibfnamefont {A.}~\bibnamefont {Sidorenko}}, \bibinfo {author}
  {\bibfnamefont {H.}~\bibnamefont {Winkler}}, \bibinfo {author} {\bibfnamefont
  {D.~A.}\ \bibnamefont {Zocco}}, \bibinfo {author} {\bibfnamefont
  {A.}~\bibnamefont {Prokofiev}}, \bibinfo {author} {\bibfnamefont
  {S.}~\bibnamefont {Paschen}}, \bibinfo {author} {\bibfnamefont {M.~W.}\
  \bibnamefont {Haverkort}}, \bibinfo {author} {\bibfnamefont {L.~H.}\
  \bibnamefont {Tjeng}},\ and\ \bibinfo {author} {\bibfnamefont
  {A.}~\bibnamefont {Severing}},\ }\bibfield  {title} {\bibinfo {title}
  {{Determining the local low-energy excitations in the {K}ondo semimetal
  CeRu$_{4}$Sn$_{6}$ using resonant inelastic x-ray scattering}},\ }\href
  {https://doi.org/10.1103/PhysRevB.98.081116} {\bibfield  {journal} {\bibinfo
  {journal} {Phys. Rev. B}\ }\textbf {\bibinfo {volume} {98}},\ \bibinfo
  {pages} {081116(R)} (\bibinfo {year} {2018}{\natexlab{b}})}\BibitemShut
  {NoStop}%
\bibitem [{\citenamefont {Amorese}\ \emph
  {et~al.}(2019{\natexlab{a}})\citenamefont {Amorese}, \citenamefont
  {Stockert}, \citenamefont {Kummer}, \citenamefont {Brookes}, \citenamefont
  {Kim}, \citenamefont {Fisk}, \citenamefont {Haverkort}, \citenamefont
  {Thalmeier}, \citenamefont {Tjeng},\ and\ \citenamefont
  {Severing}}]{AmoreseSmB6}%
  \BibitemOpen
  \bibfield  {author} {\bibinfo {author} {\bibfnamefont {A.}~\bibnamefont
  {Amorese}}, \bibinfo {author} {\bibfnamefont {O.}~\bibnamefont {Stockert}},
  \bibinfo {author} {\bibfnamefont {K.}~\bibnamefont {Kummer}}, \bibinfo
  {author} {\bibfnamefont {N.~B.}\ \bibnamefont {Brookes}}, \bibinfo {author}
  {\bibfnamefont {D.-J.}\ \bibnamefont {Kim}}, \bibinfo {author} {\bibfnamefont
  {Z.}~\bibnamefont {Fisk}}, \bibinfo {author} {\bibfnamefont {M.~W.}\
  \bibnamefont {Haverkort}}, \bibinfo {author} {\bibfnamefont {P.}~\bibnamefont
  {Thalmeier}}, \bibinfo {author} {\bibfnamefont {L.~H.}\ \bibnamefont
  {Tjeng}},\ and\ \bibinfo {author} {\bibfnamefont {A.}~\bibnamefont
  {Severing}},\ }\bibfield  {title} {\bibinfo {title} {Resonant inelastic x-ray
  scattering investigation of the crystal-field splitting of {Sm}$^{3+}$ in
  {SmB}$_{6}$},\ }\href {https://doi.org/10.1103/PhysRevB.100.241107}
  {\bibfield  {journal} {\bibinfo  {journal} {Phys. Rev. B}\ }\textbf {\bibinfo
  {volume} {100}},\ \bibinfo {pages} {241107(R)} (\bibinfo {year}
  {2019}{\natexlab{a}})}\BibitemShut {NoStop}%
\bibitem [{\citenamefont {Givord}\ \emph
  {et~al.}(2007{\natexlab{a}})\citenamefont {Givord}, \citenamefont
  {Boucherle}, \citenamefont {Murani}, \citenamefont {Bewley}, \citenamefont
  {Gal{\'{e}}ra},\ and\ \citenamefont {Lejay}}]{Givord_2007}%
  \BibitemOpen
  \bibfield  {author} {\bibinfo {author} {\bibfnamefont {F.}~\bibnamefont
  {Givord}}, \bibinfo {author} {\bibfnamefont {J.-X.}\ \bibnamefont
  {Boucherle}}, \bibinfo {author} {\bibfnamefont {A.~P.}\ \bibnamefont
  {Murani}}, \bibinfo {author} {\bibfnamefont {R.}~\bibnamefont {Bewley}},
  \bibinfo {author} {\bibfnamefont {R.-M.}\ \bibnamefont {Gal{\'{e}}ra}},\ and\
  \bibinfo {author} {\bibfnamefont {P.}~\bibnamefont {Lejay}},\ }\bibfield
  {title} {\bibinfo {title} {Crystal electric field excitations in the cerium
  compound {CeRh}$_3${B}$_2$ studied by inelastic neutron scattering},\ }\href
  {https://doi.org/10.1088/0953-8984/19/50/506210} {\bibfield  {journal}
  {\bibinfo  {journal} {Journal of Physics: Condensed Matter}\ }\textbf
  {\bibinfo {volume} {19}},\ \bibinfo {pages} {506210} (\bibinfo {year}
  {2007}{\natexlab{a}})}\BibitemShut {NoStop}%
\bibitem [{\citenamefont {Shigetoh}\ \emph {et~al.}(2007)\citenamefont
  {Shigetoh}, \citenamefont {Ishida}, \citenamefont {Ayabe}, \citenamefont
  {Onimaru}, \citenamefont {Umeo}, \citenamefont {Muro}, \citenamefont
  {Motoya}, \citenamefont {Sera},\ and\ \citenamefont {Takabatake}}]{Shigetoh}%
  \BibitemOpen
  \bibfield  {author} {\bibinfo {author} {\bibfnamefont {K.}~\bibnamefont
  {Shigetoh}}, \bibinfo {author} {\bibfnamefont {A.}~\bibnamefont {Ishida}},
  \bibinfo {author} {\bibfnamefont {Y.}~\bibnamefont {Ayabe}}, \bibinfo
  {author} {\bibfnamefont {T.}~\bibnamefont {Onimaru}}, \bibinfo {author}
  {\bibfnamefont {K.}~\bibnamefont {Umeo}}, \bibinfo {author} {\bibfnamefont
  {Y.}~\bibnamefont {Muro}}, \bibinfo {author} {\bibfnamefont {K.}~\bibnamefont
  {Motoya}}, \bibinfo {author} {\bibfnamefont {M.}~\bibnamefont {Sera}},\ and\
  \bibinfo {author} {\bibfnamefont {T.}~\bibnamefont {Takabatake}},\ }\bibfield
   {title} {\bibinfo {title} {Easy-plane magnetocrystalline anisotropy in the
  multistep metamagnet {CeIr}$_3${Si}$_2$},\ }\href
  {https://doi.org/10.1103/PhysRevB.76.184429} {\bibfield  {journal} {\bibinfo
  {journal} {Phys. Rev. B}\ }\textbf {\bibinfo {volume} {76}},\ \bibinfo
  {pages} {184429} (\bibinfo {year} {2007})}\BibitemShut {NoStop}%
\bibitem [{\citenamefont {Muro}\ \emph {et~al.}(2007)\citenamefont {Muro},
  \citenamefont {Ohno}, \citenamefont {Okada},\ and\ \citenamefont
  {Motoya}}]{Muro}%
  \BibitemOpen
  \bibfield  {author} {\bibinfo {author} {\bibfnamefont {Y.}~\bibnamefont
  {Muro}}, \bibinfo {author} {\bibfnamefont {Y.}~\bibnamefont {Ohno}}, \bibinfo
  {author} {\bibfnamefont {T.}~\bibnamefont {Okada}},\ and\ \bibinfo {author}
  {\bibfnamefont {K.}~\bibnamefont {Motoya}},\ }\bibfield  {title} {\bibinfo
  {title} {Multi-step metamagnetism in {CeIr}$_3${Si}$_2$},\ }\href
  {https://doi.org/10.1016/j.jmmm.2006.10.097} {\bibfield  {journal} {\bibinfo
  {journal} {Journal of Magnetism and Magnetic Materials}\ }\textbf {\bibinfo
  {volume} {310}},\ \bibinfo {pages} {389} (\bibinfo {year}
  {2007})}\BibitemShut {NoStop}%
\bibitem [{\citenamefont {Bewley}\ \emph {et~al.}(2006)\citenamefont {Bewley},
  \citenamefont {Eccleston}, \citenamefont {McEwen}, \citenamefont {Hayden},
  \citenamefont {Dove}, \citenamefont {Bennington}, \citenamefont {Treadgold},\
  and\ \citenamefont {Coleman}}]{BEWLEY20061029}%
  \BibitemOpen
  \bibfield  {author} {\bibinfo {author} {\bibfnamefont {R.~I.}\ \bibnamefont
  {Bewley}}, \bibinfo {author} {\bibfnamefont {R.~S.}\ \bibnamefont
  {Eccleston}}, \bibinfo {author} {\bibfnamefont {K.~A.}\ \bibnamefont
  {McEwen}}, \bibinfo {author} {\bibfnamefont {S.~M.}\ \bibnamefont {Hayden}},
  \bibinfo {author} {\bibfnamefont {M.~T.}\ \bibnamefont {Dove}}, \bibinfo
  {author} {\bibfnamefont {S.~M.}\ \bibnamefont {Bennington}}, \bibinfo
  {author} {\bibfnamefont {J.~R.}\ \bibnamefont {Treadgold}},\ and\ \bibinfo
  {author} {\bibfnamefont {R.~L.~S.}\ \bibnamefont {Coleman}},\ }\bibfield
  {title} {\bibinfo {title} {{MERLIN}, a new high count rate spectrometer at
  {ISIS}},\ }\href
  {https://doi.org/https://doi.org/10.1016/j.physb.2006.05.328} {\bibfield
  {journal} {\bibinfo  {journal} {Physica B: Condensed Matter}\ }\textbf
  {\bibinfo {volume} {385-386}},\ \bibinfo {pages} {1029} (\bibinfo {year}
  {2006})}\BibitemShut {NoStop}%
\bibitem [{\citenamefont {Brookes}\ \emph {et~al.}(2018)\citenamefont
  {Brookes}, \citenamefont {Yakhou-Harris}, \citenamefont {Kummer},
  \citenamefont {Fondacaro}, \citenamefont {Cezar}, \citenamefont {Betto},
  \citenamefont {Velez-Fort}, \citenamefont {Amorese}, \citenamefont
  {Ghiringhelli}, \citenamefont {Braicovich}, \citenamefont {Barrett},
  \citenamefont {Berruyer}, \citenamefont {Cianciosi}, \citenamefont {Eybert},
  \citenamefont {Marion}, \citenamefont {{van der Linden}},\ and\ \citenamefont
  {Zhang}}]{ID32}%
  \BibitemOpen
  \bibfield  {author} {\bibinfo {author} {\bibfnamefont {N.~B.}\ \bibnamefont
  {Brookes}}, \bibinfo {author} {\bibfnamefont {F.}~\bibnamefont
  {Yakhou-Harris}}, \bibinfo {author} {\bibfnamefont {K.}~\bibnamefont
  {Kummer}}, \bibinfo {author} {\bibfnamefont {A.}~\bibnamefont {Fondacaro}},
  \bibinfo {author} {\bibfnamefont {J.~C.}\ \bibnamefont {Cezar}}, \bibinfo
  {author} {\bibfnamefont {D.}~\bibnamefont {Betto}}, \bibinfo {author}
  {\bibfnamefont {E.}~\bibnamefont {Velez-Fort}}, \bibinfo {author}
  {\bibfnamefont {A.}~\bibnamefont {Amorese}}, \bibinfo {author} {\bibfnamefont
  {G.}~\bibnamefont {Ghiringhelli}}, \bibinfo {author} {\bibfnamefont
  {L.}~\bibnamefont {Braicovich}}, \bibinfo {author} {\bibfnamefont
  {R.}~\bibnamefont {Barrett}}, \bibinfo {author} {\bibfnamefont
  {G.}~\bibnamefont {Berruyer}}, \bibinfo {author} {\bibfnamefont
  {F.}~\bibnamefont {Cianciosi}}, \bibinfo {author} {\bibfnamefont
  {L.}~\bibnamefont {Eybert}}, \bibinfo {author} {\bibfnamefont
  {P.}~\bibnamefont {Marion}}, \bibinfo {author} {\bibfnamefont
  {P.}~\bibnamefont {{van der Linden}}},\ and\ \bibinfo {author} {\bibfnamefont
  {L.}~\bibnamefont {Zhang}},\ }\bibfield  {title} {\bibinfo {title} {{The
  beamline ID32 at the ESRF for soft X-ray high energy resolution resonant
  inelastic X-ray scattering and polarisation dependent X-ray absorption
  spectroscopy}},\ }\href
  {https://doi.org/https://doi.org/10.1016/j.nima.2018.07.001} {\bibfield
  {journal} {\bibinfo  {journal} {Nuclear Instruments and Methods in Physics
  Research Section A: Accelerators, Spectrometers, Detectors and Associated
  Equipment}\ }\textbf {\bibinfo {volume} {903}},\ \bibinfo {pages} {175}
  (\bibinfo {year} {2018})}\BibitemShut {NoStop}%
\bibitem [{\citenamefont {Kummer}\ \emph {et~al.}(2017)\citenamefont {Kummer},
  \citenamefont {Tamborrino}, \citenamefont {Amorese}, \citenamefont {Minola},
  \citenamefont {Braicovich}, \citenamefont {Brookes},\ and\ \citenamefont
  {Ghiringhelli}}]{RixsToolBox}%
  \BibitemOpen
  \bibfield  {author} {\bibinfo {author} {\bibfnamefont {K.}~\bibnamefont
  {Kummer}}, \bibinfo {author} {\bibfnamefont {A.}~\bibnamefont {Tamborrino}},
  \bibinfo {author} {\bibfnamefont {A.}~\bibnamefont {Amorese}}, \bibinfo
  {author} {\bibfnamefont {M.}~\bibnamefont {Minola}}, \bibinfo {author}
  {\bibfnamefont {L.}~\bibnamefont {Braicovich}}, \bibinfo {author}
  {\bibfnamefont {N.~B.}\ \bibnamefont {Brookes}},\ and\ \bibinfo {author}
  {\bibfnamefont {G.}~\bibnamefont {Ghiringhelli}},\ }\bibfield  {title}
  {\bibinfo {title} {{RixsToolBox: software for the analysis of soft X-ray RIXS
  data acquired with 2D detectors}},\ }\href
  {https://doi.org/10.1107/S1600577517000832} {\bibfield  {journal} {\bibinfo
  {journal} {Journal of synchrotron radiation}\ }\textbf {\bibinfo {volume}
  {24}},\ \bibinfo {pages} {531} (\bibinfo {year} {2017})}\BibitemShut
  {NoStop}%
\bibitem [{\citenamefont {Amorese}\ \emph
  {et~al.}(2019{\natexlab{b}})\citenamefont {Amorese}, \citenamefont {Langini},
  \citenamefont {Dellea}, \citenamefont {Kummer}, \citenamefont {Brookes},
  \citenamefont {Braicovich},\ and\ \citenamefont
  {Ghiringhelli}}]{AmoreseCCD1}%
  \BibitemOpen
  \bibfield  {author} {\bibinfo {author} {\bibfnamefont {A.}~\bibnamefont
  {Amorese}}, \bibinfo {author} {\bibfnamefont {C.}~\bibnamefont {Langini}},
  \bibinfo {author} {\bibfnamefont {G.}~\bibnamefont {Dellea}}, \bibinfo
  {author} {\bibfnamefont {K.}~\bibnamefont {Kummer}}, \bibinfo {author}
  {\bibfnamefont {N.~B.}\ \bibnamefont {Brookes}}, \bibinfo {author}
  {\bibfnamefont {L.}~\bibnamefont {Braicovich}},\ and\ \bibinfo {author}
  {\bibfnamefont {G.}~\bibnamefont {Ghiringhelli}},\ }\bibfield  {title}
  {\bibinfo {title} {Enhanced spatial resolution of commercial soft {X-ray CCD}
  detectors by single-photon centroid reconstruction},\ }\href
  {https://doi.org/https://doi.org/10.1016/j.nima.2019.03.010} {\bibfield
  {journal} {\bibinfo  {journal} {Nuclear Instruments and Methods in Physics
  Research Section A: Accelerators, Spectrometers, Detectors and Associated
  Equipment}\ }\textbf {\bibinfo {volume} {935}},\ \bibinfo {pages} {222}
  (\bibinfo {year} {2019}{\natexlab{b}})}\BibitemShut {NoStop}%
\bibitem [{\citenamefont {Amorese}\ \emph
  {et~al.}(2019{\natexlab{c}})\citenamefont {Amorese}, \citenamefont {Langini},
  \citenamefont {Dellea}, \citenamefont {Kummer}, \citenamefont {Brookes},
  \citenamefont {Braicovich},\ and\ \citenamefont
  {Ghiringhelli}}]{AmoreseCCD2}%
  \BibitemOpen
  \bibfield  {author} {\bibinfo {author} {\bibfnamefont {A.}~\bibnamefont
  {Amorese}}, \bibinfo {author} {\bibfnamefont {C.}~\bibnamefont {Langini}},
  \bibinfo {author} {\bibfnamefont {G.}~\bibnamefont {Dellea}}, \bibinfo
  {author} {\bibfnamefont {K.}~\bibnamefont {Kummer}}, \bibinfo {author}
  {\bibfnamefont {N.~B.}\ \bibnamefont {Brookes}}, \bibinfo {author}
  {\bibfnamefont {L.}~\bibnamefont {Braicovich}},\ and\ \bibinfo {author}
  {\bibfnamefont {G.}~\bibnamefont {Ghiringhelli}},\ }\bibfield  {title}
  {\bibinfo {title} {{Enhanced spatial resolution of commercial soft X-ray CCD
  detectors by single-photon centroid reconstruction: Technical Note}},\ }\href
  {https://doi.org/https://doi.org/10.1016/j.nima.2019.02.044} {\bibfield
  {journal} {\bibinfo  {journal} {Nuclear Instruments and Methods in Physics
  Research Section A: Accelerators, Spectrometers, Detectors and Associated
  Equipment}\ }\textbf {\bibinfo {volume} {935}},\ \bibinfo {pages} {227}
  (\bibinfo {year} {2019}{\natexlab{c}})}\BibitemShut {NoStop}%
 \bibitem []{SuppMat}%
  \BibitemOpen
  \bibfield  {title}
  {\bibinfo {title} {{See Supplemental Material at [URL to be inserted by publisher] for the full set of RIXS spectra acquired and the corresponding crystal field calculations, for more information on the crystal field scheme proposed in this work, and for comparison with the crystal field scheme and reference system used in previous works}}\ }\BibitemShut {NoStop}%
\bibitem [{\citenamefont {Chapon}\ \emph {et~al.}(2011)\citenamefont {Chapon},
  \citenamefont {Manuel}, \citenamefont {Radaelli}, \citenamefont {Benson},
  \citenamefont {Perrott}, \citenamefont {Ansell}, \citenamefont {Rhodes},
  \citenamefont {Raspino}, \citenamefont {Duxbury}, \citenamefont {Spill},\
  and\ \citenamefont {Norris}}]{doi:10.1080/10448632.2011.569650}%
  \BibitemOpen
  \bibfield  {author} {\bibinfo {author} {\bibfnamefont {L.~C.}\ \bibnamefont
  {Chapon}}, \bibinfo {author} {\bibfnamefont {P.}~\bibnamefont {Manuel}},
  \bibinfo {author} {\bibfnamefont {P.~G.}\ \bibnamefont {Radaelli}}, \bibinfo
  {author} {\bibfnamefont {C.}~\bibnamefont {Benson}}, \bibinfo {author}
  {\bibfnamefont {L.}~\bibnamefont {Perrott}}, \bibinfo {author} {\bibfnamefont
  {S.}~\bibnamefont {Ansell}}, \bibinfo {author} {\bibfnamefont {N.~J.}\
  \bibnamefont {Rhodes}}, \bibinfo {author} {\bibfnamefont {D.}~\bibnamefont
  {Raspino}}, \bibinfo {author} {\bibfnamefont {D.}~\bibnamefont {Duxbury}},
  \bibinfo {author} {\bibfnamefont {E.}~\bibnamefont {Spill}},\ and\ \bibinfo
  {author} {\bibfnamefont {J.}~\bibnamefont {Norris}},\ }\bibfield  {title}
  {\bibinfo {title} {{WISH}: The new powder and single crystal magnetic
  diffractometer on the second target station},\ }\href
  {https://doi.org/10.1080/10448632.2011.569650} {\bibfield  {journal}
  {\bibinfo  {journal} {Neutron News}\ }\textbf {\bibinfo {volume} {22}},\
  \bibinfo {pages} {22} (\bibinfo {year} {2011})},\ \Eprint
  {https://arxiv.org/abs/https://doi.org/10.1080/10448632.2011.569650}
  {https://doi.org/10.1080/10448632.2011.569650} \BibitemShut {NoStop}%
\bibitem [{\citenamefont {Carvajal}(1993)}]{FullProf}%
  \BibitemOpen
  \bibfield  {author} {\bibinfo {author} {\bibfnamefont {J.~R.}\ \bibnamefont
  {Carvajal}},\ }\bibfield  {title} {\bibinfo {title} {Recent advances in
  magnetic structure determination by neutron powder diffraction},\ }\href
  {https://doi.org/10.1016/0921-4526(93)90108-I} {\bibfield  {journal}
  {\bibinfo  {journal} {Physica B}\ }\textbf {\bibinfo {volume} {192}},\
  \bibinfo {pages} {55} (\bibinfo {year} {1993})}\BibitemShut {NoStop}%
\bibitem [{\citenamefont {Wybourne}\ and\ \citenamefont
  {Meggers}(1965)}]{Wybourne1965}%
  \BibitemOpen
  \bibfield  {author} {\bibinfo {author} {\bibfnamefont {B.~G.}\ \bibnamefont
  {Wybourne}}\ and\ \bibinfo {author} {\bibfnamefont {W.~F.}\ \bibnamefont
  {Meggers}},\ }\bibfield  {title} {\bibinfo {title} {Spectroscopic properties
  of rare earths},\ }\href {https://doi.org/10.1063/1.3047727} {\bibfield
  {journal} {\bibinfo  {journal} {Physics Today}\ }\textbf {\bibinfo {volume}
  {18}},\ \bibinfo {pages} {70} (\bibinfo {year} {1965})}\BibitemShut {NoStop}%
\bibitem [{\citenamefont {Rotter}()}]{McPhase}%
  \BibitemOpen
  \bibfield  {author} {\bibinfo {author} {\bibfnamefont {M.}~\bibnamefont
  {Rotter}},\ }\href
  {http://www2.cpfs.mpg.de/~rotter/homepage_mcphase/manual/manual.html} {\emph
  {\bibinfo {title} {McPhase User Manual}}}\BibitemShut {NoStop}%
\bibitem [{\citenamefont {Boothroyd}(2020)}]{Boothroyd}%
  \BibitemOpen
  \bibfield  {author} {\bibinfo {author} {\bibfnamefont {A.~T.}\ \bibnamefont
  {Boothroyd}},\ }\href {https://doi.org/10.1093/oso/9780198862314.001.0001}
  {\emph {\bibinfo {title} {Principles of Neutron Scattering from Condensed
  Matter}}}\ (\bibinfo  {publisher} {Oxford University Press},\ \bibinfo {year}
  {2020})\BibitemShut {NoStop}%
\bibitem [{\citenamefont {Haverkort}\ \emph {et~al.}(2012)\citenamefont
  {Haverkort}, \citenamefont {Zwierzycki},\ and\ \citenamefont
  {Andersen}}]{Haverkort2012}%
  \BibitemOpen
  \bibfield  {author} {\bibinfo {author} {\bibfnamefont {M.~W.}\ \bibnamefont
  {Haverkort}}, \bibinfo {author} {\bibfnamefont {M.}~\bibnamefont
  {Zwierzycki}},\ and\ \bibinfo {author} {\bibfnamefont {O.~K.}\ \bibnamefont
  {Andersen}},\ }\bibfield  {title} {\bibinfo {title} {Multiplet ligand-field
  theory using {W}annier orbitals},\ }\href
  {https://doi.org/10.1103/PhysRevB.85.165113} {\bibfield  {journal} {\bibinfo
  {journal} {Phys. Rev. B}\ }\textbf {\bibinfo {volume} {85}},\ \bibinfo
  {pages} {165113} (\bibinfo {year} {2012})}\BibitemShut {NoStop}%
\bibitem [{\citenamefont {Rotter}(2004)}]{ROTTER2004E481}%
  \BibitemOpen
  \bibfield  {author} {\bibinfo {author} {\bibfnamefont {M.}~\bibnamefont
  {Rotter}},\ }\bibfield  {title} {\bibinfo {title} {Using {McPhase} to
  calculate magnetic phase diagrams of rare earth compounds},\ }\href
  {https://doi.org/https://doi.org/10.1016/j.jmmm.2003.12.1394} {\bibfield
  {journal} {\bibinfo  {journal} {Journal of Magnetism and Magnetic Materials}\
  }\textbf {\bibinfo {volume} {272-276}},\ \bibinfo {pages} {E481} (\bibinfo
  {year} {2004})},\ \bibinfo {note} {proceedings of the International
  Conference on Magnetism (ICM 2003)}\BibitemShut {NoStop}%
\bibitem [{SPE()}]{SPECTRE}%
  \BibitemOpen
  \href@noop {} {}\bibinfo {howpublished} {A. T. Boothroyd, SPECTRE: a program
  for calculating spectroscopic properties of rare-earth ions in crystals
  (1990--2014). Available at
  https://xray.physics.ox.ac.uk/software.htm}\BibitemShut {NoStop}%
\bibitem [{\citenamefont {Zaliznyak}\ and\ \citenamefont
  {Lee}(2005)}]{Zaliznyak}%
  \BibitemOpen
  \bibfield  {author} {\bibinfo {author} {\bibfnamefont {I.~A.}\ \bibnamefont
  {Zaliznyak}}\ and\ \bibinfo {author} {\bibfnamefont {S.-H.}\ \bibnamefont
  {Lee}},\ }\bibinfo {title} {Modern techniques for characterizing magnetic
  materials}\ (\bibinfo  {publisher} {Springer Science \& Business Media},\
  \bibinfo {year} {2005})\ Chap.\ \bibinfo {chapter} {Chapter 1: Magnetic
  neutron scattering}, pp.\ \bibinfo {pages} {1 -- 85}\BibitemShut {NoStop}%
\bibitem [{\citenamefont {Campbell}\ \emph {et~al.}(2006)\citenamefont
  {Campbell}, \citenamefont {Stokes}, \citenamefont {Tanner},\ and\
  \citenamefont {Hatch}}]{ISODISPLACE}%
  \BibitemOpen
  \bibfield  {author} {\bibinfo {author} {\bibfnamefont {B.~J.}\ \bibnamefont
  {Campbell}}, \bibinfo {author} {\bibfnamefont {H.~T.}\ \bibnamefont
  {Stokes}}, \bibinfo {author} {\bibfnamefont {D.~E.}\ \bibnamefont {Tanner}},\
  and\ \bibinfo {author} {\bibfnamefont {D.~M.}\ \bibnamefont {Hatch}},\
  }\bibfield  {title} {\bibinfo {title} {{{\it ISODISPLACE}: a web-based tool
  for exploring structural distortions}},\ }\href
  {https://doi.org/10.1107/S0021889806014075} {\bibfield  {journal} {\bibinfo
  {journal} {Journal of Applied Crystallography}\ }\textbf {\bibinfo {volume}
  {39}},\ \bibinfo {pages} {607} (\bibinfo {year} {2006})}\BibitemShut
  {NoStop}%
\bibitem [{\citenamefont {Perez-Mato}\ \emph {et~al.}(2015)\citenamefont
  {Perez-Mato}, \citenamefont {Gallego}, \citenamefont {Tasci}, \citenamefont
  {Elcoro}, \citenamefont {de~la Flor},\ and\ \citenamefont {Aroyo}}]{MAXSYM}%
  \BibitemOpen
  \bibfield  {author} {\bibinfo {author} {\bibfnamefont {J.}~\bibnamefont
  {Perez-Mato}}, \bibinfo {author} {\bibfnamefont {S.}~\bibnamefont {Gallego}},
  \bibinfo {author} {\bibfnamefont {E.}~\bibnamefont {Tasci}}, \bibinfo
  {author} {\bibfnamefont {L.}~\bibnamefont {Elcoro}}, \bibinfo {author}
  {\bibfnamefont {G.}~\bibnamefont {de~la Flor}},\ and\ \bibinfo {author}
  {\bibfnamefont {M.}~\bibnamefont {Aroyo}},\ }\bibfield  {title} {\bibinfo
  {title} {Symmetry-based computational tools for magnetic crystallography},\
  }\href {https://doi.org/10.1146/annurev-matsci-070214-021008} {\bibfield
  {journal} {\bibinfo  {journal} {Annual Review of Materials Research}\
  }\textbf {\bibinfo {volume} {45}},\ \bibinfo {pages} {217} (\bibinfo {year}
  {2015})},\ \Eprint
  {https://arxiv.org/abs/Https://doi.org/10.1146/annurev-matsci-070214-021008}
  {Https://doi.org/10.1146/annurev-matsci-070214-021008} \BibitemShut {NoStop}%
\bibitem [{\citenamefont {Anand}\ \emph
  {et~al.}(2018{\natexlab{b}})\citenamefont {Anand}, \citenamefont {Hillier},
  \citenamefont {Adroja}, \citenamefont {Khalyavin}, \citenamefont {Manuel},
  \citenamefont {Andre}, \citenamefont {Rols},\ and\ \citenamefont
  {Koza}}]{Anand2018}%
  \BibitemOpen
  \bibfield  {author} {\bibinfo {author} {\bibfnamefont {V.~K.}\ \bibnamefont
  {Anand}}, \bibinfo {author} {\bibfnamefont {A.~D.}\ \bibnamefont {Hillier}},
  \bibinfo {author} {\bibfnamefont {D.~T.}\ \bibnamefont {Adroja}}, \bibinfo
  {author} {\bibfnamefont {D.~D.}\ \bibnamefont {Khalyavin}}, \bibinfo {author}
  {\bibfnamefont {P.}~\bibnamefont {Manuel}}, \bibinfo {author} {\bibfnamefont
  {G.}~\bibnamefont {Andre}}, \bibinfo {author} {\bibfnamefont
  {S.}~\bibnamefont {Rols}},\ and\ \bibinfo {author} {\bibfnamefont {M.~M.}\
  \bibnamefont {Koza}},\ }\bibfield  {title} {\bibinfo {title} {{Understanding
  the magnetism in noncentrosymmetric CeIrGe}$_3$: {M}uon spin relaxation and
  neutron scattering studies},\ }\href
  {https://doi.org/10.1103/PhysRevB.97.184422} {\bibfield  {journal} {\bibinfo
  {journal} {Phys. Rev. B}\ }\textbf {\bibinfo {volume} {97}},\ \bibinfo
  {pages} {184422} (\bibinfo {year} {2018}{\natexlab{b}})}\BibitemShut
  {NoStop}%
\bibitem [{\citenamefont {Hillier}\ \emph {et~al.}(2012)\citenamefont
  {Hillier}, \citenamefont {Adroja}, \citenamefont {Manuel}, \citenamefont
  {Anand}, \citenamefont {Taylor}, \citenamefont {McEwen},\ and\ \citenamefont
  {Rainford}}]{Hillier2012}%
  \BibitemOpen
  \bibfield  {author} {\bibinfo {author} {\bibfnamefont {A.~D.}\ \bibnamefont
  {Hillier}}, \bibinfo {author} {\bibfnamefont {D.~T.}\ \bibnamefont {Adroja}},
  \bibinfo {author} {\bibfnamefont {P.}~\bibnamefont {Manuel}}, \bibinfo
  {author} {\bibfnamefont {V.~K.}\ \bibnamefont {Anand}}, \bibinfo {author}
  {\bibfnamefont {J.~W.}\ \bibnamefont {Taylor}}, \bibinfo {author}
  {\bibfnamefont {K.~A.}\ \bibnamefont {McEwen}},\ and\ \bibinfo {author}
  {\bibfnamefont {M.~M.}\ \bibnamefont {Rainford}, \bibfnamefont
  {B.~D.~Koza}},\ }\bibfield  {title} {\bibinfo {title} {Muon spin relaxation
  and neutron scattering investigations of the noncentrosymmetric heavy-fermion
  antiferromagnet {CeRhGe}$_ 3$},\ }\href
  {https://doi.org/10.1103/PhysRevB.85.134405} {\bibfield  {journal} {\bibinfo
  {journal} {Phys. Rev. B}\ }\textbf {\bibinfo {volume} {85}},\ \bibinfo
  {pages} {134405} (\bibinfo {year} {2012})}\BibitemShut {NoStop}%
\bibitem [{\citenamefont {Aroyo}\ \emph {et~al.}(2006)\citenamefont {Aroyo},
  \citenamefont {Kirov}, \citenamefont {Capillas}, \citenamefont {Perez-Mato},\
  and\ \citenamefont {Wondratschek}}]{Aroyo2006}%
  \BibitemOpen
  \bibfield  {author} {\bibinfo {author} {\bibfnamefont {M.~I.}\ \bibnamefont
  {Aroyo}}, \bibinfo {author} {\bibfnamefont {A.}~\bibnamefont {Kirov}},
  \bibinfo {author} {\bibfnamefont {C.}~\bibnamefont {Capillas}}, \bibinfo
  {author} {\bibfnamefont {J.~M.}\ \bibnamefont {Perez-Mato}},\ and\ \bibinfo
  {author} {\bibfnamefont {H.}~\bibnamefont {Wondratschek}},\ }\bibfield
  {title} {\bibinfo {title} {{Bilbao Crystallographic Server. II.
  Representations of crystallographic point groups and space groups}},\ }\href
  {https://doi.org/10.1107/S0108767305040286} {\bibfield  {journal} {\bibinfo
  {journal} {Acta Cryst.}\ }\textbf {\bibinfo {volume} {A62}},\ \bibinfo
  {pages} {115} (\bibinfo {year} {2006})}\BibitemShut {NoStop}%
\bibitem [{\citenamefont {Izyumov}\ \emph {et~al.}(1991)\citenamefont
  {Izyumov}, \citenamefont {Naish},\ and\ \citenamefont {Ozerov}}]{IzumovBook}%
  \BibitemOpen
  \bibfield  {author} {\bibinfo {author} {\bibfnamefont {Y.~A.}\ \bibnamefont
  {Izyumov}}, \bibinfo {author} {\bibfnamefont {V.~E.}\ \bibnamefont {Naish}},\
  and\ \bibinfo {author} {\bibfnamefont {R.~P.}\ \bibnamefont {Ozerov}},\
  }\href@noop {} {\emph {\bibinfo {title} {Neutron diffraction of magnetic
  materials}}}\ (\bibinfo  {publisher} {New York: Consulting Bureau},\ \bibinfo
  {year} {1991})\BibitemShut {NoStop}%
\bibitem [{\citenamefont {Givord}\ \emph
  {et~al.}(2007{\natexlab{b}})\citenamefont {Givord}, \citenamefont
  {Boucherle}, \citenamefont {Gal{\'{e}}ra}, \citenamefont {Fillion},\ and\
  \citenamefont {Lejay}}]{Givord2007a}%
  \BibitemOpen
  \bibfield  {author} {\bibinfo {author} {\bibfnamefont {F.}~\bibnamefont
  {Givord}}, \bibinfo {author} {\bibfnamefont {J.-X.}\ \bibnamefont
  {Boucherle}}, \bibinfo {author} {\bibfnamefont {R.-M.}\ \bibnamefont
  {Gal{\'{e}}ra}}, \bibinfo {author} {\bibfnamefont {G.}~\bibnamefont
  {Fillion}},\ and\ \bibinfo {author} {\bibfnamefont {P.}~\bibnamefont
  {Lejay}},\ }\bibfield  {title} {\bibinfo {title} {Ferromagnetism and crystal
  electric field in the cerium compound {CeRh}$_3${B}$_2$},\ }\href
  {https://doi.org/10.1088/0953-8984/19/35/356208} {\bibfield  {journal}
  {\bibinfo  {journal} {J. Phys.: Cond. Matter}\ }\textbf {\bibinfo {volume}
  {19}},\ \bibinfo {pages} {356208} (\bibinfo {year}
  {2007}{\natexlab{b}})}\BibitemShut {NoStop}%
\bibitem [{\citenamefont {Givord}\ \emph
  {et~al.}(2007{\natexlab{c}})\citenamefont {Givord}, \citenamefont
  {Boucherle}, \citenamefont {Murani}, \citenamefont {Bewley}, \citenamefont
  {Gal{\'{e}}ra},\ and\ \citenamefont {Lejay}}]{Givord2007}%
  \BibitemOpen
  \bibfield  {author} {\bibinfo {author} {\bibfnamefont {F.}~\bibnamefont
  {Givord}}, \bibinfo {author} {\bibfnamefont {J.-X.}\ \bibnamefont
  {Boucherle}}, \bibinfo {author} {\bibfnamefont {A.~P.}\ \bibnamefont
  {Murani}}, \bibinfo {author} {\bibfnamefont {R.}~\bibnamefont {Bewley}},
  \bibinfo {author} {\bibfnamefont {R.-M.}\ \bibnamefont {Gal{\'{e}}ra}},\ and\
  \bibinfo {author} {\bibfnamefont {P.}~\bibnamefont {Lejay}},\ }\bibfield
  {title} {\bibinfo {title} {Crystal electric field excitations in the cerium
  compound {CeRh}$_3${B}$_2$ studied by inelastic neutron scattering},\ }\href
  {https://doi.org/10.1088/0953-8984/19/50/506210} {\bibfield  {journal}
  {\bibinfo  {journal} {Journal of Physics: Condensed Matter}\ }\textbf
  {\bibinfo {volume} {19}},\ \bibinfo {pages} {506210} (\bibinfo {year}
  {2007}{\natexlab{c}})}\BibitemShut {NoStop}%
\bibitem [{\citenamefont {Ball}\ \emph {et~al.}(1993)\citenamefont {Ball},
  \citenamefont {Gignoux}, \citenamefont {Murani},\ and\ \citenamefont
  {Schmitt}}]{PrGa2}%
  \BibitemOpen
  \bibfield  {author} {\bibinfo {author} {\bibfnamefont {A.~R.}\ \bibnamefont
  {Ball}}, \bibinfo {author} {\bibfnamefont {D.}~\bibnamefont {Gignoux}},
  \bibinfo {author} {\bibfnamefont {A.~P.}\ \bibnamefont {Murani}},\ and\
  \bibinfo {author} {\bibfnamefont {D.}~\bibnamefont {Schmitt}},\ }\bibfield
  {title} {\bibinfo {title} {Crystal field and complex phase diagram in
  hexagonal {PrGa}$_2$},\ }\href
  {https://doi.org/https://doi.org/10.1016/0921-4526(93)90468-L} {\bibfield
  {journal} {\bibinfo  {journal} {Physica B: Condensed Matter}\ }\textbf
  {\bibinfo {volume} {190}},\ \bibinfo {pages} {214} (\bibinfo {year}
  {1993})}\BibitemShut {NoStop}%
\bibitem [{\citenamefont {Ball}\ \emph {et~al.}(1994)\citenamefont {Ball},
  \citenamefont {Gignoux}, \citenamefont {Rodriguez~Fernandez},\ and\
  \citenamefont {Schmitt}}]{NdGa2}%
  \BibitemOpen
  \bibfield  {author} {\bibinfo {author} {\bibfnamefont {A.~R.}\ \bibnamefont
  {Ball}}, \bibinfo {author} {\bibfnamefont {D.}~\bibnamefont {Gignoux}},
  \bibinfo {author} {\bibfnamefont {J.}~\bibnamefont {Rodriguez~Fernandez}},\
  and\ \bibinfo {author} {\bibfnamefont {D.}~\bibnamefont {Schmitt}},\
  }\bibfield  {title} {\bibinfo {title} {Magnetic properties and complex phase
  diagram of hexagonal {NdGa}$_2$},\ }\href
  {https://doi.org/https://doi.org/10.1016/0304-8853(94)90714-5} {\bibfield
  {journal} {\bibinfo  {journal} {Journal of Magnetism and Magnetic Materials}\
  }\textbf {\bibinfo {volume} {137}},\ \bibinfo {pages} {281} (\bibinfo {year}
  {1994})}\BibitemShut {NoStop}%
\bibitem [{\citenamefont {Bud'ko}\ \emph {et~al.}(1999)\citenamefont {Bud'ko},
  \citenamefont {Islam}, \citenamefont {Wiener}, \citenamefont {Fisher},
  \citenamefont {Lacerda},\ and\ \citenamefont {Canfield}}]{TbNi2Ge2}%
  \BibitemOpen
  \bibfield  {author} {\bibinfo {author} {\bibfnamefont {S.~L.}\ \bibnamefont
  {Bud'ko}}, \bibinfo {author} {\bibfnamefont {Z.}~\bibnamefont {Islam}},
  \bibinfo {author} {\bibfnamefont {T.~A.}\ \bibnamefont {Wiener}}, \bibinfo
  {author} {\bibfnamefont {I.~R.}\ \bibnamefont {Fisher}}, \bibinfo {author}
  {\bibfnamefont {A.~H.}\ \bibnamefont {Lacerda}},\ and\ \bibinfo {author}
  {\bibfnamefont {P.~C.}\ \bibnamefont {Canfield}},\ }\bibfield  {title}
  {\bibinfo {title} {{Anisotropy and metamagnetism in the RNi$_2$Ge$_2$ (R=Y,
  La–Nd, Sm–Lu) series}},\ }\href
  {https://doi.org/https://doi.org/10.1016/S0304-8853(99)00486-2} {\bibfield
  {journal} {\bibinfo  {journal} {Journal of Magnetism and Magnetic Materials}\
  }\textbf {\bibinfo {volume} {205}},\ \bibinfo {pages} {53} (\bibinfo {year}
  {1999})}\BibitemShut {NoStop}%
\bibitem [{\citenamefont {Shigeoka}\ \emph {et~al.}(1992)\citenamefont
  {Shigeoka}, \citenamefont {Fujii}, \citenamefont {Nishi}, \citenamefont
  {Uwatoko}, \citenamefont {Takabatake}, \citenamefont {Oguro}, \citenamefont
  {Motoya}, \citenamefont {Iwata},\ and\ \citenamefont {Ito}}]{TbNi2Si2}%
  \BibitemOpen
  \bibfield  {author} {\bibinfo {author} {\bibfnamefont {T.}~\bibnamefont
  {Shigeoka}}, \bibinfo {author} {\bibfnamefont {H.}~\bibnamefont {Fujii}},
  \bibinfo {author} {\bibfnamefont {M.}~\bibnamefont {Nishi}}, \bibinfo
  {author} {\bibfnamefont {Y.}~\bibnamefont {Uwatoko}}, \bibinfo {author}
  {\bibfnamefont {T.}~\bibnamefont {Takabatake}}, \bibinfo {author}
  {\bibfnamefont {I.}~\bibnamefont {Oguro}}, \bibinfo {author} {\bibfnamefont
  {K.}~\bibnamefont {Motoya}}, \bibinfo {author} {\bibfnamefont
  {N.}~\bibnamefont {Iwata}},\ and\ \bibinfo {author} {\bibfnamefont
  {Y.}~\bibnamefont {Ito}},\ }\bibfield  {title} {\bibinfo {title}
  {Metamagnetism in {TbNi}$_2${Si}$_2$ single crystal},\ }\href
  {https://doi.org/10.1143/JPSJ.61.4559} {\bibfield  {journal} {\bibinfo
  {journal} {Journal of the Physical Society of Japan}\ }\textbf {\bibinfo
  {volume} {61}},\ \bibinfo {pages} {4559} (\bibinfo {year}
  {1992})}\BibitemShut {NoStop}%
\bibitem [{\citenamefont {Morosan}\ \emph {et~al.}(2004)\citenamefont
  {Morosan}, \citenamefont {Bud’ko}, \citenamefont {Canfield}, \citenamefont
  {Torikachvili},\ and\ \citenamefont {Lacerda}}]{TmAgGe}%
  \BibitemOpen
  \bibfield  {author} {\bibinfo {author} {\bibfnamefont {E.}~\bibnamefont
  {Morosan}}, \bibinfo {author} {\bibfnamefont {S.~L.}\ \bibnamefont
  {Bud’ko}}, \bibinfo {author} {\bibfnamefont {P.~C.}\ \bibnamefont
  {Canfield}}, \bibinfo {author} {\bibfnamefont {M.~S.}\ \bibnamefont
  {Torikachvili}},\ and\ \bibinfo {author} {\bibfnamefont {A.~H.}\ \bibnamefont
  {Lacerda}},\ }\bibfield  {title} {\bibinfo {title} {Thermodynamic and
  transport properties of {RAgGe (R=Tb–Lu)} single crystals},\ }\href
  {https://doi.org/https://doi.org/10.1016/j.jmmm.2003.11.014} {\bibfield
  {journal} {\bibinfo  {journal} {Journal of Magnetism and Magnetic Materials}\
  }\textbf {\bibinfo {volume} {277}},\ \bibinfo {pages} {298} (\bibinfo {year}
  {2004})}\BibitemShut {NoStop}%
\bibitem [{\citenamefont {Goddard}\ \emph {et~al.}(2007)\citenamefont
  {Goddard}, \citenamefont {Singleton}, \citenamefont {Lima~Sharma},
  \citenamefont {Morosan}, \citenamefont {Blundell}, \citenamefont {Bud'ko},\
  and\ \citenamefont {Canfield}}]{TmAgGeFrustration}%
  \BibitemOpen
  \bibfield  {author} {\bibinfo {author} {\bibfnamefont {P.~A.}\ \bibnamefont
  {Goddard}}, \bibinfo {author} {\bibfnamefont {J.}~\bibnamefont {Singleton}},
  \bibinfo {author} {\bibfnamefont {A.~L.}\ \bibnamefont {Lima~Sharma}},
  \bibinfo {author} {\bibfnamefont {E.}~\bibnamefont {Morosan}}, \bibinfo
  {author} {\bibfnamefont {S.~J.}\ \bibnamefont {Blundell}}, \bibinfo {author}
  {\bibfnamefont {S.~L.}\ \bibnamefont {Bud'ko}},\ and\ \bibinfo {author}
  {\bibfnamefont {P.~C.}\ \bibnamefont {Canfield}},\ }\bibfield  {title}
  {\bibinfo {title} {{Separation of energy scales in the kagome antiferromagnet
  TmAgGe: A magnetic-field-orientation study up to 55T}},\ }\href
  {https://doi.org/10.1103/PhysRevB.75.094426} {\bibfield  {journal} {\bibinfo
  {journal} {Phys. Rev. B}\ }\textbf {\bibinfo {volume} {75}},\ \bibinfo
  {pages} {094426} (\bibinfo {year} {2007})}\BibitemShut {NoStop}%
\bibitem [{\citenamefont {Abliz}\ \emph {et~al.}(2003)\citenamefont {Abliz},
  \citenamefont {Kindo}, \citenamefont {Kadowaki},\ and\ \citenamefont
  {Takeya}}]{HoNi2B2C}%
  \BibitemOpen
  \bibfield  {author} {\bibinfo {author} {\bibfnamefont {M.}~\bibnamefont
  {Abliz}}, \bibinfo {author} {\bibfnamefont {K.}~\bibnamefont {Kindo}},
  \bibinfo {author} {\bibfnamefont {K.}~\bibnamefont {Kadowaki}},\ and\
  \bibinfo {author} {\bibfnamefont {H.}~\bibnamefont {Takeya}},\ }\bibfield
  {title} {\bibinfo {title} {High-field magnetization measurements and
  crystalline electric-field effect in {HoNi}$_2${B}$_2${C}},\ }\href
  {https://doi.org/10.1143/jpsj.72.2599} {\bibfield  {journal} {\bibinfo
  {journal} {Journal of the Physical Society of Japan}\ }\textbf {\bibinfo
  {volume} {72}},\ \bibinfo {pages} {2599} (\bibinfo {year}
  {2003})}\BibitemShut {NoStop}%
\bibitem [{\citenamefont {Gabáni}\ \emph {et~al.}(2008)\citenamefont
  {Gabáni}, \citenamefont {Maťaš}, \citenamefont {Priputen}, \citenamefont
  {Flachbart}, \citenamefont {Siemensmeyer}, \citenamefont {Wulf},
  \citenamefont {Evdokimova},\ and\ \citenamefont {Shitsevalova}}]{TmB4}%
  \BibitemOpen
  \bibfield  {author} {\bibinfo {author} {\bibfnamefont {S.}~\bibnamefont
  {Gabáni}}, \bibinfo {author} {\bibfnamefont {S.}~\bibnamefont {Maťaš}},
  \bibinfo {author} {\bibfnamefont {P.}~\bibnamefont {Priputen}}, \bibinfo
  {author} {\bibfnamefont {K.}~\bibnamefont {Flachbart}}, \bibinfo {author}
  {\bibfnamefont {K.}~\bibnamefont {Siemensmeyer}}, \bibinfo {author}
  {\bibfnamefont {E.}~\bibnamefont {Wulf}}, \bibinfo {author} {\bibfnamefont
  {A.}~\bibnamefont {Evdokimova}},\ and\ \bibinfo {author} {\bibfnamefont
  {N.}~\bibnamefont {Shitsevalova}},\ }\bibfield  {title} {\bibinfo {title}
  {Magnetic structure and phase diagram of {TmB}$_4$},\ }\href
  {https://doi.org/10.12693/APhysPolA.113.227} {\bibfield  {journal} {\bibinfo
  {journal} {Acta Physica Polonica A}\ }\textbf {\bibinfo {volume} {113}},\
  \bibinfo {pages} {227} (\bibinfo {year} {2008})}\BibitemShut {NoStop}%
\bibitem [{\citenamefont {Rossat-Mignod}\ \emph {et~al.}(1983)\citenamefont
  {Rossat-Mignod}, \citenamefont {Burlet}, \citenamefont {Quezel},
  \citenamefont {Effantin}, \citenamefont {Delacôte}, \citenamefont
  {Bartholin}, \citenamefont {Vogt},\ and\ \citenamefont {Ravot}}]{CeSb}%
  \BibitemOpen
  \bibfield  {author} {\bibinfo {author} {\bibfnamefont {J.}~\bibnamefont
  {Rossat-Mignod}}, \bibinfo {author} {\bibfnamefont {P.}~\bibnamefont
  {Burlet}}, \bibinfo {author} {\bibfnamefont {S.}~\bibnamefont {Quezel}},
  \bibinfo {author} {\bibfnamefont {J.~M.}\ \bibnamefont {Effantin}}, \bibinfo
  {author} {\bibfnamefont {D.}~\bibnamefont {Delacôte}}, \bibinfo {author}
  {\bibfnamefont {H.}~\bibnamefont {Bartholin}}, \bibinfo {author}
  {\bibfnamefont {O.}~\bibnamefont {Vogt}},\ and\ \bibinfo {author}
  {\bibfnamefont {D.}~\bibnamefont {Ravot}},\ }\bibfield  {title} {\bibinfo
  {title} {Magnetic properties of cerium monopnictides},\ }\href
  {https://doi.org/https://doi.org/10.1016/0304-8853(83)90295-0} {\bibfield
  {journal} {\bibinfo  {journal} {Journal of Magnetism and Magnetic Materials}\
  }\textbf {\bibinfo {volume} {31-34}},\ \bibinfo {pages} {398} (\bibinfo
  {year} {1983})}\BibitemShut {NoStop}%
\bibitem [{\citenamefont {Gignoux}\ and\ \citenamefont
  {Schmitt}(1991)}]{ReviewGignoux}%
  \BibitemOpen
  \bibfield  {author} {\bibinfo {author} {\bibfnamefont {D.}~\bibnamefont
  {Gignoux}}\ and\ \bibinfo {author} {\bibfnamefont {D.}~\bibnamefont
  {Schmitt}},\ }\bibfield  {title} {\bibinfo {title} {Rare earth
  intermetallics},\ }\href
  {https://doi.org/https://doi.org/10.1016/0304-8853(91)90815-R} {\bibfield
  {journal} {\bibinfo  {journal} {Journal of Magnetism and Magnetic Materials}\
  }\textbf {\bibinfo {volume} {100}},\ \bibinfo {pages} {99} (\bibinfo {year}
  {1991})}\BibitemShut {NoStop}%
\bibitem [{\citenamefont {Date}(1990)}]{ReviewDate}%
  \BibitemOpen
  \bibfield  {author} {\bibinfo {author} {\bibfnamefont {M.}~\bibnamefont
  {Date}},\ }\bibfield  {title} {\bibinfo {title} {Field-induced magnetic phase
  transitions},\ }\href
  {https://doi.org/https://doi.org/10.1016/S0304-8853(10)80005-8} {\bibfield
  {journal} {\bibinfo  {journal} {Journal of Magnetism and Magnetic Materials}\
  }\textbf {\bibinfo {volume} {90-91}},\ \bibinfo {pages} {1} (\bibinfo {year}
  {1990})}\BibitemShut {NoStop}%
\bibitem [{\citenamefont {Kuroda}\ \emph {et~al.}(2020)\citenamefont {Kuroda},
  \citenamefont {Arai}, \citenamefont {Rezaei}, \citenamefont {Kunisada},
  \citenamefont {Sakuragi}, \citenamefont {Alaei}, \citenamefont {Kinoshita},
  \citenamefont {Bareille}, \citenamefont {Noguchi}, \citenamefont {Nakayama},
  \citenamefont {Akebi}, \citenamefont {Sakano}, \citenamefont {Kawaguchi},
  \citenamefont {Arita}, \citenamefont {Ideta}, \citenamefont {Tanaka},
  \citenamefont {Kitazawa}, \citenamefont {Okazaki}, \citenamefont {Tokunaga},
  \citenamefont {Haga}, \citenamefont {Shin}, \citenamefont {Suzuki},
  \citenamefont {Arita},\ and\ \citenamefont {Kondo}}]{CeSb_Electronic}%
  \BibitemOpen
  \bibfield  {author} {\bibinfo {author} {\bibfnamefont {K.}~\bibnamefont
  {Kuroda}}, \bibinfo {author} {\bibfnamefont {Y.}~\bibnamefont {Arai}},
  \bibinfo {author} {\bibfnamefont {N.}~\bibnamefont {Rezaei}}, \bibinfo
  {author} {\bibfnamefont {S.}~\bibnamefont {Kunisada}}, \bibinfo {author}
  {\bibfnamefont {S.}~\bibnamefont {Sakuragi}}, \bibinfo {author}
  {\bibfnamefont {M.}~\bibnamefont {Alaei}}, \bibinfo {author} {\bibfnamefont
  {Y.}~\bibnamefont {Kinoshita}}, \bibinfo {author} {\bibfnamefont
  {C.}~\bibnamefont {Bareille}}, \bibinfo {author} {\bibfnamefont
  {R.}~\bibnamefont {Noguchi}}, \bibinfo {author} {\bibfnamefont
  {M.}~\bibnamefont {Nakayama}}, \bibinfo {author} {\bibfnamefont
  {S.}~\bibnamefont {Akebi}}, \bibinfo {author} {\bibfnamefont
  {M.}~\bibnamefont {Sakano}}, \bibinfo {author} {\bibfnamefont
  {K.}~\bibnamefont {Kawaguchi}}, \bibinfo {author} {\bibfnamefont
  {M.}~\bibnamefont {Arita}}, \bibinfo {author} {\bibfnamefont
  {S.}~\bibnamefont {Ideta}}, \bibinfo {author} {\bibfnamefont
  {K.}~\bibnamefont {Tanaka}}, \bibinfo {author} {\bibfnamefont
  {H.}~\bibnamefont {Kitazawa}}, \bibinfo {author} {\bibfnamefont
  {K.}~\bibnamefont {Okazaki}}, \bibinfo {author} {\bibfnamefont
  {M.}~\bibnamefont {Tokunaga}}, \bibinfo {author} {\bibfnamefont
  {Y.}~\bibnamefont {Haga}}, \bibinfo {author} {\bibfnamefont {S.}~\bibnamefont
  {Shin}}, \bibinfo {author} {\bibfnamefont {H.~S.}\ \bibnamefont {Suzuki}},
  \bibinfo {author} {\bibfnamefont {R.}~\bibnamefont {Arita}},\ and\ \bibinfo
  {author} {\bibfnamefont {T.}~\bibnamefont {Kondo}},\ }\bibfield  {title}
  {\bibinfo {title} {Devil's staircase transition of the electronic structures
  in {CeSb}},\ }\href {https://doi.org/10.1038/s41467-020-16707-6} {\bibfield
  {journal} {\bibinfo  {journal} {Nature Communications}\ }\textbf {\bibinfo
  {volume} {11}},\ \bibinfo {pages} {2888} (\bibinfo {year}
  {2020})}\BibitemShut {NoStop}%
\bibitem [{\citenamefont {Bak}(1986)}]{Bak}%
  \BibitemOpen
  \bibfield  {author} {\bibinfo {author} {\bibfnamefont {P.}~\bibnamefont
  {Bak}},\ }\bibfield  {title} {\bibinfo {title} {The devil{\textquotesingle}s
  staircase},\ }\href {https://doi.org/10.1063/1.881047} {\bibfield  {journal}
  {\bibinfo  {journal} {Physics Today}\ }\textbf {\bibinfo {volume} {39}},\
  \bibinfo {pages} {38} (\bibinfo {year} {1986})}\BibitemShut {NoStop}%
\bibitem [{\citenamefont {Bak}\ and\ \citenamefont
  {Bruinsma}(1982)}]{BakIsing}%
  \BibitemOpen
  \bibfield  {author} {\bibinfo {author} {\bibfnamefont {P.}~\bibnamefont
  {Bak}}\ and\ \bibinfo {author} {\bibfnamefont {R.}~\bibnamefont {Bruinsma}},\
  }\bibfield  {title} {\bibinfo {title} {One-dimensional ising model and the
  complete devil's staircase},\ }\href
  {https://doi.org/10.1103/PhysRevLett.49.249} {\bibfield  {journal} {\bibinfo
  {journal} {Phys. Rev. Lett.}\ }\textbf {\bibinfo {volume} {49}},\ \bibinfo
  {pages} {249} (\bibinfo {year} {1982})}\BibitemShut {NoStop}%
\bibitem [{\citenamefont {Bak}(1982)}]{BakIncommensurate}%
  \BibitemOpen
  \bibfield  {author} {\bibinfo {author} {\bibfnamefont {P.}~\bibnamefont
  {Bak}},\ }\bibfield  {title} {\bibinfo {title} {Commensurate phases,
  incommensurate phases and the devil{\textquotesingle}s staircase},\ }\href
  {https://doi.org/10.1088/0034-4885/45/6/001} {\bibfield  {journal} {\bibinfo
  {journal} {Reports on Progress in Physics}\ }\textbf {\bibinfo {volume}
  {45}},\ \bibinfo {pages} {587} (\bibinfo {year} {1982})}\BibitemShut
  {NoStop}%
\bibitem [{\citenamefont {Kaplan}(2009)}]{Kaplan}%
  \BibitemOpen
  \bibfield  {author} {\bibinfo {author} {\bibfnamefont {T.~A.}\ \bibnamefont
  {Kaplan}},\ }\bibfield  {title} {\bibinfo {title} {Frustrated classical
  {H}eisenberg model in one dimension with nearest-neighbor biquadratic
  exchange: Exact solution for the ground-state phase diagram},\ }\href
  {https://doi.org/10.1103/PhysRevB.80.012407} {\bibfield  {journal} {\bibinfo
  {journal} {Phys. Rev. B}\ }\textbf {\bibinfo {volume} {80}},\ \bibinfo
  {pages} {012407} (\bibinfo {year} {2009})}\BibitemShut {NoStop}%
\bibitem [{\citenamefont {Pikul}\ \emph {et~al.}(2020)\citenamefont {Pikul},
  \citenamefont {Amorese}, \citenamefont {Severing}, \citenamefont {Adroja},\
  and\ \citenamefont {Kaczorowski}}]{Proposal1}%
  \BibitemOpen
  \bibfield  {author} {\bibinfo {author} {\bibfnamefont {A.}~\bibnamefont
  {Pikul}}, \bibinfo {author} {\bibfnamefont {A.}~\bibnamefont {Amorese}},
  \bibinfo {author} {\bibfnamefont {A.}~\bibnamefont {Severing}}, \bibinfo
  {author} {\bibfnamefont {D.~T.}\ \bibnamefont {Adroja}},\ and\ \bibinfo
  {author} {\bibfnamefont {D.}~\bibnamefont {Kaczorowski}},\ }\href
  {https://doi.org/10.5286/ISIS.E.RB2010784} {\bibinfo {title} {Inelastic
  neutron scattering study to investigate giant crystal-electric-field and
  intermultiplet splitting in {CeRh}$_3${Si}$_2$}} (\bibinfo {year}
  {2020})\BibitemShut {NoStop}%
\bibitem [{\citenamefont {Adroja}\ \emph {et~al.}(2020)\citenamefont {Adroja},
  \citenamefont {Khalyavin}, \citenamefont {Pikul},\ and\ \citenamefont
  {Kaczorowski}}]{Proposal2}%
  \BibitemOpen
  \bibfield  {author} {\bibinfo {author} {\bibfnamefont {D.~T.}\ \bibnamefont
  {Adroja}}, \bibinfo {author} {\bibfnamefont {D.}~\bibnamefont {Khalyavin}},
  \bibinfo {author} {\bibfnamefont {A.}~\bibnamefont {Pikul}},\ and\ \bibinfo
  {author} {\bibfnamefont {D.}~\bibnamefont {Kaczorowski}},\ }\href
  {https://doi.org/10.5286/ISIS.E.RB2010798} {\bibinfo {title} {{Investigation
  of Multiple Magnetic Structures of CeRh$_3$Si$_2$ using Neutron Diffraction
  Study}}} (\bibinfo {year} {2020})\BibitemShut {NoStop}%
\bibitem [{\citenamefont {Wojdyr}(2010)}]{fityk}%
  \BibitemOpen
  \bibfield  {author} {\bibinfo {author} {\bibfnamefont {M.}~\bibnamefont
  {Wojdyr}},\ }\bibfield  {title} {\bibinfo {title} {{{\it Fityk}: a
  general-purpose peak fitting program}},\ }\href
  {https://doi.org/10.1107/S0021889810030499} {\bibfield  {journal} {\bibinfo
  {journal} {Journal of Applied Crystallography}\ }\textbf {\bibinfo {volume}
  {43}},\ \bibinfo {pages} {1126} (\bibinfo {year} {2010})}\BibitemShut
  {NoStop}%
\bibitem [{\citenamefont {Cowan}(1981)}]{CowanBook}%
  \BibitemOpen
  \bibfield  {author} {\bibinfo {author} {\bibfnamefont {R.}~\bibnamefont
  {Cowan}},\ }\href@noop {} {\emph {\bibinfo {title} {The Theory of Atomic
  Structure and Spectra.}}}\ (\bibinfo  {publisher} {University of California,
  Berkeley},\ \bibinfo {year} {1981})\BibitemShut {NoStop}%
\end{thebibliography}

\newpage



\begin{thebibliography}{61}%
\makeatletter
\providecommand \@ifxundefined [1]{%
 \@ifx{#1\undefined}
}%
\providecommand \@ifnum [1]{%
 \ifnum #1\expandafter \@firstoftwo
 \else \expandafter \@secondoftwo
 \fi
}%
\providecommand \@ifx [1]{%
 \ifx #1\expandafter \@firstoftwo
 \else \expandafter \@secondoftwo
 \fi
}%
\providecommand \natexlab [1]{#1}%
\providecommand \enquote  [1]{``#1''}%
\providecommand \bibnamefont  [1]{#1}%
\providecommand \bibfnamefont [1]{#1}%
\providecommand \citenamefont [1]{#1}%
\providecommand \href@noop [0]{\@secondoftwo}%
\providecommand \href [0]{\begingroup \@sanitize@url \@href}%
\providecommand \@href[1]{\@@startlink{#1}\@@href}%
\providecommand \@@href[1]{\endgroup#1\@@endlink}%
\providecommand \@sanitize@url [0]{\catcode `\\12\catcode `\$12\catcode
  `\&12\catcode `\#12\catcode `\^12\catcode `\_12\catcode `\%12\relax}%
\providecommand \@@startlink[1]{}%
\providecommand \@@endlink[0]{}%
\providecommand \url  [0]{\begingroup\@sanitize@url \@url }%
\providecommand \@url [1]{\endgroup\@href {#1}{\urlprefix }}%
\providecommand \urlprefix  [0]{URL }%
\providecommand \Eprint [0]{\href }%
\providecommand \doibase [0]{https://doi.org/}%
\providecommand \selectlanguage [0]{\@gobble}%
\providecommand \bibinfo  [0]{\@secondoftwo}%
\providecommand \bibfield  [0]{\@secondoftwo}%
\providecommand \translation [1]{[#1]}%
\providecommand \BibitemOpen [0]{}%
\providecommand \bibitemStop [0]{}%
\providecommand \bibitemNoStop [0]{.\EOS\space}%
\providecommand \EOS [0]{\spacefactor3000\relax}%
\providecommand \BibitemShut  [1]{\csname bibitem#1\endcsname}%
\let\auto@bib@innerbib\@empty

\bibitem [{\citenamefont {Pikul}\ \emph {et~al.}(2010)\citenamefont {Pikul},
  \citenamefont {Kaczorowski}, \citenamefont {Gajek}, \citenamefont
  {St\c{e}pie\'{n}-Damm}, \citenamefont {\'{S}lebarski}, \citenamefont
  {Werwi\'{n}ski},\ and\ \citenamefont {Szajek}}]{Pikul2010}%
  \BibitemOpen
  \bibfield  {author} {\bibinfo {author} {\bibfnamefont {A.~P.}\ \bibnamefont
  {Pikul}}, \bibinfo {author} {\bibfnamefont {D.}~\bibnamefont {Kaczorowski}},
  \bibinfo {author} {\bibfnamefont {Z.}~\bibnamefont {Gajek}}, \bibinfo
  {author} {\bibfnamefont {J.}~\bibnamefont {St\c{e}pie\'{n}-Damm}}, \bibinfo
  {author} {\bibfnamefont {A.}~\bibnamefont {\'{S}lebarski}}, \bibinfo {author}
  {\bibfnamefont {M.}~\bibnamefont {Werwi\'{n}ski}},\ and\ \bibinfo {author}
  {\bibfnamefont {A.}~\bibnamefont {Szajek}},\ }\bibfield  {title} {\bibinfo
  {title} {Giant crystal-electric-field effect and complex magnetic behavior in
  single-crystalline {CeRh}$_3${Si}$_2$},\ }\href
  {https://doi.org/10.1103/PhysRevB.81.174408} {\bibfield  {journal} {\bibinfo
  {journal} {PhysRevB}\ }\textbf {\bibinfo {volume} {81}},\ \bibinfo
  {pages} {174408} (\bibinfo {year} {2010})}\BibitemShut {NoStop}%

\end{thebibliography}
%
%
%
%
\section{Supplemental Material}

\subsection{Experimental RIXS spectra and crystal field calculation}
\begin{figure*}
    \centering
    \includegraphics[width=1.00\columnwidth]{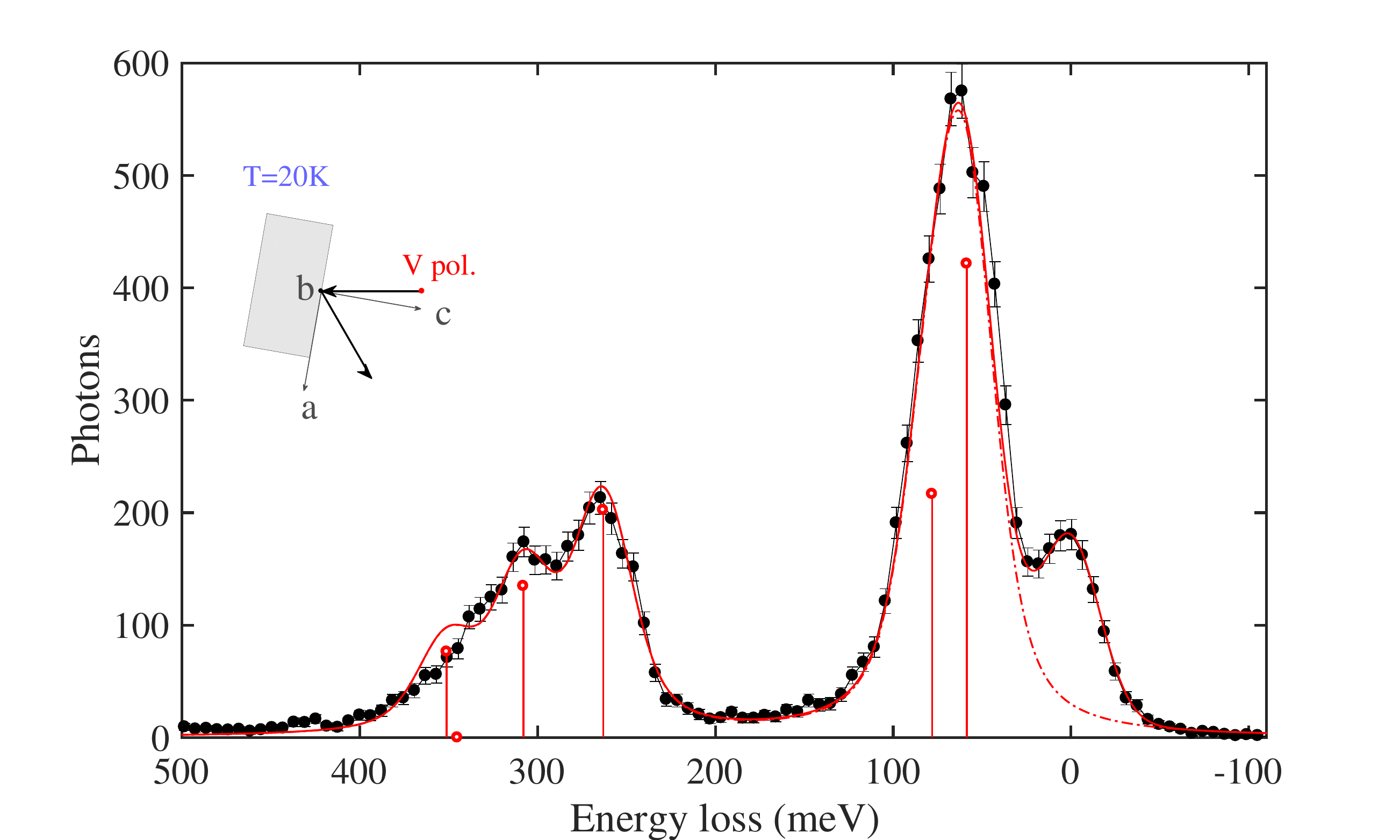}
    \includegraphics[width=1.00\columnwidth]{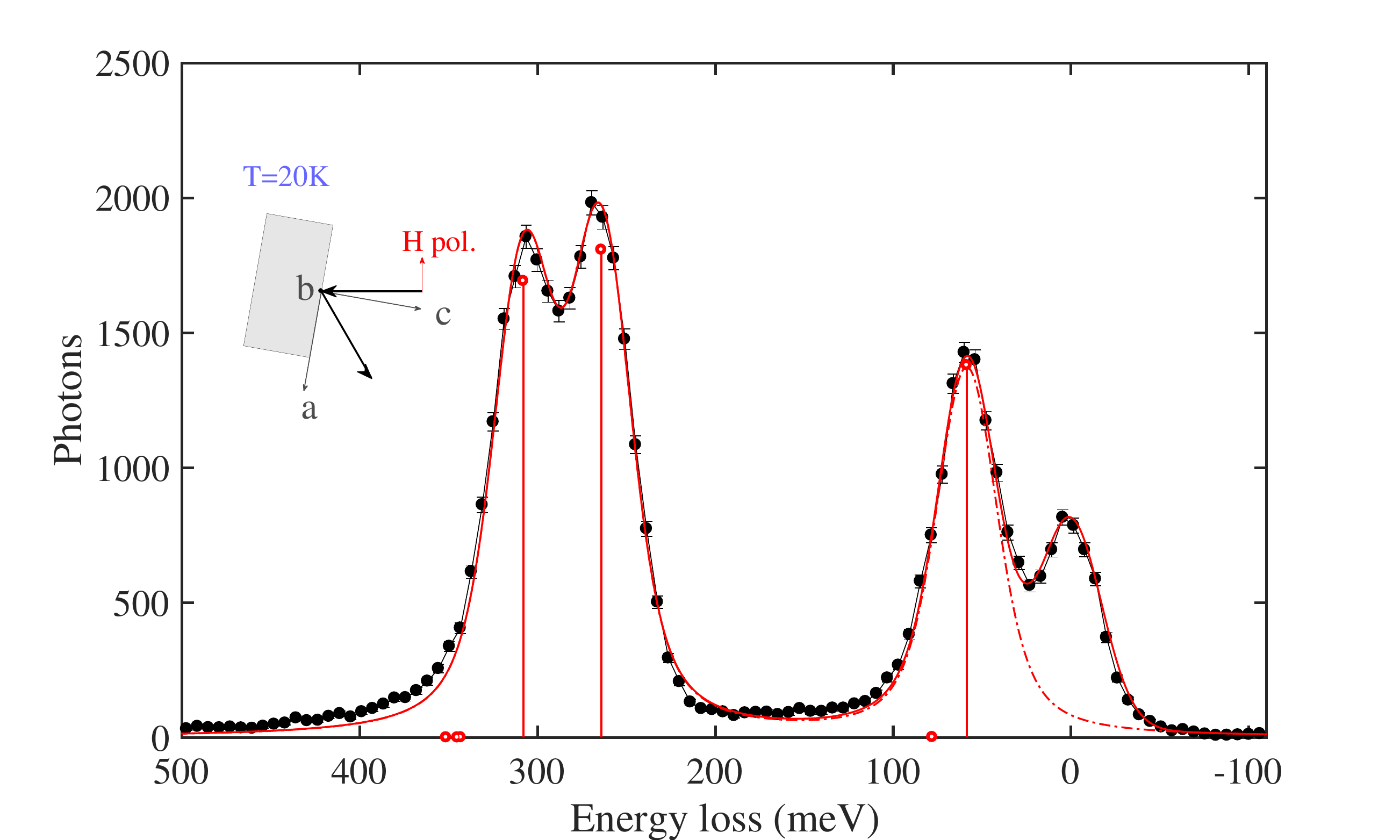}\\	   
    \includegraphics[width=1.00\columnwidth]{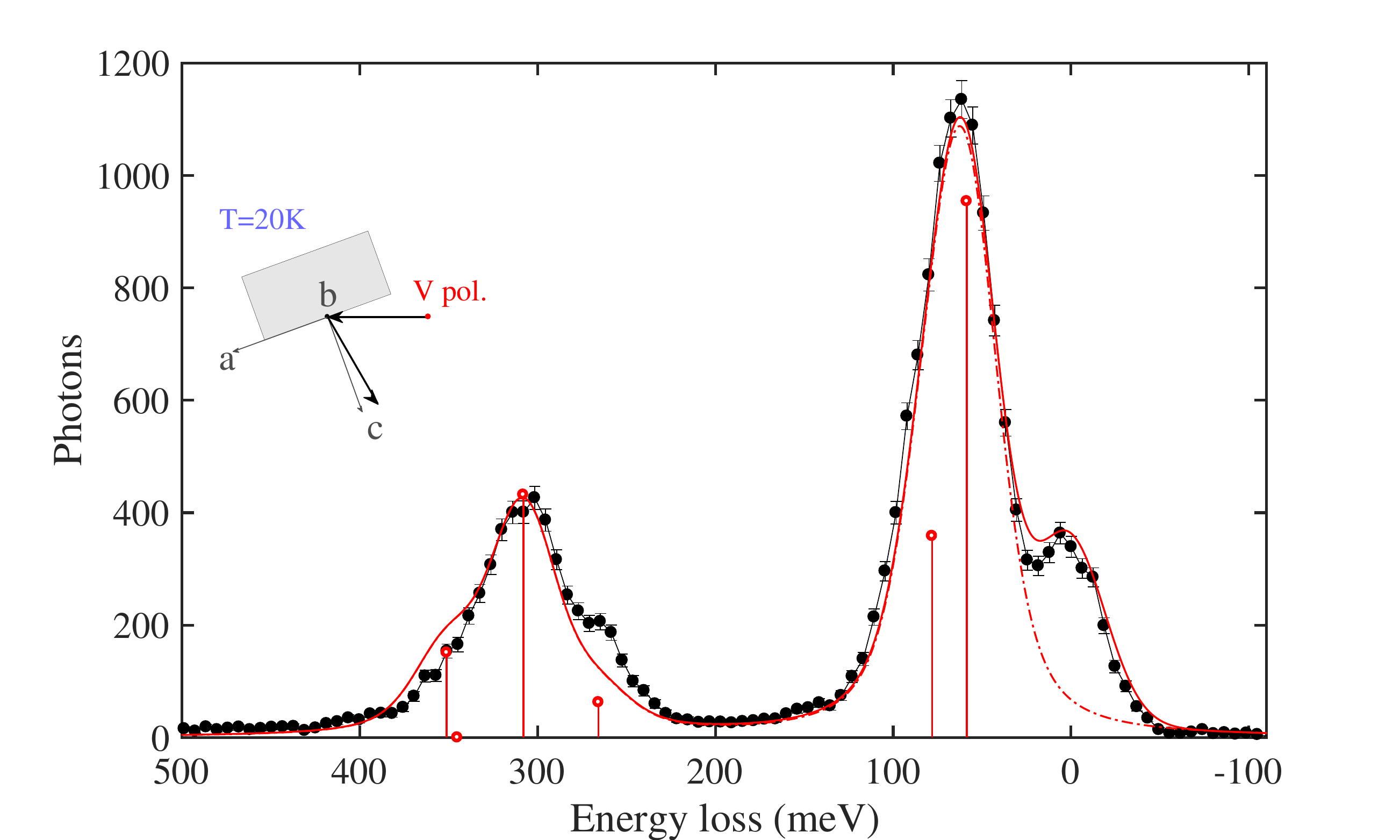}
     \includegraphics[width=1.00\columnwidth]{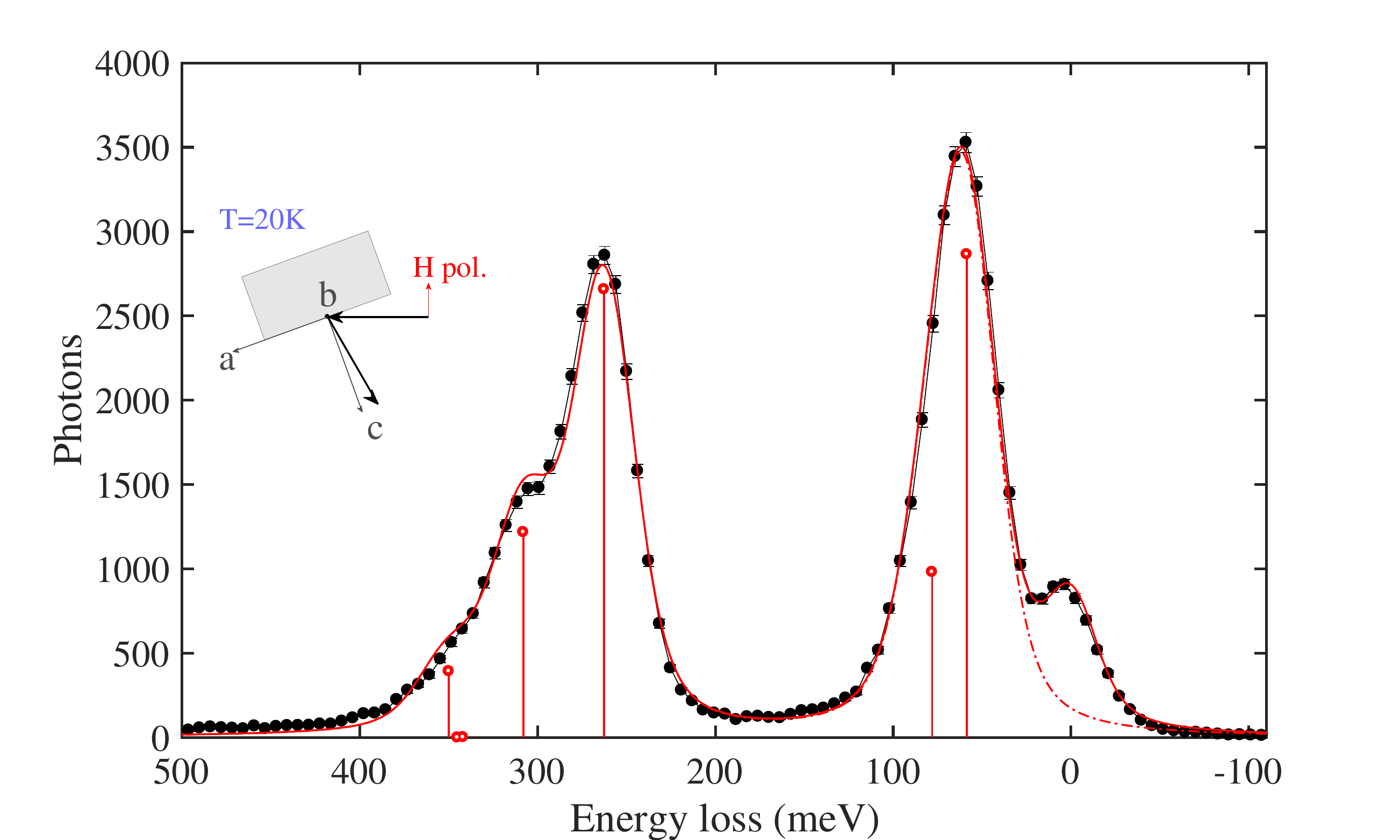}\\
     \includegraphics[width=1.00\columnwidth]{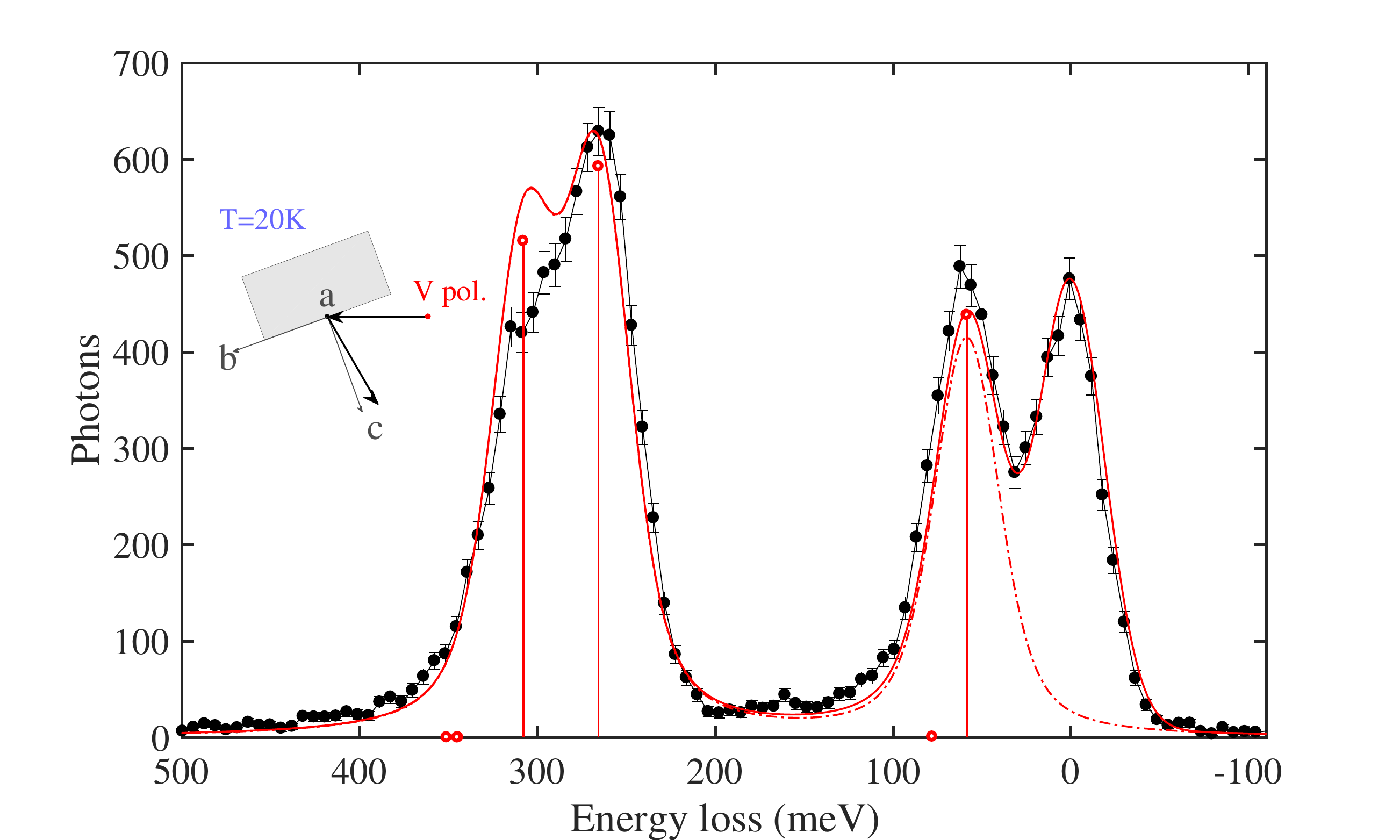}
    \includegraphics[width=1.00\columnwidth]{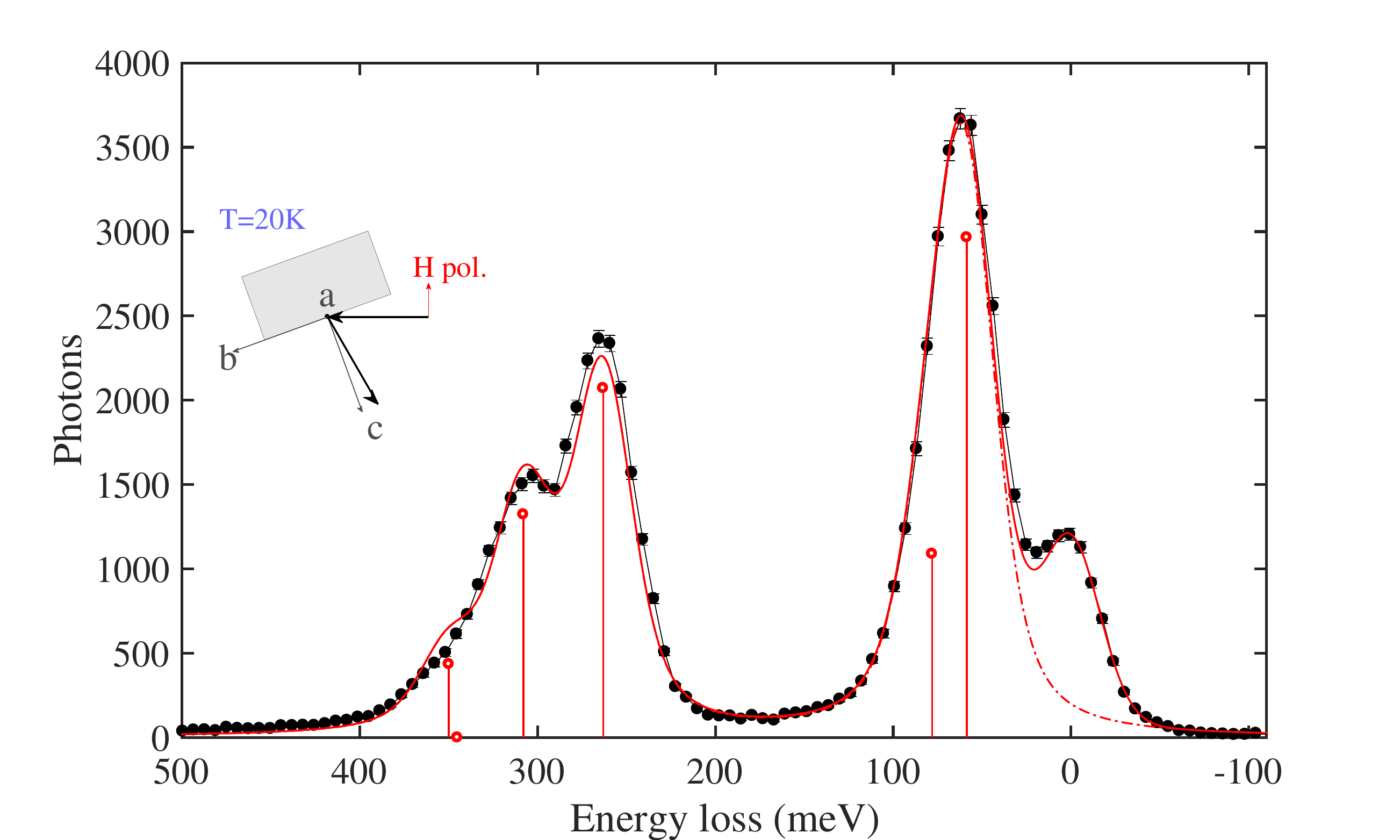}\\
            \includegraphics[width=1.00\columnwidth]{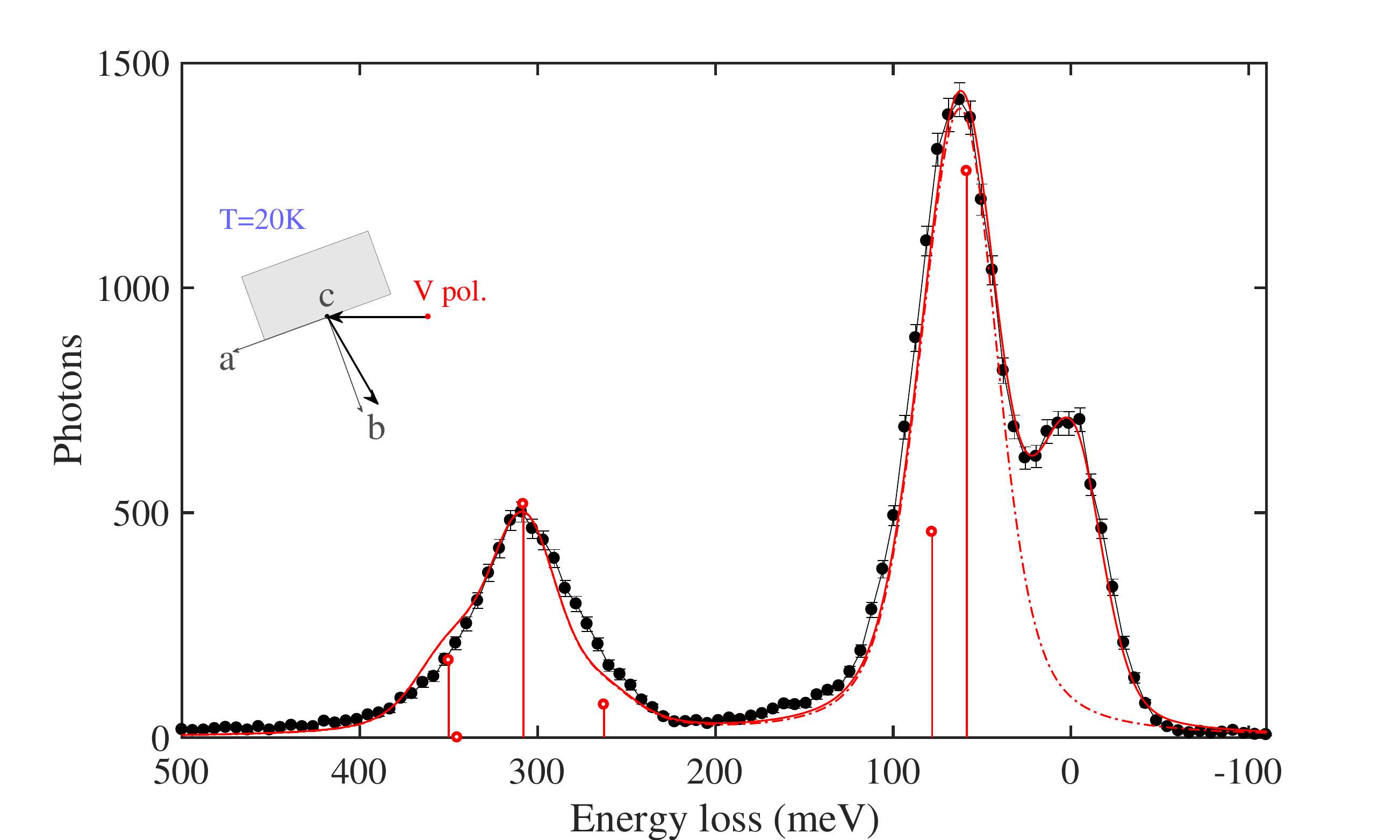} 
    \includegraphics[width=1.00\columnwidth]{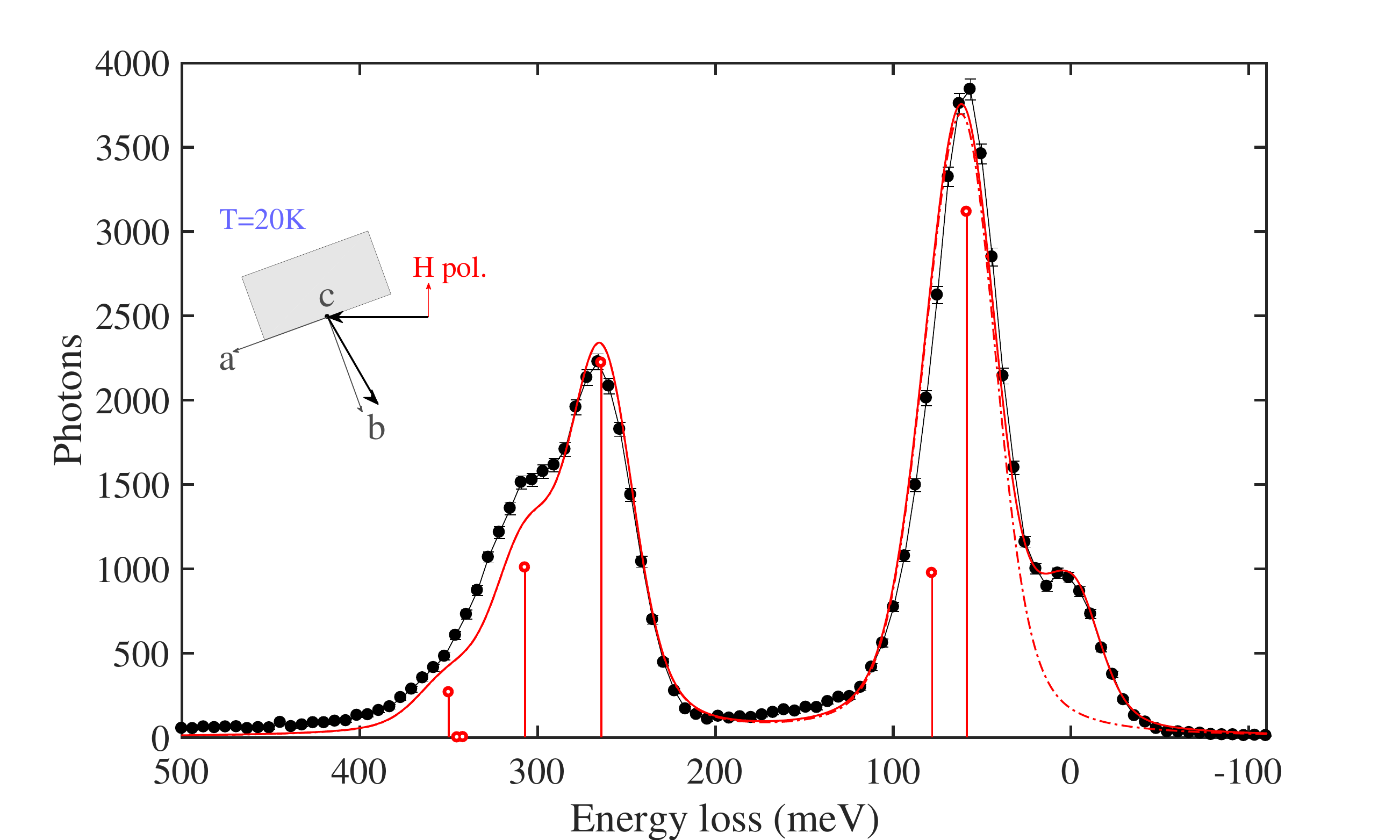}  \\
    \caption{Set of RIXS spectra for various experimental settings (on the same row two spectra with different incident polarizations) and corresponding simulations.}
\label{FigRIXS2}
\end{figure*}

\begin{figure*}
    \centering
 \includegraphics[width=1.00\columnwidth]{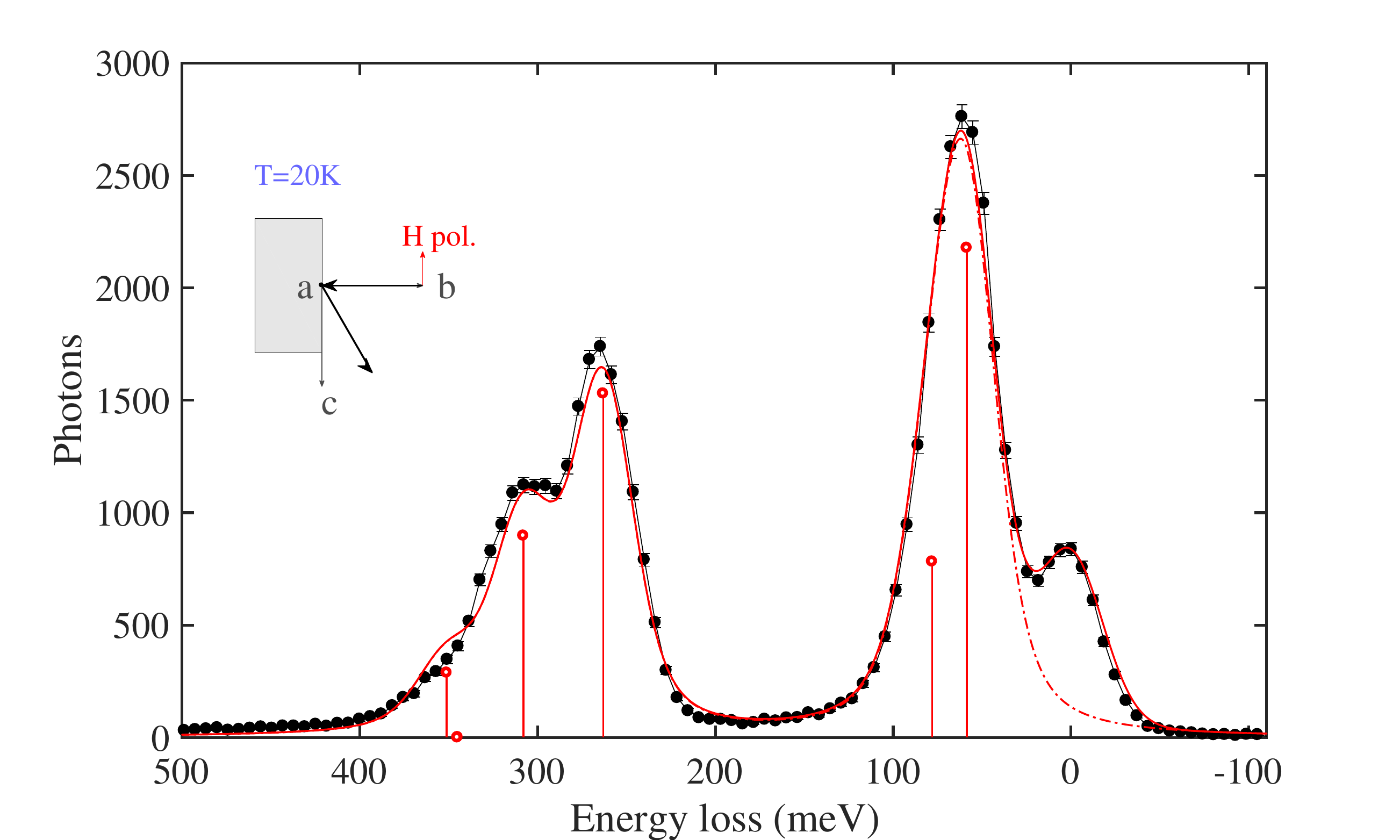}
        \includegraphics[width=1.00\columnwidth]{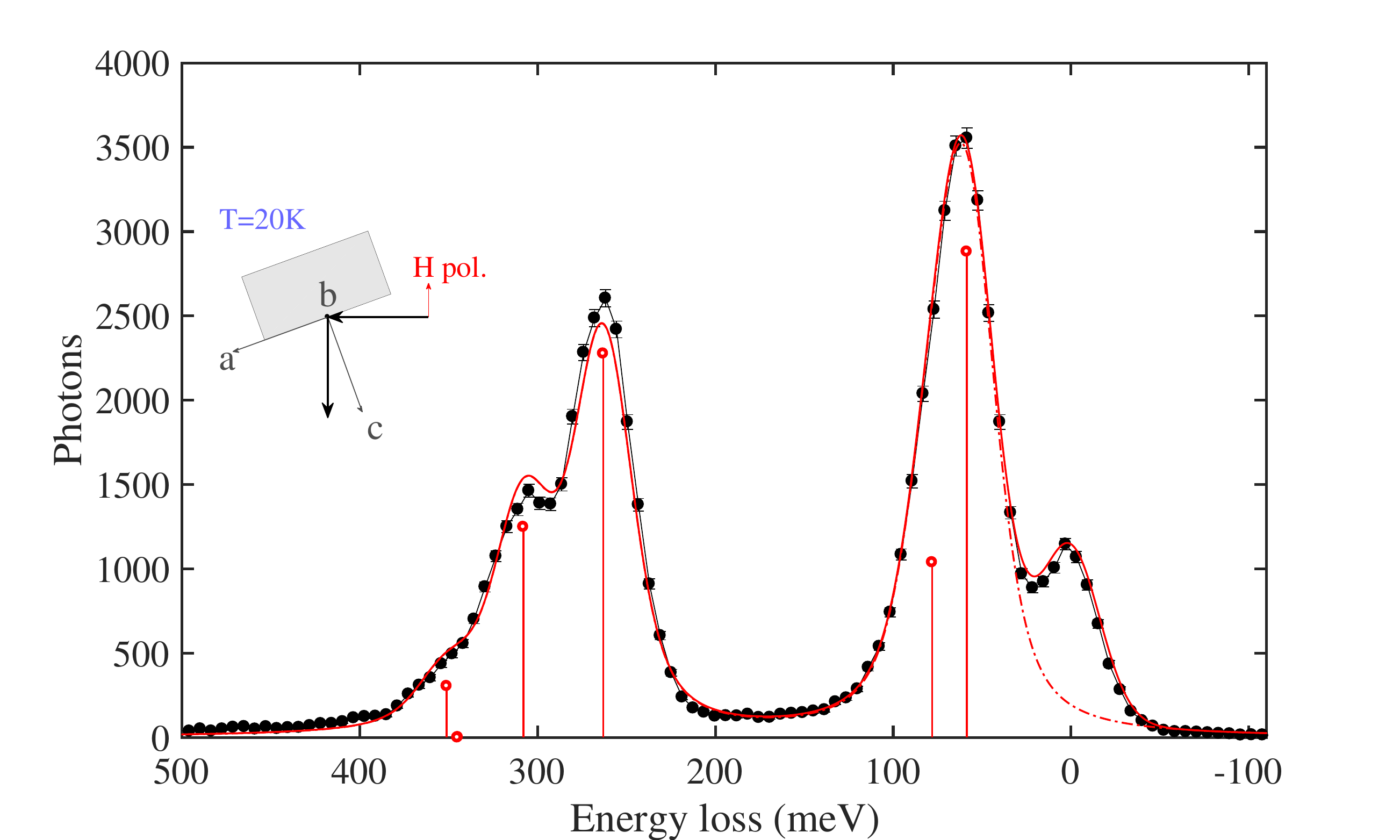}   
            \caption{Other RIXS spectra with various experimental settings, and corresponding simulations.}
\label{FigRIXS3}
\end{figure*}

\begin{figure*}
    \centering
    \includegraphics[width=1.00\columnwidth]{SUPPLEMENTAL_S0002.pdf}
    \includegraphics[width=1.00\columnwidth]{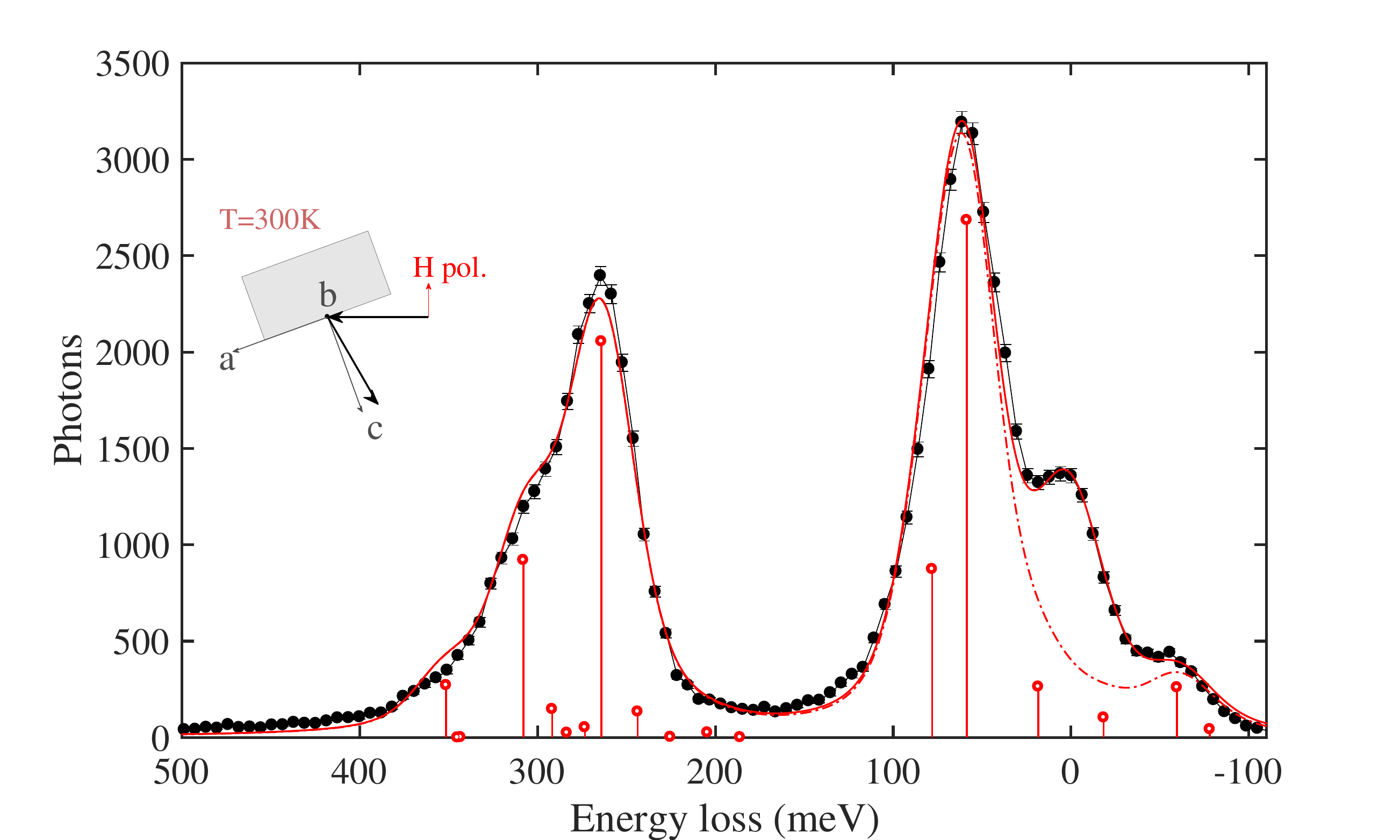}\\
    \includegraphics[width=1.00\columnwidth]{SUPPLEMENTAL_S0018.pdf}
    \includegraphics[width=1.00\columnwidth]{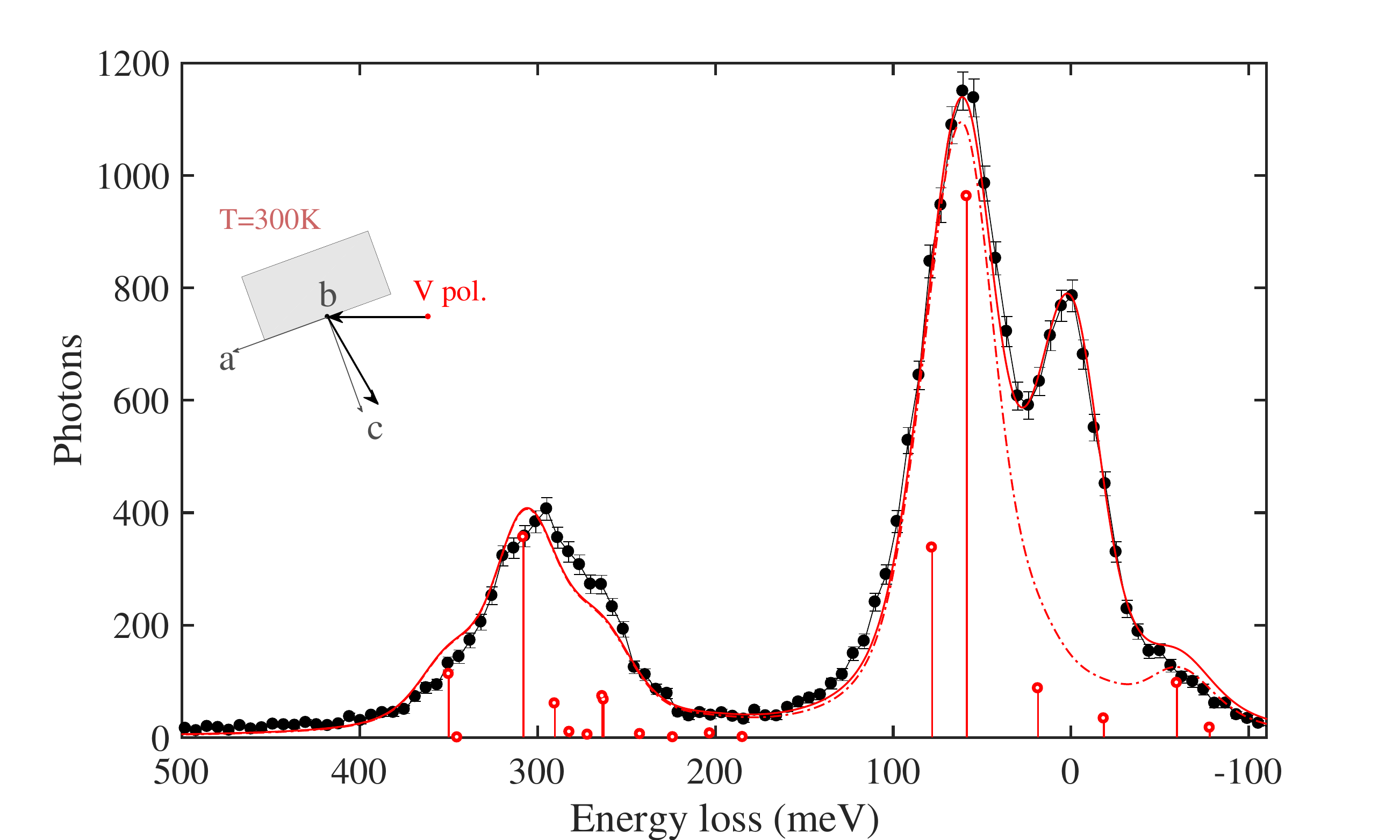}\\
    \includegraphics[width=1.00\columnwidth]{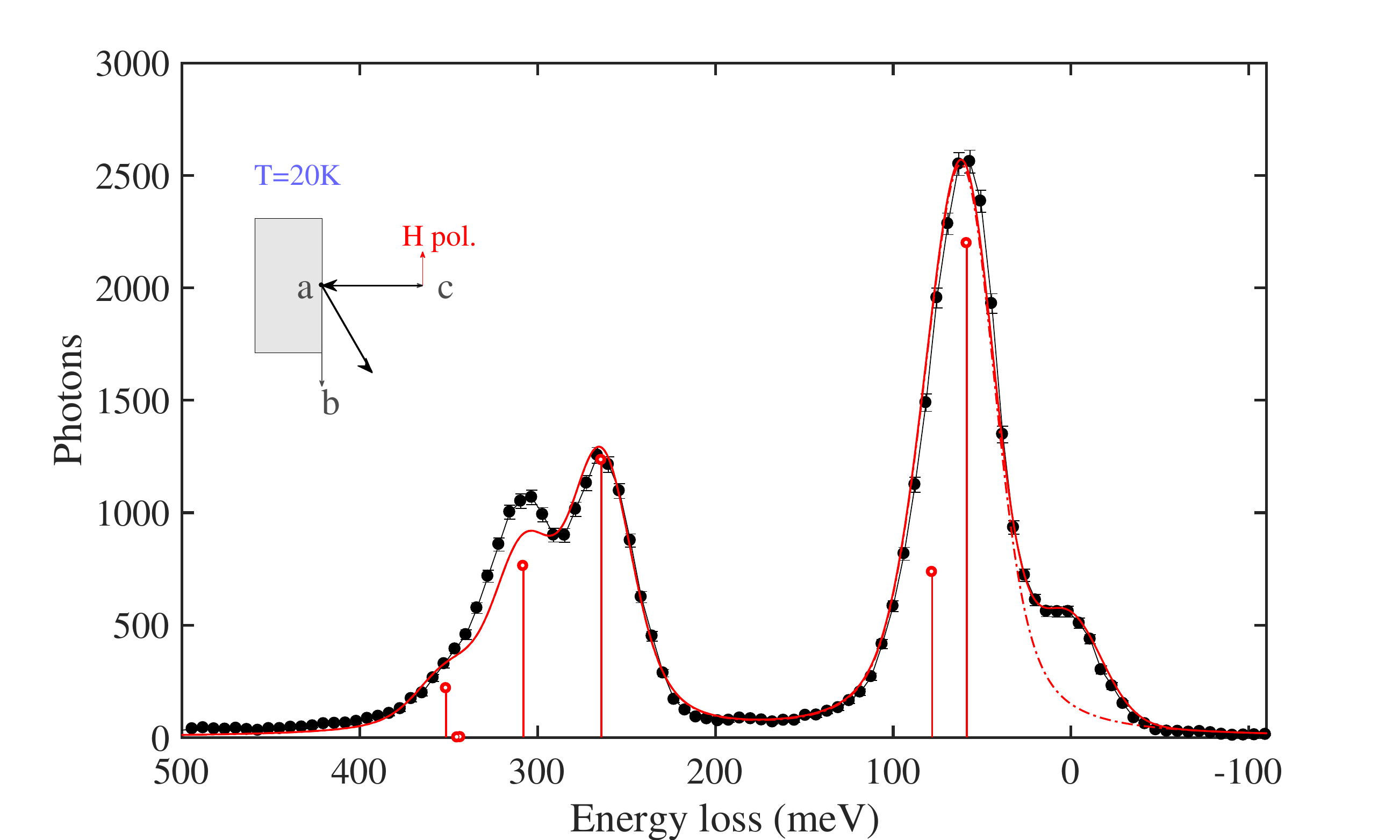}
    \includegraphics[width=1.00\columnwidth]{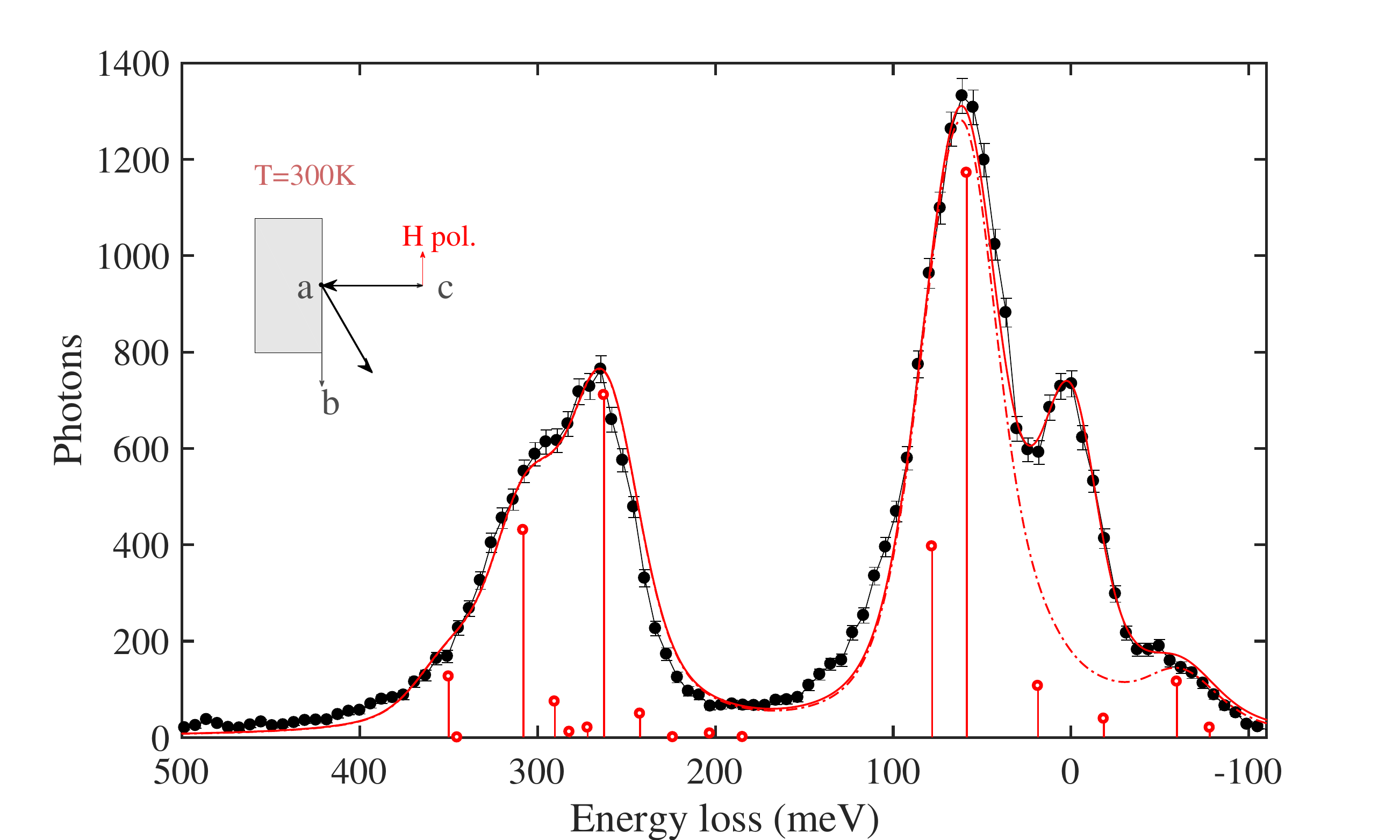}\\
    \includegraphics[width=1.00\columnwidth]{SUPPLEMENTAL_S0039.pdf}
    \includegraphics[width=1.00\columnwidth]{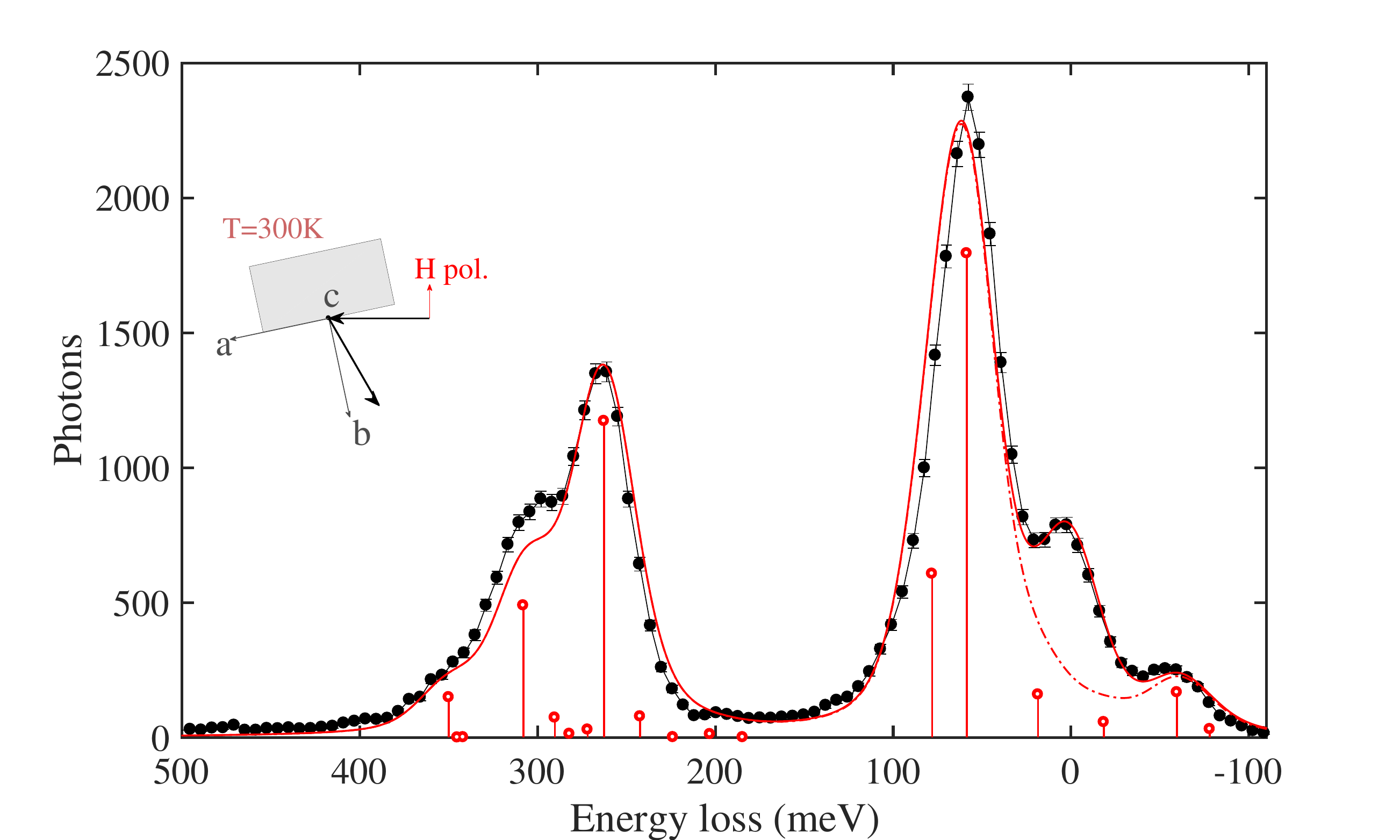}
    \caption{Temperature dependence of RIXS spectra (20\,K in the left panels, 300\,K in the right panels), acquired with different scattering geometries}
\label{FigRIXS4}
\end{figure*}

\begin{figure*}
    \centering
    \includegraphics[width=1.00\columnwidth]{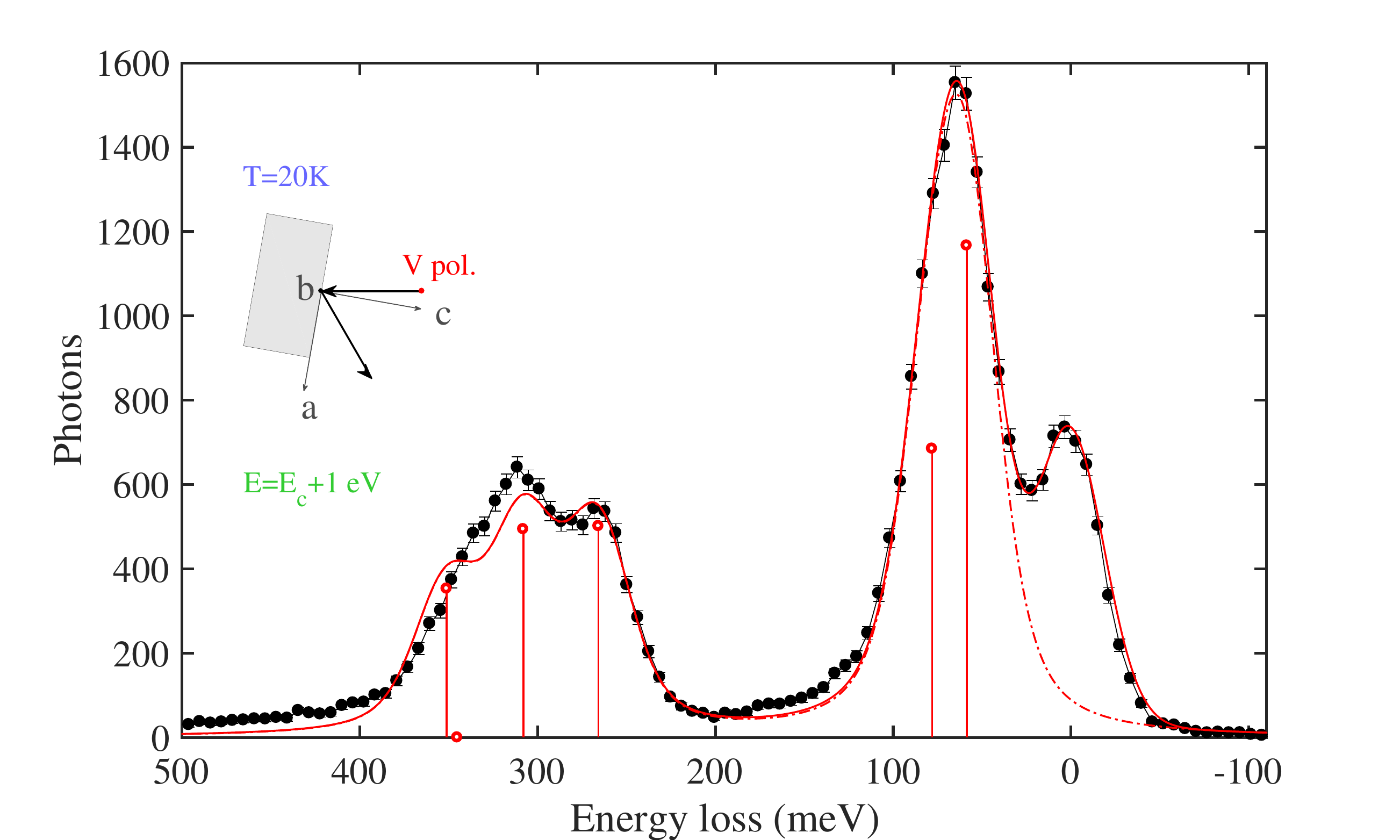}
    \includegraphics[width=1.00\columnwidth]{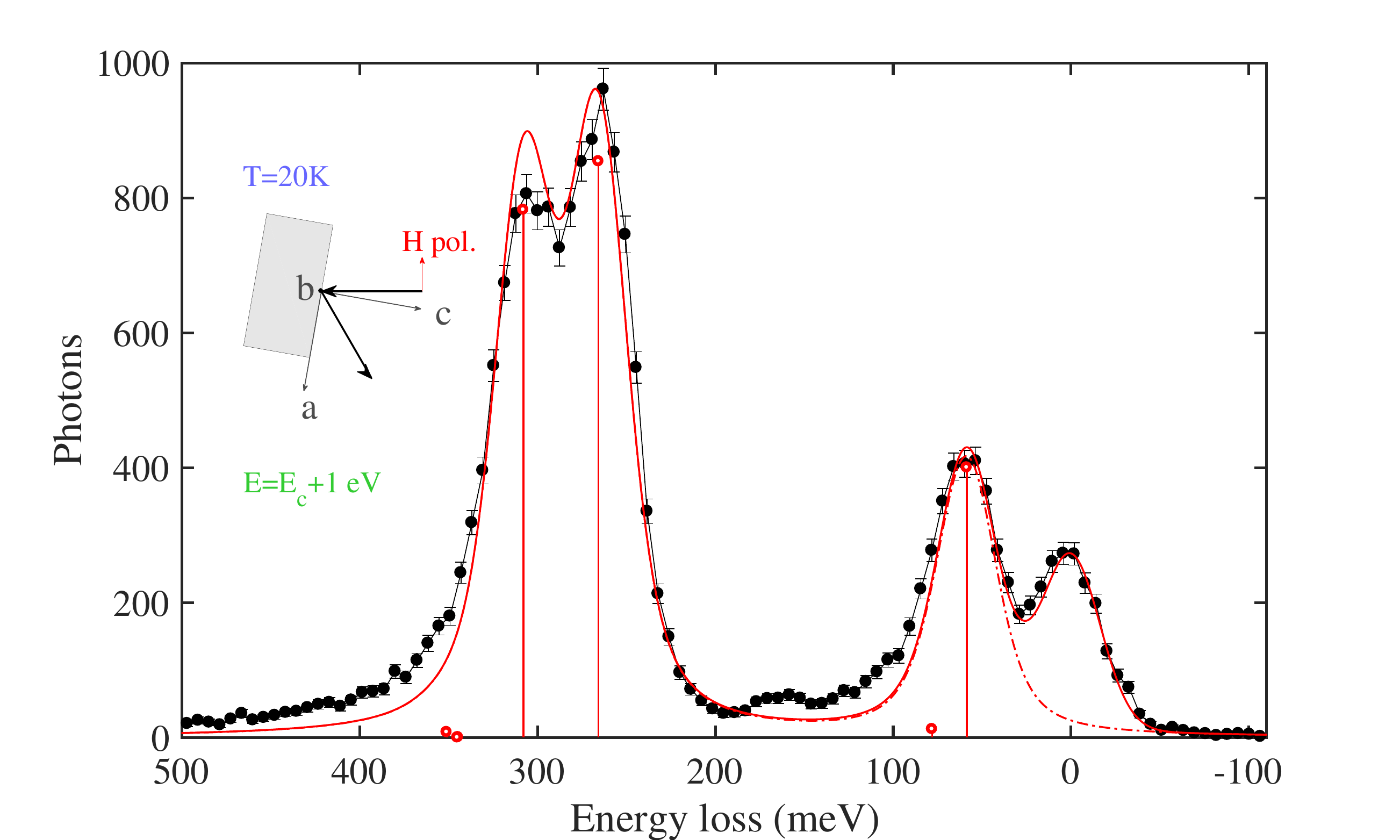}
    \includegraphics[width=1.00\columnwidth]{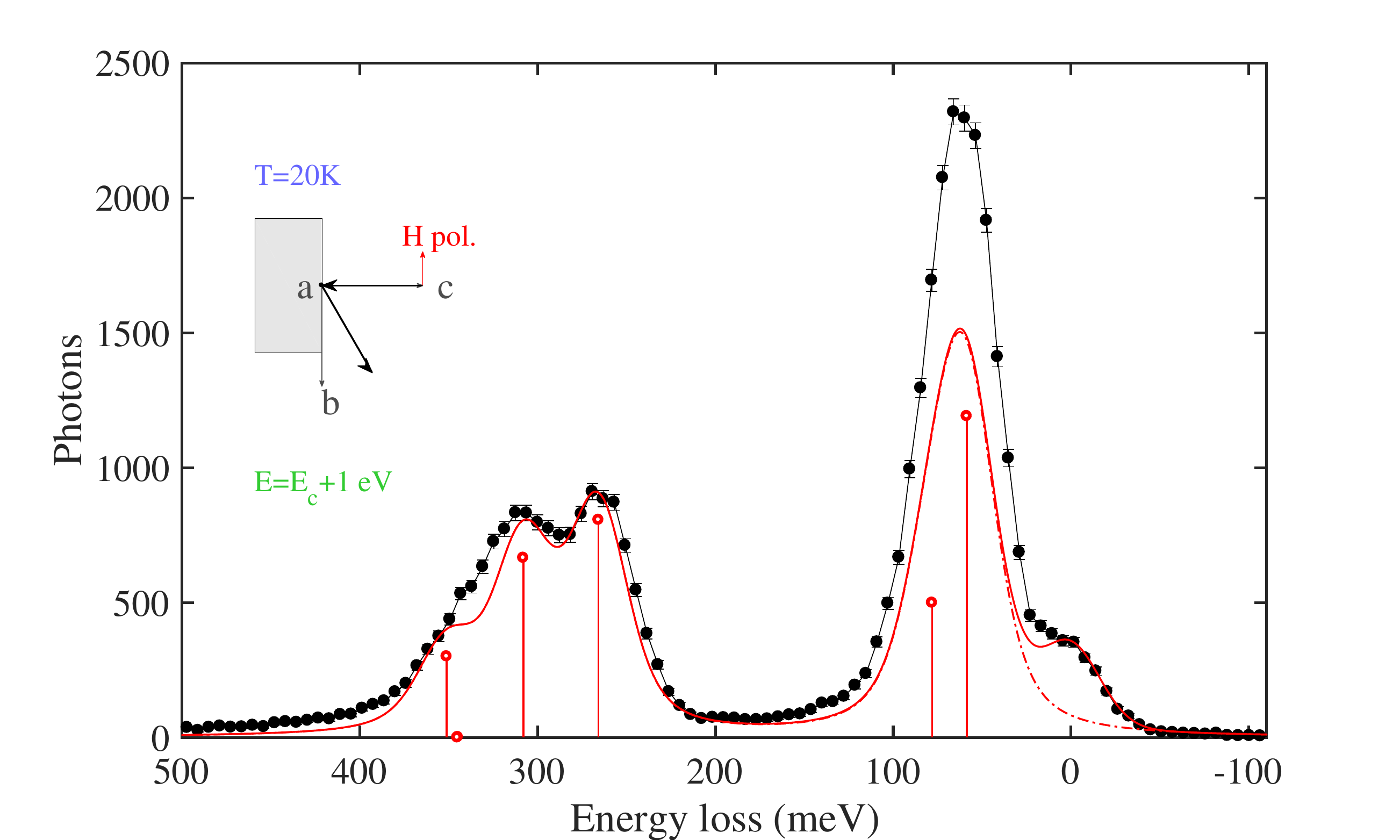}
    \includegraphics[width=1.00\columnwidth]{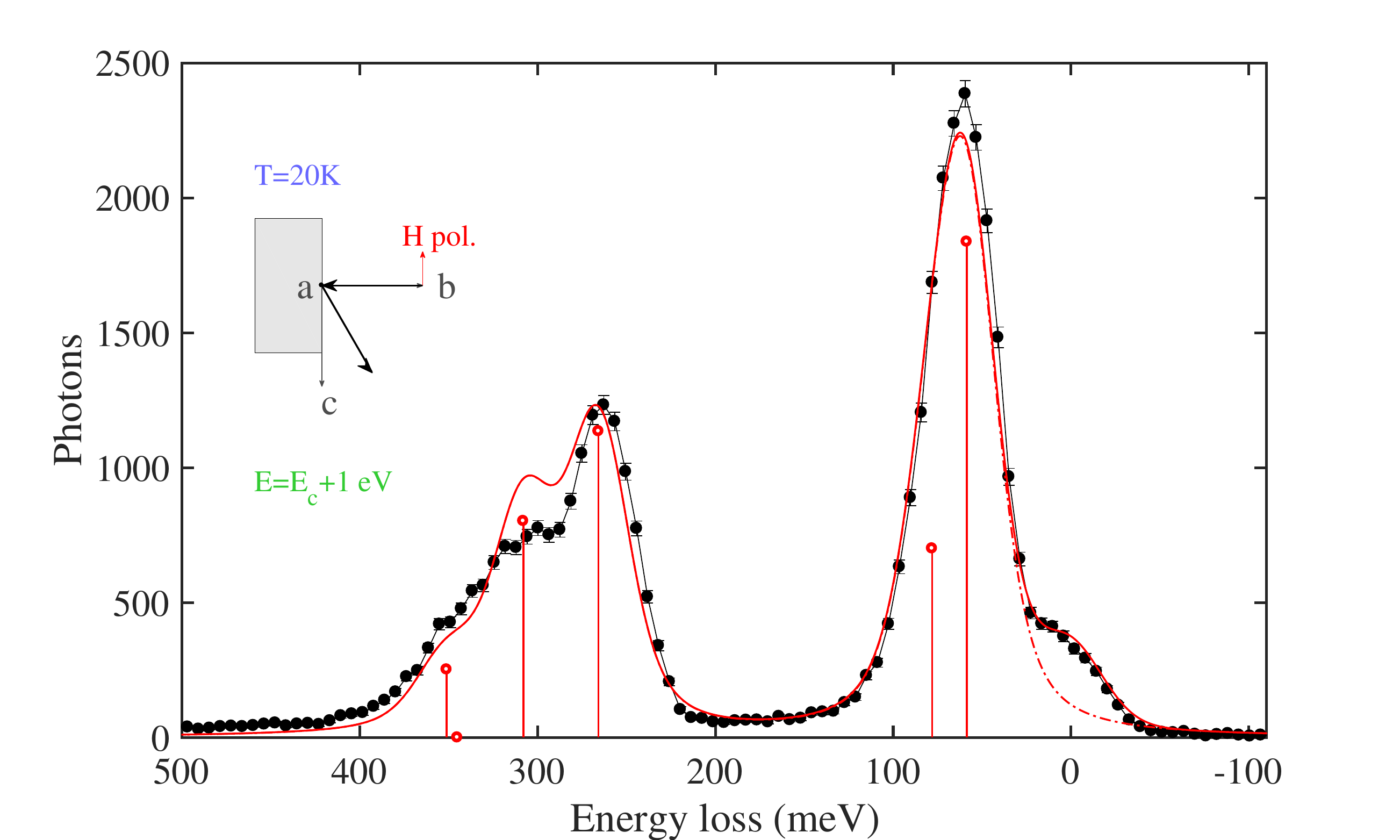}
    \caption{Collection of RIXS spectra acquired with higher incident photon energy: $E_{c}$+1\,eV}
\label{FigRIXS5}
\end{figure*}

We show here in Figs \ref{FigRIXS2}, \ref{FigRIXS3}, \ref{FigRIXS4} and \ref{FigRIXS5} the full set of experimental RIXS spectra of CeRh$_3$Si$_2$ acquired on ID32 at ESRF (black circles) The red dots indicate the transition intensities resulting from our crystal-field analysis presented in the main text, and the full red line is the calculated spectrum with the present experimental resolution The elastic intensity at 0\,eV, not obtainable with simulations, was adjusted to fit the spectra.

\subsection{Crystal field scheme in different reference systems and comparison with previous literature}
Tables \ref{T1b}, \ref{T2b}, \ref{T3b} and \ref{T4b} contain energies, wave functions, and moments for all states as obtained from the crystal-field calculations based on the two different quantization axis ($ z’ \parallel a$ and $z \parallel c$), for the present set of crystal-field parameters obtained from fits to RIXS and INS data, and for the parameters obtained by Pikul \textit{et al.} \cite{Pikul2010} from the fit of the static susceptibility From the tables it is evident that the two works find essentially the same crystal field ground state, which expressed in the reference system of this work ($x’ \parallel c$, $y’ \parallel b$, $z’ \parallel a$, see Tables \ref{T1b} and \ref{T3b}) is an almost pure $\ket{J=\frac{5}{2},J_{z'}=\pm\frac{1}{2}}$  doublet It is to be noted that the choice of quantization axis does not change the energy eigenvalues (i.e the energy of crystal field levels), but it changes the eigenvectors, because the basis functions change.

\newpage

\newpage
 \thispagestyle{empty}
\begin{sidewaystable*}[p]
\centering
\def\arraystretch{1.5}
\noindent
\begin{tabular}{|l|l|l|l|l|l|l|l|l|l|l|l|l|l|l|l|l|l|l|}
\hline
    \multicolumn{2}{|l|}{Energy}    &    
    \multicolumn{14}{|l|}{ $\ket{J,J_{z' \parallel a}}$ components}   & 
    \multicolumn{3}{|l|}{Moments}\\ \hline
	meV & 
	Kelvin & 
	$\ket{\frac{5}{2},-\frac{5}{2}}$ & 
	$\ket{\frac{5}{2},-\frac{3}{2}}$ & 
	$\ket{\frac{5}{2},-\frac{1}{2}}$ & 
	$\ket{\frac{5}{2},+\frac{1}{2}}$ & 
	$\ket{\frac{5}{2},+\frac{3}{2}}$ & 
	$\ket{\frac{5}{2},+\frac{5}{2}}$ & 
	$\ket{\frac{7}{2},-\frac{7}{2}}$ & 
	$\ket{\frac{7}{2},-\frac{5}{2}}$ & 
	$\ket{\frac{7}{2},-\frac{3}{2}}$ & 
	$\ket{\frac{7}{2},-\frac{1}{2}}$ & 
	$\ket{\frac{7}{2},+\frac{1}{2}}$ & 
	$\ket{\frac{7}{2},+\frac{3}{2}}$ & 
	$\ket{\frac{7}{2},+\frac{5}{2}}$ & 
	$\ket{\frac{7}{2},+\frac{7}{2}}$ & 
 	$ \mu_{a \parallel z'}$  &
  	$\mu_{b \parallel y'}$  &
  	$\mu_{c \parallel x'}$  \\ \hline
0    & 0    & 0    & 0       & 0.9991  & 0       & 0       & 0       & 0    & 0      & 0      & -0.0426 & 0      & 0      & 0      & 0 & -0.47   & -1.26  & -1.26  \\ \hline
0    & 0    & 0    & 0       & 0      & 0.9991   & 0       & 0       & 0    & 0      & 0      & 0      & 0.0426  & 0      & 0      & 0 & 0.47    & 1.26   & 1.26   \\ \hline
59.7  & 692  & 0    & 0.99899  & 0      & 0       & 0       & 0       & 0    & 0      & -0.1415 & 0      & 0      & 0      & 0      & 0 & -1.42   & 0     & 0     \\ \hline
59.7  & 692  & 0    & 0       & 0      & 0       & 0.99899  & 0       & 0    & 0      & 0      & 0      & 0      & 0.1415  & 0      & 0 & 1.42    & 0     & 0     \\ \hline
78.1  & 906  & 0.9999& 0       & 0      & 0       & 0       & 0       & 0    & 0.0122  & 0      & 0      & 0      & 0      & 0      & 0 & -2.13   & 0     & 0     \\ \hline
78.1  & 906  & 0    & 0       & 0      & 0       & 0       & 0.9999   & 0    & 0      & 0      & 0      & 0      & 0      & -0.0122 & 0 & 2.13    & 0     & 0     \\ \hline
265.7 & 3082 & 0    & 0       & 0.0426  & 0        & 0      & 0       & 0    & 0      & 0     & 0.9991  & 0     & 0      & 0      & 0 & -0.53    & -2.26   & -2.26  \\ \hline
265.7 & 3082 & 0    & 0       & 0      & -0.0426  & 0       & 0       & 0    & 0      & 0      & 0      & 0.9991  & 0      & 0      & 0 & 0.53   & 2.26  & 2.26 \\ \hline
304.9 & 3537 & 0    & 0.1415   & 0      & 0       & 0       & 0       & 0    & 0      & 0.9899  & 0      & 0      & 0      & 0      & 0 & -1.58   & 0     & 0     \\ \hline
304.9 & 3537 & 0    & 0       & 0      & 0       & -0.1415  & 0       & 0    & 0      & 0      & 0      & 0      & 0.9899  & 0      & 0 & 1.58    & 0     & 0     \\ \hline
344.7 & 3999 & 0    & 0       & 0      & 0       & 0       & 0       & 1    & 0      & 0      & 0      & 0      & 0      & 0      & 0 & -4     & 0  & 0  \\ \hline
344.7 & 3999 & 0    & 0       & 0      & 0       & 0       & 0       & 0    & 0      & 0      & 0      & 0      & 0      & 0      & 1 & 4      & 0  & 0  \\ \hline
352.9 & 4094 & -0.0122 & 0    & 0       & 0       & 0       & 0       & 0    & 0.9999   & 0     & 0      & 0      & 0      & 0      & 0 & -2.87    & 0   & 0 \\ \hline
352.9 & 4094 & 0    & 0       & 0      & 0       & 0       & 0.0122   & 0    & 0      & 0      & 0      & 0      & 0      & 0.9999  & 0 & 2.87   & 0   & 0   \\ \hline
\end{tabular}
\caption{Crystal field scheme corresponding to the crystal field parameters determined in this work, expressed in the reference system with the quantization axis  $z' \parallel a$ ($x' \parallel c$ and $y'\parallel b$).\\
Crystal field parameters:\\
$A_{2}^{0}$:  	-135\,meV (-1566\,K), $A_{4}^{0}$: -172\,meV (-1995\,K),  $A_{6}^{0}$: 30\,meV (348\,K), $\zeta_{SO}$: 	 76.6\,meV (888\,K).}\label{T1b}
\end{sidewaystable*}

\newpage
 \thispagestyle{empty}
\begin{sidewaystable*}[p]
\centering
\def\arraystretch{1.5}
\noindent
\begin{tabular}{|l|l|l|l|l|l|l|l|l|l|l|l|l|l|l|l|l|l|l|}
\hline
    \multicolumn{2}{|l|}{Energy}    &    
    \multicolumn{14}{|l|}{ $\ket{J,J_{z \parallel c}}$ components}   & 
    \multicolumn{3}{|l|}{Moments}\\ \hline
	meV & 
	Kelvin & 
	$\ket{\frac{5}{2},-\frac{5}{2}}$ & 
	$\ket{\frac{5}{2},-\frac{3}{2}}$ & 
	$\ket{\frac{5}{2},-\frac{1}{2}}$ & 
	$\ket{\frac{5}{2},+\frac{1}{2}}$ & 
	$\ket{\frac{5}{2},+\frac{3}{2}}$ & 
	$\ket{\frac{5}{2},+\frac{5}{2}}$ & 
	$\ket{\frac{7}{2},-\frac{7}{2}}$ & 
	$\ket{\frac{7}{2},-\frac{5}{2}}$ & 
	$\ket{\frac{7}{2},-\frac{3}{2}}$ & 
	$\ket{\frac{7}{2},-\frac{1}{2}}$ & 
	$\ket{\frac{7}{2},+\frac{1}{2}}$ & 
	$\ket{\frac{7}{2},+\frac{3}{2}}$ & 
	$\ket{\frac{7}{2},+\frac{5}{2}}$ & 
	$\ket{\frac{7}{2},+\frac{7}{2}}$ & 
 	$ \mu_{a \parallel x}$  &
  	$\mu_{b \parallel y}$  &
  	$\mu_{c \parallel z}$  \\ \hline
0    & 0    & 0    & 0.353 &  0   & -0.499& 0    & 0.790 & 0.032 & 0    & -0.021& 0    & 0.016 & 0    & -0.012& 0    & -0.47 & -1.26 & -1.26 \\ \hline
0    & 0    & 0.790 & 0    &-0.499 & 0    & 0.353 & 0    & 0    & 0.012 & 0    & -0.016& 0    & 0.021 & 0    & -0.032& 0.47  & 1.26  & 1.26  \\ \hline
59.7  & 692  & 0    & 0.742 & 0    & -0.350& 0    & -0.553& 0.081 & 0    & 0.018 & 0    & -0.068& 0    & 0.092 & 0    & -1.42 & 0    & 0    \\ \hline
59.7  & 692  & -0.553& 0    & -0.35 & 0    & 0.74  & 0    & 0    & -0.092& 0    & 0.068 & 0    & -0.018& 0    & -0.081& 1.42  & 0    & 0    \\ \hline
78.1  & 906  & 0    & 0.559 & 0    & 0.791 & 0    & 0.250 & -0.004& 0    & -0.008& 0    & 0.003 & 0    & 0.008 & 0    & -2.13 & 0    & 0    \\ \hline
78.1  & 906  & 0.250 & 0    & 0.791 & 0    & 0.559 & 0    & 0    & -0.008& 0    & -0.003& 0    & 0.008 & 0    & 0.004 & 2.13  & 0    & 0    \\ \hline
265.7 & 3082 & 0    & -0.015& 0    & 0.021 & 0    & -0.034& 0.739 & 0    & -0.484& 0    & 0.375 & 0    & -0.279& 0    & -0.53 & -2.26 & -2.26 \\ \hline
265.7 & 3082 & 0.034 & 0    & -0.021& 0    & 0.015 & 0    & 0    & -0.279& 0    & 0.375 & 0    & -0.484& 0    & 0.739 & 0.53  & 2.26  & 2.26  \\ \hline
304.9 & 3537 & -0.079& 0    & -0.05 & 0    & 0.106 & 0    & 0    & 0.643 & 0    & -0.479& 0    & 0.124 & 0    & 0.567 & -1.58  & 0    & 0   \\ \hline
304.9 & 3537 & 0    & -0.106& 0    & 0.050 & 0    & 0.079 & 0.567 & 0    & 0.124 & 0    & -0.479& 0    & 0.643 & 0    & 1.58 & 0    & 0     \\ \hline
344.7 & 3999 & 0    & 0    & 0    & 0    & 0    & 0    & 0.125 & 0    & 0.573 & 0    & 0.740 & 0    & 0.331 & 0     & -4   & 0 & 0 	  \\ \hline
344.7 & 3999 & 0    & 0    & 0    & 0    & 0    & 0    & 0    & 0.331 & 0    & 0.740 & 0    & 0.573 & 0    & 0.125  & 4    & 0 & 0 	  \\ \hline
352.9 & 4095 & 0    & 0.007 & 0    & 0.010 & 0   & 0.003  & 0.331 & 0    & 0.650 & 0    & -0.279  & 0    & -0.625& 0   & -2.87  & 0  & 0    \\  \hline
352.9 & 4095 & -0.003& 0    & -0.010& 0    & -0.007& 0    & 0    & -0.625& 0    & -0.279& 0      & 0.650 & 0    & 0.331 & 2.87 & 0  & 0     \\ \hline
\end{tabular}
\caption{Crystal field scheme corrisponding to the crystal field parameters determined in this work, expressed in the reference system with the quantization axis $z \parallel c$ ($x \parallel a$ and $y \parallel b$), the same reference system used in Ref \cite{Pikul2010}).\\
Crystal field parameters:\\
$A_{2}^{0}$:  	67.5\,meV (783\,K), $A_{2}^{2}$: -82.6\,meV (-959\,K), $A_{4}^{0}$: -64.5\,meV (-748\,K), $A_{4}^{2}$:  68\,meV (788\,K), $A_{4}^{4}$: -90\,meV (-1043\,K), $A_{6}^{0}$: -9.4\,meV (-109\,K), $A_{6}^{2}$: 9.6\,meV (112\,K), $A_{6}^{4}$: -10.5\,meV (-122\,K), $A_{6}^{6}$: 14.2\,meV (165\,K),  $\zeta_{SO}$: 	 76.6\,meV (888\,K).}\label{T2b}
\end{sidewaystable*}
\newpage

\thispagestyle{empty}
\begin{sidewaystable*}[p]
\centering
\def\arraystretch{1.5}
\noindent
\begin{tabular}{|l|l|l|l|l|l|l|l|l|l|l|l|l|l|l|l|l|l|l|}
\hline
    \multicolumn{2}{|l|}{Energy}    &    
    \multicolumn{14}{|l|}{ $\ket{J,J_{z' \parallel a}}$ components}   & 
    \multicolumn{3}{|l|}{Moments}\\ \hline
	meV & 
	Kelvin & 
	$\ket{\frac{5}{2},-\frac{5}{2}}$ & 
	$\ket{\frac{5}{2},-\frac{3}{2}}$ & 
	$\ket{\frac{5}{2},-\frac{1}{2}}$ & 
	$\ket{\frac{5}{2},+\frac{1}{2}}$ & 
	$\ket{\frac{5}{2},+\frac{3}{2}}$ & 
	$\ket{\frac{5}{2},+\frac{5}{2}}$ & 
	$\ket{\frac{7}{2},-\frac{7}{2}}$ & 
	$\ket{\frac{7}{2},-\frac{5}{2}}$ & 
	$\ket{\frac{7}{2},-\frac{3}{2}}$ & 
	$\ket{\frac{7}{2},-\frac{1}{2}}$ & 
	$\ket{\frac{7}{2},+\frac{1}{2}}$ & 
	$\ket{\frac{7}{2},+\frac{3}{2}}$ & 
	$\ket{\frac{7}{2},+\frac{5}{2}}$ & 
	$\ket{\frac{7}{2},+\frac{7}{2}}$ & 
 	$\mu_{a \parallel z'}$  &
  	$\mu_{b \parallel y'}$  &
  	$\mu_{c \parallel x'}$  \\ \hline
0    & 0    & 0.09    & 0     &  0.991   & 0       & -0.009  & 0     & 0     & 0.024   & 0       & -0.021 & 0     & 0.042    & 0     & 0.085  & -0.43  & -1.3    & -1.19   \\  \hline
0    & 0    & 0      & -0.009 &  0      & 0.991    & 0      & 0.09   & -0.085 & 0      & -0.042   & 0     & 0.021  & 0       & -0.024 & 0     & 0.43 & 1.3   & 1.19  \\  \hline
56.4  & 655  & -0.602  & 0     & 0.039    & 0.000    & 0.736   & 0     & 0     & 0.209   &  0       & 0.008  & 0    & -0.032   & 0     & 0.221  & -0.12 & -0.75  & -0.64  \\  \hline
56.4  & 655  &  0     & 0.736  & 0       & 0.039    & 0      & -0.602 & -0.221 & 0      & 0.032    & 0     & -0.008 & 0       & -0.209  & 0.0    & 0.12  & 0.75   & 0.64   \\  \hline
59.7  & 693  & 0.695   & 0     & -0.052    & 0       & 0.591    & 0     & 0     &  0.213  & 0       &  0.05  & 0    & 0.252    & 0      & -0.232    & -0.15 & -0.54  & -0.87  \\  \hline
59.7  & 693  & 0      & 0.591  & 0       & -0.052    & 0       & 0.695  & 0.232  & 0      & -0.252   &  0    & -0.05 &  0      & -0.213  & 0        & 0.15  & 0.54   & 0.87   \\  \hline
267.1 & 3099 & 0      & 0.01   & 0        & 0.024    & 0      & -0.034  & 0.371  & 0      & 0.304   & 0     & 0.843  & 0      & -0.240 & 0     & -0.11 & -0.48  & -2.21  \\  \hline
267.1 & 3099 & 0.034   & 0     & -0.024    & 0      & -0.01   &  0     & 0      & -0.240  & 0      & 0.843  & 0     & 0.304   & 0     & 0.371  & 0.11  & 0.48   & 2.21   \\  \hline
300.4 & 3485 & -0.058  & 0     & -0.048   &  0      & 0.028   & 0     & 0    & -0.389  & 0       & -0.486 & 0     & 0.738    & 0     & 0.249  & -0.67 & -1.33  & -2.51 \\  \hline
300.4 & 3485 & 0      & -0.028 & 0       &  0.048   & 0      & 0.058  &  0.249 & 0      &  0.738   & 0     &-0.486  &  0      & -0.389  &  0    & 0.67  & 1.33   & 2.51  \\  \hline
471.8 & 5473 & 0      &-0.166  &  0      &  0.065   &  0     & -0.366 & 0.67  & 0       & -0.503    &  0 &-0.212  & 0       & -0.295  & 0     & -1.56  & -0.5    & -0.77   \\  \hline
471.8 & 5473 & 0.366   & 0     & -0.065   & 0       &  0.166  & 0     & 0     & -0.295  & 0       & -0.212 & 0     & -0.503   & 0     & 0.67   & 1.56 & 0.5   & 0.77  \\  \hline
507.7 & 5889 & 0.088   & 0     &-0.083    & 0       & -0.284 & 0      & 0      & 0.784   &  0     & -0.078 & 0     & 0.209  & 0     & 0.491  & -0.63 & -0.21   & -1.94  \\  \hline
507.7 & 5889 & 0      & 0.284  & 0       & 0.083    & 0     & -0.088  & 0.491   & 0      & 0.209   & 0     & -0.078 &  0    & 0.784  & 0    & 0.63  & 0.21   & 1.94   \\  \hline
\end{tabular}
\caption{Crystal field scheme corrisponding to the crystal field parameters from reference \cite{Pikul2010}, expressed in the reference system with quantization axis $z' \parallel a$ ($x' \parallel c$ and $y'\parallel b$).\\
Crystal field parameters:\\
$A_{2}^{0}$:  	-190.5\,meV (-2210\,K), $A_{2}^{2}$: 15.3\,meV (177\,K), $A_{4}^{0}$: -278.3\,meV (-3229\,K), $A_{4}^{2}$:  -101.3\,meV (-1175\,K), $A_{4}^{4}$: 83.2\,meV (965\,K), $A_{6}^{0}$: 78.4\,meV (910\,K), $A_{6}^{2}$: 73.5\,meV (853\,K), $A_{6}^{4}$: 457\,meV (5302\,K), $A_{6}^{6}$: -364.8\,meV (-4232\,K), $\zeta_{SO}$: 	 80.5\,meV (934\,K) \\
}\label{T3b}
\end{sidewaystable*}

\newpage
 \thispagestyle{empty}
\begin{sidewaystable*}[p]
\centering
\def\arraystretch{1.5}
\noindent
\begin{tabular}{|l|l|l|l|l|l|l|l|l|l|l|l|l|l|l|l|l|l|l|}
\hline
    \multicolumn{2}{|l|}{Energy}    &    
    \multicolumn{14}{|l|}{ $\ket{J,J_{z \parallel c}}$ components}   & 
    \multicolumn{3}{|l|}{Moments}\\ \hline
	meV & 
	Kelvin & 
	$\ket{\frac{5}{2},-\frac{5}{2}}$ & 
	$\ket{\frac{5}{2},-\frac{3}{2}}$ & 
	$\ket{\frac{5}{2},-\frac{1}{2}}$ & 
	$\ket{\frac{5}{2},+\frac{1}{2}}$ & 
	$\ket{\frac{5}{2},+\frac{3}{2}}$ & 
	$\ket{\frac{5}{2},+\frac{5}{2}}$ & 
	$\ket{\frac{7}{2},-\frac{7}{2}}$ & 
	$\ket{\frac{7}{2},-\frac{5}{2}}$ & 
	$\ket{\frac{7}{2},-\frac{3}{2}}$ & 
	$\ket{\frac{7}{2},-\frac{1}{2}}$ & 
	$\ket{\frac{7}{2},+\frac{1}{2}}$ & 
	$\ket{\frac{7}{2},+\frac{3}{2}}$ & 
	$\ket{\frac{7}{2},+\frac{5}{2}}$ & 
	$\ket{\frac{7}{2},+\frac{7}{2}}$ & 
	$\mu_{a \parallel x}$ &
	$ \mu_{b \parallel y}$ & 
	$\mu_{c \parallel z}$ \\ \hline
0    & 0    & 0.801   & 0     & -0.427& 0     & 0.408  & 0     & 0     & 0.046   & 0     & 0.028  & 0     & 0.080   & 0      	& 0.027 	& -0.43 & -1.3   & -1.19  \\ \hline
0    & 0    & 0      & 0.408  & 0    & -0.427 & 0     & 0.801  & -0.027 & 0      & -0.080 & 0     & -0.028 & 0      & -0.046  	& 0 		& 0.43  & 1.3    & 1.19   \\ \hline
56.5  & 655  & -0.290  & 0     & 0.239 & 0     & 0.874  & 0     & 0     & 0.08    & 0     & -0.126 & 0     & -0.255  & 0     	& -0.083	& -0.12 & -0.75  & -0.64  \\ \hline
56.5  & 655  & 0      & 0.874  & 0    & 0.239  & 0     & -0.290 & 0.083  & 0      & 0.255  & 0     & 0.126  & 0      & -0.080 	& 0    	& 0.12  & 0.75   & 0.64   \\ \hline
59.7  & 693  & 0.464   & 0     & 0.783 & 0     & -0.077 &0      & 0     & -0.061  & 0     & -0.334 & 0     & 0.014 	& 0     	& 0.223 	& -0.15 & -0.54  & -0.87  \\ \hline
59.7  & 693  & 0      & -0.077 & 0    & 0.783  & 0     & 0.464  & -0.223 & 0      & -0.014 & 0     & 0.334  & 0    	& 0.061  	& 0    	& 0.15  & 0.54   & 0.87   \\ \hline
267.1 & 3099 & -0.016  & 0     & 0.035 & 0     & 0.018  & 0     & 0     & 0.237   & 0     & 0.510  & 0     & -0.313	& 0     	& 0.764 	& -0.11 & -0.48  & -2.21  \\ \hline
267.1 & 3099 & 0      & -0.018 & 0    & -0.035 & 0     & 0.016  & 0.764  & 0      & -0.313 & 0     & 0.510  & 0    	& 0.237  	& 0    	& 0.11  & 0.48   & 2.21   \\ \hline
300.4 & 3485 & 0      & 0.071  & 0    & 0.011  & 0     & 0.036  & -0.035& 0       & 0.218  & 0     & -0.247 & 0    	& 0.940  	& 0    	& -0.67 & -1.33  & -2.51  \\ \hline
300.4 & 3485 & -0.036  & 0     & -0.011& 0     & -0.071 & 0     & 0     & 0.940   & 0     & -0.247 & 0     & 0.218 	& 0     	& -0.035  	& 0.67  & 1.33   & 2.51   \\ \hline
471.8 & 5474 & 0.133   & 0     & 0.381 & 0     & 0.056  & 0     & 0     & 0.139   & 0     & 0.742  & 0     & 0.228 	& 0     	& -0.461	& -1.56 & -0.5   & -0.77  \\ \hline
471.8 & 5474 & 0      & -0.056 & 0    & -0.381 & 0     & -0.133 & -0.461 & 0      & 0.228  & 0     & 0.742  & 0    	& 0.139  	& 0    	& 1.56  & 0.5    & 0.77   \\ \hline
507.7 & 5890 & -0.201  & 0     & 0.013 & 0     & 0.233  & 0     & 0     &-0.170   & 0     & 0.017  & 0     & 0.855 	& 0     	& 0.381 	& -0.63 & -0.21  & -1.94  \\ \hline
507.7 & 5890 & 0      & -0.233 & 0    & -0.013 & 0     & 0.201  &  0.381 & 0      & 0.855  & 0     & 0.017  & 0    	&	-0.170 & 0    		& 0.63  & 0.21   & 1.94   \\ \hline
\end{tabular}
\caption{Crystal field scheme corrisponding to the crystal field parameters from reference \cite{Pikul2010}, expressed in the reference system with the quantization axis $z \parallel c$ ($x \parallel a$ and $y \parallel b$)\\
Crystal field parameters:\\
$A_{2}^{0}$:  	114\,meV (1322\,K), $A_{2}^{2}$: -109\,meV (-1265\,K), $A_{4}^{0}$: 63\,meV (727\,K), $A_{4}^{2}$:  114\,meV (1327\,K), $A_{4}^{4}$: -202\,meV (-2345\,K), $A_{6}^{0}$: -645\,meV (-7478\,K), $A_{6}^{2}$: -190\,meV (-2197\,K), $A_{6}^{4}$: 264\,meV (3060\,K), $A_{6}^{6}$: 193\,meV (2239\,K), $\zeta_{SO}$: 	 80.5\,meV (934\,K) \\}\label{T4b}
\end{sidewaystable*}
%


\end{document}